\documentclass[pra,onecolumn,superscriptaddress,aps,5pt]{revtex4}
\pdfoutput=1
\usepackage{amsfonts}
\usepackage{amssymb,latexsym,amsmath}
\usepackage{amsthm}
\usepackage{graphicx}
\usepackage{float}
\usepackage{dcolumn}
\usepackage{times}
\usepackage{subfigure}
\usepackage[toc,page,title,titletoc,header]{appendix}
\usepackage[usenames]{color}
\usepackage{ulem}
\usepackage{epstopdf}

\newcommand{\rmc}{\mathrm{c}}

\newcommand{\rmi}{\mathrm{i}}

\newcommand{\rmn}{\mathrm{n}}

\newcommand{\rmr}{\mathrm{r}}

\newcommand{\re}{\mathrm{Re}}

\newtheorem{theorem}{Theorem}[section]
\newtheorem{proposition}[theorem]{Proposition}
\newtheorem{remark}{Remark}

\begin{document}

\title{Existence, Stability and Dynamics
of Harmonically Trapped One-Dimensional Multi-Component Solitary Waves:
The Near-Linear Limit}

\author{H. Xu \thanks{%
Email: haitao@math.umass.edu}}
\affiliation{Department of Mathematics and Statistics, University of Massachusetts,
Amherst, MA 01003-4515, USA}

\author{P.G. Kevrekidis \thanks{%
Email: kevrekid@math.umass.edu}}
\affiliation{Department of Mathematics and Statistics, University of Massachusetts,
Amherst, MA 01003-4515, USA}

\author{T. Kapitula}
\affiliation{Department of Mathematics and Statistics,
 Calvin College,
        Grand Rapids, MI 49546}

\date{\today}

\begin{abstract}
In the present work, we consider a variety of two-component,
one-dimensional states in nonlinear Schr{\"o}dinger equations in the
presence of a parabolic trap, inspired by the atomic physics context of
Bose-Einstein condensates. The use of Lyapunov-Schmidt reduction methods
allows us to identify persistence criteria for the different families of
solutions which we classify as $(m,n)$, in accordance with the number of
nodes in each component. Upon developing the existence theory, we turn to a
stability analysis of the different configurations, using the Krein
signature and the Hamiltonian-Krein index as topological tools identifying
the number of potentially unstable eigendirections for each branch. A
systematic expansion of suitably reduced eigenvalue problems when
perturbing off of the linear limit permits us to obtain explicit
expressions for the eigenvalues of each of the states considered. Finally,
when the states are found to be unstable, typically by virtue of
Hamiltonian Hopf bifurcations, their dynamics is studied in order to
identify the nature of the respective instability. The dynamics is
generally found to lead to a vibrational evolution over long time scales.
\end{abstract}

\maketitle

\section{Introduction}

Models of the nonlinear Schr{\"o}dinger (NLS)
type~\cite{sulem,ablowitz,siambook}
have proven to be rather universal in describing envelope
nonlinear wave structures in dispersive media. Such
structures emerge in fields ranging from water waves
 and
nonlinear optics~\cite{kivshar} to plasmas~\cite{infeld}
and atomic Bose-Einstein
condensates (BECs)~\cite{emergent}.
A particularly interesting setting, recognized early
on (i.e., since the 1970's) in nonlinear optics in the context of
interaction of waves of different frequency
is that of multi-component NLS models~\cite{manakov}.
Among these, arguably, the most prototypical one
is the integrable~\cite{intman2} so-called Manakov model, which is
characterized by equal nonlinear
interactions within and across components.

Two decades after these initial developments within nonlinear
optics, a renewed interest has emerged for such multi-component
systems through the advent of ultra-cold atomic Bose-Einstein
condensates (BECs)~\cite{stringari,roman}. Numerous experiments
since then have focused on realizing such multi-component
BECs as mixtures of, e.g.,
different spin states of the same atom species
(so-called pseudo-spinor condensates)~\cite{Hall1998a,chap01:stamp}, or
different Zeeman sub-levels of the same hyperfine level
(spinor condensates)~\cite{Stenger1998a,kawueda,stampueda}.
A remarkable feature of these atomic systems when they pertain
to the same atomic species is that the so-called scattering lengths
controlling the inter-atomic interactions and hence effectively
the nonlinear prefactors are nearly equal within and across
components both in settings of, e.g., $^{87}$Rb and of $^{23}$Na.
This, in turn, translates in models well approximated by
the Manakov nonlinearity, enabling the experimental
realization not only of ground states, but also of numerous
solitonic excitations, most notably
of dark-bright solitons and their
variants~\cite{hamburg,pe1,pe2,pe3,azu,pe4,pe5}. These
developments have been recently summarized in a number of
reviews and books~\cite{siambook,emergent,revip}.

Our aim in the present work is to consider the harmonically trapped setting
of atomic Bose-Einstein condensates in the context of the multi-component
models discussed above. In earlier work, both a subset of the present
authors~\cite{todd1,todd2}, as well as other researchers~\cite{feder}
explored the use of analytical techniques in order to examine the existence
and stability of solutions in the vicinity of the well-understood linear
(quantum harmonic oscillator) limit of single-component models featuring one
atomic species. The relevant methods included, e.g., among others the use of
Lyapunov-Schmidt conditions for persistence of solutions near this limit, as
well as the use of the Krein signature and related topological index
tools~\cite{toddbook} to characterize the stability of the resulting
excitations. The topological tools are used to determine the potential number
of unstable directions associated with an excitation, while the analysis
allows us to determine which of the potential instabilities are realized.

Here, we extend such considerations to the  more involved
setting of two-component systems. While we will not do so in this paper, the
ideas presented herein can be used to consider the existence and spectral
stability of solutions to systems with three or more components. As in the
one-component setting, nonlinear states emanate (bifurcate) from a
corresponding linear state. In the one-component setting the linear state
corresponds to an eigenfunction with a specified number of nodes which is
associated with a simple eigenvalue. In the two-component case the
eigenvalues are semi-simple, and in the cases considered herein will be of
multiplicity two. At the linear level the first component will have $m$
nodes (i.e., we will denote the number of nodes
of that component by $m$),
while the second component will have $n$ nodes. The value of such
topological and analytical tools in uncovering the potential number of
unstable eigendirections of each such pair $(m,n)$ can be considerable in
shaping the expectation of the potential experimental observability of
different states.

It should be highlighted here that it does not escape us that
low atom numbers are more prone to effects of quantum fluctuations
potentially detrimental to the existence of the states
(although it is our understanding that the study of such effects
in multi-component systems is fairly limited). Nevertheless, our
argument is that the value of considerations such as the topological
ones presented herein is that they are of broader value in
uncovering potential eigendirections beyond the vicinity of
the linear limit and hence of relevance to regimes where the
states could be observable as described by the mean-field Manakov-like
limit discussed herein and as has been revealed
experimentally e.g. in~\cite{hamburg,pe1,pe2,pe3}. For instance, an intriguing example of a finding
that we present herein is that even the very robust (and experimentally
identified) dark-bright soliton not only possesses a potentially unstable
eigendirection, but this instability is realized provided that the
inter-component interaction is increased sufficiently (an experimentally
feasible scenario via the tuning of the inter-component
scattering length by means of  so-called
Feshbach resonances~\cite{siambook}).

Our presentation will be structured as follows. In section II, we will
briefly present the theoretical setup and the analysis of the
existence of the different solutions $(m,n)$. In section III, we will
present a general framework for considering the stability of these
states. In section IV, we catalogue the different possible
states $(m,n)$ with $0 \leq m,n \leq 2$. Finally, in section V
we summarize our findings and present our conclusions, as well
as a number of directions for future study.

\section{Theoretical setup and Existence Results}
We consider the following two-component system, bearing in mind
the setting of two hyperfine states of, e.g., $^{87}$Rb~\cite{revip}
\begin{eqnarray}
\label{eqn00_1}
i\partial_t {\phi}_1(x,t) &=& (-\frac{1}{2}\partial_{xx}+\frac{1}{2}\Omega^2 x^2-\mu_1){\phi}_1+ ( g_{11}|{\phi}_1|^2 + g_{12}|\phi_2|^2 )\phi_1, \\
\label{eqn00_2}
i\partial_t \phi_2(x,t) &=& (-\frac{1}{2}\partial_{xx}+\frac{1}{2}\Omega^2 x^2-\mu_2)\phi_2+ ( g_{21}|\phi_1|^2 + g_{22}|\phi_2|^2 )\phi_2
\end{eqnarray}
where $\phi_j\in \mathbb{C}$ is the mean-field wave-function of species
$j,\,g_{jk}\in \mathbb{R}^+$ with $g_{12} = g_{21}$, $\mu_j\in \mathbb{R}$
represents the chemical potential for species $j$, and parabolic trapping
potentials are considered here with the same trapping frequency $\Omega$ for
both species; $\Omega$ effectively represents the ratio of the trapping
strengths along the longitudinal (elongated) and transverse (strongly
trapped) directions.
Focusing on the wave functions such that $\int_{\mathbb{R}}|\phi_j(x)|^2 dx = O(\epsilon)$ where $\epsilon\ll 1$ (i.e., the small amplitude, near-linear
limit discussed in the previous section), we introduce the scaling $\phi_j = \epsilon^{1/2} \psi_j$ and obtain the following equations:
\begin{eqnarray}
\label{eqn0_1}
i\partial_t {\psi}_1(x,t) &=& f_1= (-\frac{1}{2}\partial_{xx}+\frac{1}{2}\Omega^2 x^2-\mu_1) {\psi}_1+\epsilon( g_{11}| {\psi}_1|^2 + g_{12}| {\psi}_2|^2 ) {\psi}_1, \\
\label{eqn0_2}
i\partial_t {\psi}_2(x,t) &=& f_2= (-\frac{1}{2}\partial_{xx}+\frac{1}{2}\Omega^2 x^2-\mu_2) {\psi}_2+\epsilon( g_{12}| {\psi}_1|^2 + g_{22}| {\psi}_2|^2 ) {\psi}_2
\end{eqnarray}
where $\int_{\mathbb{R}}| {\psi}_j(x)|^2 dx = O(1)$ now. Due to the gauge
invariance of the system, if $\{\psi_1, \psi_2\}$ are a solution, then
$\{\psi_1 e^{i\theta_1}, \psi_2 e^{i\theta_2} \}$ will also be a solution for
any real $\theta_1$ and $\theta_2$. In this paper we will focus on the existence
and spectral stability of real-valued steady-state solutions for $\psi_1$ and
$\psi_2$.


Set
\[
\boldsymbol{\psi}=\left(\begin{array}{c}\psi_1\\\psi_2\end{array}\right),\quad
\boldsymbol{f}=\left(\begin{array}{c}f_1\\f_2\end{array}\right),\quad
\boldsymbol{\mu}=\left(\begin{array}{c}\mu_1\\\mu_2\end{array}\right).
\]
We seek the stationary solutions
$\boldsymbol{\psi}(x,t)=\boldsymbol{\psi}(x)$ through the continuation of a
nontrivial solution for $\epsilon=0$. For the moment assume the asymptotic
expansions,
\begin{equation}\label{e:expand}
\boldsymbol{\psi}=\boldsymbol{\psi}^{(0)}+\epsilon\boldsymbol{\psi}^{(1)}+O(\epsilon^2),\,\,
\boldsymbol{\mu}=\boldsymbol{\mu}^{(0)}+\epsilon\boldsymbol{\mu}^{(1)}+O(\epsilon^2).
\end{equation}
These expansions will be verified through a Lyapunov-Schmidt reduction, which
requires a detailed understanding of the linearized problem associated with
(\ref{eqn0_1})-(\ref{eqn0_2})~\cite{nirenberg}. We linearize about the steady
state solution by taking the Fr\'{e}chet derivative of $\boldsymbol{f}(
{\boldsymbol{\psi}}, {\boldsymbol{\psi}}^*, \epsilon)$ and $\boldsymbol{f^*}(
{\boldsymbol{\psi}}, {\boldsymbol{\psi}}^*, \epsilon)$ with respect to
${\boldsymbol{\psi}}$ and ${\boldsymbol{\psi}}^*$; star here stands for
complex conjugation. Let $\mathcal{L}$ denote the operator associated with
the linearization
having the asymptotic expansion
$\mathcal{L}=\mathcal{L}^{0)}+\epsilon\mathcal{L}^{(1)}+O(\epsilon^2)$, where
\begin{align}
\label{eqn0_L0}
\mathcal{L}^{(0)}=
\left(
    \begin{array}{cccc}
        -\frac{1}{2}\partial_{xx} + \frac{1}{2}\Omega^2 x^2 - \mu_1^{(0)} & 0 & 0 & 0\\
        0 & -\frac{1}{2}\partial_{xx} + \frac{1}{2}\Omega^2 x^2 - \mu_1^{(0)} & 0 & 0\\
        0 & 0 & -\frac{1}{2}\partial_{xx} + \frac{1}{2}\Omega^2 x^2 - \mu_2^{(0)} & 0\\
        0 & 0 & 0 & -\frac{1}{2}\partial_{xx} + \frac{1}{2}\Omega^2 x^2 - \mu_2^{(0)}
    \end{array}
\right),
\end{align}
and
\begin{align}
\label{eqn0_L}
\begin{split}
\mathcal{L}^{(1)}=
\left(
    \begin{array}{cccc}
        2g_{11}|{\psi}_1^{(0)}|^2+g_{12}|{\psi}_2^{(0)}|^2  & g_{11}({\psi}_1^{(0)})^2 & g_{12}{\psi}_1^{(0)} {\psi}_2^{(0)} & g_{12}{\psi}_1^{(0)} {\psi}_2^{(0)}\\
        g_{11}({\psi}_1^{(0)})^2 & 2g_{11}|{\psi}_1^{(0)}|^2+g_{12}|{\psi}_2^{(0)}|^2 & g_{12}{\psi}_1^{(0)}{\psi}_2^{(0)} & g_{12}{\psi}_1^{(0)}{\psi}_2^{(0)}\\
        g_{12}{\psi}_1^{(0)}{\psi}_2^{(0)} & g_{12}{\psi}_1^{(0)}{\psi}_2^{(0)} & 2g_{22}|\psi_2^{(0)}|^2+g_{21}|\psi_1^{(0)}|^2 & g_{22}({\psi}_2^{(0)})^2\\
        g_{12}{\psi}_1^{(0)}{\psi}_2^{(0)} & g_{12}{\psi}_1^{(0)}{\psi}_2^{(0)} & g_{22}({\psi}_2^{(0)})^2 & 2g_{22}|{\psi}_2^{(0)}|^2+g_{21}|{\psi}_1^{(0)}|^2
    \end{array}
\right).
\end{split}
\end{align}
%
%
Focusing on real solutions, we can directly see a symmetry of $\mathcal{L}$:
\begin{equation}\label{e:L-property}
J_1^T \mathcal{L} J_1 = \mathcal{L},\quad
J_1 =
\left(
    \begin{array}{cccc}
        0 & 1 & 0 & 0\\
        1 & 0 & 0 & 0\\
        0 & 0 & 0 & 1\\
        0 & 0 & 1 & 0
    \end{array}
\right).
\end{equation}
%
Using the linear eigenvalues of the quantum harmonic oscillator,
\[
\mu_1^{(0)}=\Omega(m+\frac{1}{2}),\quad
\mu_2^{(0)}=\Omega(n+\frac{1}{2}),
\]
it can be directly observed that
 $\mathcal{L}^{(0)}$ has a non-empty kernel spanned by $\{
(u_{m},0,0,0)^T , (0,u_{m},0,0)^T , (0,0,u_{n},0)^T, (0,0,0,u_{n})^T \}$,
where
\[
u_{k}(x) = \sqrt{\frac{1}{2^k k!
}}\left(\frac{\Omega}{\pi}\right)^{1/4} H_k(\sqrt{\Omega} x)e^{-\frac{\Omega
x^2}{2}},
\]
and $H_k$ are the Hermite polynomials. The first three states are
\[
u_0(x)=\left(\frac{\Omega}{\pi}\right)^{1/4} e^{-\Omega x^2/2},\,\,
u_1(x)=\left(\frac{\Omega}{\pi}\right)^{1/4}\sqrt{2\Omega} x e^{-\Omega x^2/2},\,\,
u_2(x)=\left(\frac{\Omega}{\pi}\right)^{1/4}\sqrt{\frac{1}{2}}(2\Omega x^2-1) e^{-\Omega x^2/2}.
\]
The collection of states $\{u_0,u_1,u_2,\dots\}$ has the properties that:
\begin{enumerate}
\item $\displaystyle{\left(-\frac{1}{2}\partial_{xx}+\frac{1}{2}\Omega^2
    x^2\right)u_{j}=\Omega(j+\frac{1}{2})u_j}$ for each $j=0,1,2,\dots$
\item $u_j(x)$ has $j$ simple zeros for each $j=0,1,2,\dots$
\item the set is orthonormal under the inner product $\displaystyle{\langle
    g,h\rangle=\int_\mathbb{R}g(x)h(x)^*\,\mathrm{d}x}$
\item the set is a basis for $L^2(\mathbb{R})$.
\end{enumerate}

%

We now consider the existence problem. We apply the Lyapunov-Schmidt
Reduction Method to Eqs.~(\ref{eqn0_1})-(\ref{eqn0_2}) with
\[
\boldsymbol{\mu}\approx\left(\begin{array}{c}\Omega(m+1/2)\\\Omega(n+1/2)\end{array}\right).
\]
The state will hereafter be denoted as $(m,n)$. Since the vector field is
smooth, and the eigenvalues are semi-simple, the reduction guarantees that
both $\boldsymbol{\mu}$ and $\boldsymbol{\psi}$ will have an asymptotic
expansion in $\epsilon$ of~(\ref{e:expand}). 
Equations~(\ref{eqn0_1})--(\ref{eqn0_2}) at order $O(1)$ are
\begin{eqnarray}
\label{eqn1_1}
0 &=& (-\frac{1}{2}\partial_{xx}+\frac{1}{2}\Omega^2 x^2-\mu_1^{(0)}){\psi}_1^{(0)}, \\
\label{eqn1_2}
0 &=& (-\frac{1}{2}\partial_{xx}+\frac{1}{2}\Omega^2 x^2-\mu_2^{(0)}){\psi}_2^{(0)}.
\end{eqnarray}
The nontrivial solution is the expected one,
\[
\boldsymbol{\psi}^{(0)}=\left(\begin{array}{c}au_m\\bu_n\end{array}\right),\quad
\boldsymbol{\mu}^{(0)}=\left(\begin{array}{c}\Omega(m+1/2)\\\Omega(n+1/2)\end{array}\right),
\]
where $a,b\in\mathbb{R}$.
%

The next set of equations at $O(\epsilon)$ will provide the definitive values
that $a$ and $b$ must assume. Equations~(\ref{eqn0_1})--(\ref{eqn0_2}) at order
$O(\epsilon)$ are
\begin{eqnarray}
\label{eqn2_1}
0 &=& (-\frac{1}{2}\partial_{xx}+\frac{1}{2}\Omega^2 x^2-\mu_1^{(0)}){\psi}_1^{(1)} + ( g_{11}|{\psi}_1^{(0)}|^2 + g_{12}|{\psi}_2^{(0)}|^2 - \mu_1^{(1)} ){\psi}_1^{(0)} , \\
\label{eqn2_2}
0 &=& (-\frac{1}{2}\partial_{xx}+\frac{1}{2}\Omega^2 x^2-\mu_2^{(0)}){\psi}_2^{(1)} + ( g_{12}|{\psi}_1^{(0)}|^2 + g_{22}|{\psi}_2^{(0)}|^2 - \mu_2^{(1)} ){\psi}_2^{(0)}.
\end{eqnarray}
Solvability requires
\begin{eqnarray}
\label{eqn1_bifur}
0 &=& a( \mu_1^{(1)} - A g_{11}a^2 - B g_{12}b^2 ) , \\
\label{eqn2_bifur}
0 &=& b( \mu_2^{(1)} - B g_{12}a^2 - C g_{22}b^2 )
\end{eqnarray}
where
\[
A=A_m=\langle u_m^2,u_m^2\rangle,\quad B=B_{m,n}=\langle u_m^2,u_n^2\rangle,\quad C=C_n=\langle u_n^2,u_n^2\rangle.
\]
Note that $A,B,C>0$ are positive real numbers; in particular, for a few
values of the indices we have
\[
A_0=C_0=B_{0,0}=\sqrt{\frac{\Omega}{2\pi}},\,\,
A_1=C_1=B_{1,1}=\frac{3}{4}\sqrt{\frac{\Omega}{2\pi}},\,\,
A_2=C_2=B_{2,2}=\frac{41}{64}\sqrt{\frac{\Omega}{2\pi}},
\]
and
\[
B_{1,0}=B_{0,1}=\frac{1}{2}\sqrt{\frac{\Omega}{2\pi}},\,\,
B_{2,0}=B_{0,2}=\frac{3}{8}\sqrt{\frac{\Omega}{2\pi}},\,\,
B_{2,1}=B_{1,2}=\frac{7}{16}\sqrt{\frac{\Omega}{2\pi}}.
\]
Solving Eqs.~(\ref{eqn1_bifur})--(\ref{eqn2_bifur}) requires that for
nontrivial solutions the pair $(a, b)$ should satisfy one of the following:
\begin{enumerate}
\item $a=0$, and $\displaystyle{b^2=\frac{\mu_2^{(1)}}{C g_{22}}}$ for
    $\mu_2^{(1)}>0$
\item $b=0$, and $\displaystyle{a^2=\frac{\mu_1^{(1)}}{A g_{11}}}$ for
    $\mu_1^{(1)}> 0$
\item $a,b\neq 0$ and $\displaystyle{\left(
    \begin{array}{cc}
        A g_{11} & B g_{12}\\
        B g_{12} & C g_{22}
    \end{array}
\right)
\left(
    \begin{array}{cc}
        a^2\\
        b^2
    \end{array}
\right)=
\left(
    \begin{array}{cc}
        \mu_1^{(1)}\\
        \mu_2^{(1)}
    \end{array}
\right)}$.
\end{enumerate}
Both the first and second case correspond to effectively single-component
solutions, and will not be further considered in what follows except in a
parenthetical manner. Regarding the third case,
\[
AC g_{11}g_{22} - B^2 g_{12}^2\neq 0\,\,\leadsto\,\,
a^2=\frac{Cg_{22}\mu_1^{(1)}-Bg_{12}\mu_2^{(1)}}{AC
g_{11}g_{22} - B^2 g_{12}^2}>0 ,\quad
b^2=\frac{Ag_{11}\mu_2^{(1)}-Bg_{12}\mu_1^{(1)}}{AC g_{11}g_{22} - B^2
g_{12}^2}>0.
\]
On the other hand, if the coefficients $g_{11},g_{12},g_{22}$ are special
enough such that $A_m C_n g_{11}g_{22} - B_{m,n}^2 g_{12}^2=0$ for some
$(m,n)$, then the nontrivial two-component solutions for the state $(m,n)$
will exist only if
\[
\frac{\mu_1^{(1)}}{\mu_2^{(1)}}=\frac{A g_{11}}{B g_{12}}.
\]
Moreover, when those two-component solutions exist, $a$ and $b$ are not
uniquely determined by $\mu_1^{(1)}$ and $\mu_2^{(1)}$ as in the above but
there exists a family of available values for them. It is worthwhile to note
that this condition is reminiscent of the phase separation criterion between
two components and, in fact, coincides with the latter when
$m=n$~\cite{siambook,emergent}.
Nevertheless, we will not focus on this singular case here and in that
light,
in all that follows we
focus on the cases where $ab\neq 0$ and $AC g_{11}g_{22} - B^2 g_{12}^2\neq 0$, and without loss of generality assume $m\leq n$. 

\section{Spectral stability}


If $\boldsymbol{{\psi}}$ is a steady-state solution to the system
(\ref{eqn0_1})--(\ref{eqn0_2}), then we consider the perturbation ansatz
${\tilde{\psi}}_j(x,t)={\psi}_j+\delta(e^{\lambda t}v_j(x)+e^{ \lambda^* t}
w_j^*(x))$ of such a solution. After substituting
$\boldsymbol{{\tilde{\psi}}}$ back into the system and linearizing around the
solution $\boldsymbol{{\psi}}$, we obtain the eigenvalue problem,
\begin{eqnarray}
\label{eqn_ev}
J \mathcal{L} \boldsymbol{\xi} = i\lambda\boldsymbol{\xi}\quad\leadsto\quad
(-iJ)\mathcal{L} \boldsymbol{\xi} = \lambda\boldsymbol{\xi},
\end{eqnarray}
where $J={\rm diag}(1,-1,1,-1)$ and $\boldsymbol{\xi}=(v_1,w_1,v_2,w_2)^T$.
The operator $-iJ$ is skew-symmetric, and the operator $\mathcal{L}$ is
self-adjoint. Consequently, this is a Hamiltonian eigenvalue problem. Because
the solutions are purely real, an important consequence is that the
eigenvalues satisfy the four-fold symmetry, $\{\pm\lambda,\pm\lambda^*\}$, which can also be explicitly seen from (\ref{e:L-property}).
Moreover, because of the unbounded potential term $\Omega^2x^2/2$ in the
operator $\mathcal{L}$, the spectrum is purely discrete, and each eigenvalue
has finite geometric and algebraic multiplicity.

\subsection{The unperturbed spectrum}

The spectrum for small $\epsilon$ will be determined via a perturbation
expansion from the $\epsilon=0$ spectrum. Consequently, it is important to
first have a detailed description of the unperturbed spectrum. For a given
eigenvalue, $\lambda$, let $E_\lambda$ denote the corresponding eigenspace.
It is straightforward to infer that given the quantum harmonic oscillator
nature of its constituents, 
the eigenvalues of $(-iJ)\mathcal{L}^{(0)}$ are
$-i\ell\Omega,\,\ell\in\mathbb{Z}$. Due to the four-fold spectral symmetry we
can focus on the lower-half complex plane in the following. For each
nonnegative $\ell$ there are three possibilities:
\begin{enumerate}
\item if $\ell\leq m$, then for $\lambda^{(0)}=-i\ell\Omega$,
\[
    E_{-i\ell\Omega}=\mathrm{Span}\left\{\left(\begin{array}{c}u_{m+\ell}\\0\\0\\0\end{array}\right),
    \left(\begin{array}{c}0\\u_{m-\ell}\\0\\0\end{array}\right),
    \left(\begin{array}{c}0\\0\\u_{n+\ell}\\0\end{array}\right),
    \left(\begin{array}{c}0\\0\\0\\u_{n-\ell}\end{array}\right)\right\}
\]
\item if $m<\ell\leq n$, then for $\lambda^{(0)}=-i\ell\Omega$,
\[
E_{-i\ell\Omega}=\mathrm{Span}\left\{\left(\begin{array}{c}u_{m+\ell}\\0\\0\\0\end{array}\right),
    \left(\begin{array}{c}0\\0\\u_{n+\ell}\\0\end{array}\right),
    \left(\begin{array}{c}0\\0\\0\\u_{n-\ell}\end{array}\right)\right\}
\]
\item If $\ell> n$, then for $\lambda^{(0)}=-i\ell\Omega$,
\[
E_{-i\ell\Omega}=\mathrm{Span}\left\{\left(\begin{array}{c}u_{m+\ell}\\0\\0\\0\end{array}\right),
    \left(\begin{array}{c}0\\0\\u_{n+\ell}\\0\end{array}\right)\right\}.
\]
\end{enumerate}
The kernel has dimension four.

\subsection{Krein Signature and Hamiltonian-Krein Index}

The spectrum of $(-iJ)\mathcal{L}$ is completely known for the unperturbed
problem. In particular, it is purely imaginary, so that the unperturbed wave
is spectrally stable. Because of the four-fold symmetry, eigenvalues which
are simple will remain purely imaginary for small $\epsilon$. However, as we
see above the unperturbed eigenvalues are semi-simple, which implies that
some could gain a nontrivial real part upon perturbation. Our first goal is
to show via the Hamiltonian-Krein index (HKI) that all but a finite number of
the eigenvalues will remain purely imaginary under small perturbation.
Moreover, the index will precisely locate which among the infinitely many
eigenvalues can gain nonzero real part under perturbation. See
~\cite{toddbook} for a more detailed exposition of what follows.

For the operator $(-iJ)\mathcal{L}$ let $k_\rmr$ denote the total number of
real positive eigenvalues (counting multiplicity), and $k_\rmc$ the total
number of eigenvalues with positive real part and nonzero imaginary part
(counting multiplicity). Regarding the purely imaginary eigenvalues, let
$\lambda$ be a purely imaginary eigenvalue with finite multiplicity, and let
$E_\lambda$ denote the associated eigenspace. The negative Krein index
associated with $E_\lambda$ is
$k_\rmi^-(\lambda)=\rmn(\mathcal{L}|_{E_\lambda})$. Here $\rmn(S)$ denotes
the number of negative eigenvalues (counting multiplicity) associated with a
Hermitian matrix $S$, and $\mathcal{L}|_{E_\lambda}$ denotes the Hermitian
matrix induced by restricting $\mathcal{L}$ to operate on $E_\lambda$. If
$\lambda$ is a simple eigenvalue with associated eigenvector
$\boldsymbol{\xi}$, then $\rmn(\mathcal{L}|_{E_\lambda})=\rmn(\langle
\boldsymbol{\xi}, \mathcal{L}\boldsymbol{\xi} \rangle)$. The eigenvalue is
said to have positive Krein signature if $\rmn(\mathcal{L}|_{E_\lambda})=0$;
otherwise, it is said to have negative Krein signature. Let $k_\rmi^-$ denote
the total negative Krein index,
\[
k_\rmi^-=\sum_{\lambda\in\sigma(-(iJ)\mathcal{L})\cap i\mathbb{R}}k_\rmi^-(\lambda).
\]
The HKI is the sum of all three indices,
\[
K_{\mathrm{Ham}}=k_\rmr+k_\rmc+k_\rmi^-.
\]
Because of the four-fold eigenvalue symmetry, $k_\rmc$ and $k_\rmi^-$ will be
even integers. In particular, there will be precisely $k_\rmc/2$ eigenvalues
with positive real part and negative imaginary part, and $k_\rmi^-/2$ purely
imaginary eigenvalues with negative imaginary part and negative Krein index.

The negative Krein index can be easily computed for the unperturbed problem.
Again, we focus only on those eigenvalues with negative imaginary part. Using
the diagonal form of $\mathcal{L}^{(0)}$, and the bases for the spectral
subspaces given in the previous subsection, we find
\begin{enumerate}
\item if $\lambda=-i\ell\Omega$ with $0<\ell\le m$, then
    $k_\rmi^-(-i\ell\Omega)=2$
\item if $\lambda=-i\ell\Omega$ with $m<\ell\le n$, then
    $k_\rmi^-(-i\ell\Omega)=1$
\item if $\lambda=-i\ell\Omega$ with $n<\ell$, then
    $k_\rmi^-(-i\ell\Omega)=0$.
\end{enumerate}
The four-fold symmetry implies that the eigenvalues with positive imaginary
part satisfy $k_i^-(i \ell \Omega)=k_i^-(-i \ell \Omega)$ for any $l\in \mathbb{N}$.
Consequently, the total negative Krein index is
\[
k_\rmi^-=4m+2(n-m)=2(m+n),
\]
so the HKI for the unperturbed problem is
\[
K_{\mathrm{Ham}}=2(m+n).
\]
Half of these eigenvalues have negative imaginary part, and half have
positive imaginary part.

Since the index is integer-valued, for operators which depend continuously on
parameters it remains unchanged for small perturbations. This statement,
however, requires that no additional eigenvalues can be added into the mix
via a bifurcation from the origin. Recall that we consider only those waves
which are nontrivial in both components. The gauge symmetry implies that the
geometric multiplicity of the origin will always be minimally two, and the
Hamiltonian structure of the spectral problem means the algebraic
multiplicity will always then be minimally four. For the unperturbed problem
the algebraic multiplicity of the origin is precisely four. Since the origin
is isolated, this then implies that for small perturbations the multiplicity
will remain four. Consequently, we know that for small $\epsilon$,
\[
K_{\mathrm{Ham}}=2(m+n),
\]
and for those eigenvalues associated with the HKI having nonzero imaginary
part, half will have positive imaginary part, and half will have negative
imaginary part.

The HKI provides for an upper bound of the number of eigenvalues with
positive real part. In order to locate those eigenvalues with small positive
real part for the perturbed problem, we do a perturbation expansion. However,
it is not necessary for us to perform an expansion for each eigenvalue.
Purely imaginary eigenvalues can leave the imaginary axis only via the
collision of eigenvalues of opposite Krein signature. This implies that for
the perturbation expansion we only need to consider those eigenvalues for which
the induced matrix $\mathcal{L}^{(0)}|_{E_{-i\ell\Omega}}$ is indefinite.

Restricting to those eigenvalues with negative imaginary part, this means we
only have $\ell>0$. If $0<\ell\le m$, the facts that
$\mathrm{dim}[E_{-i\ell\Omega}]=4$ and $k_\rmi^-(-i\ell\Omega)=2$ imply
that at most two eigenvalues can be created with positive real part
(collision of a pair of eigenvalues with negative Krein signature with a pair
with positive Krein signature). If $m<\ell\le n$, the facts that
$\mathrm{dim}[E_{-i\ell\Omega}]=3$ and $k_\rmi^-(-i\ell\Omega)=1$ imply
that at most one eigenvalue can be created with positive real part (collision
of one eigenvalue with negative Krein signature with one with positive
Krein signature). Finally, if $\ell>n$ then the unperturbed eigenvalue has
positive Krein signature, and will consequently remain purely imaginary under
small perturbation. In conclusion, when performing the perturbation expansion
we need only start with those unperturbed eigenvalues with $0<\ell\le n$.

\subsection{Reduced Eigenvalue Problem}

Knowing the Hamiltonian-Krein index, we will examine the exact number of
eigenvalue pairs with nonzero growth rates (i.e., associated with
instabilities) by finding the leading-order correction to each eigenvalue.
Since the eigenvalues are semi-simple, and the underlying solution is smooth
in $\epsilon$, the eigenvalues and associated eigenfunctions have the
expansions,
\[
\lambda=\lambda^{(0)}+\epsilon \lambda^{(1)}+O(\epsilon^2),\quad
\boldsymbol{\xi}=\boldsymbol{\xi}^{(0)}+\epsilon \boldsymbol{\xi}^{(1)}+O(\epsilon^2).
\]
The $O(1)$ and $O(\epsilon)$ reductions of Eqn.~(\ref{eqn_ev}) as
\begin{eqnarray}
\label{eqn_ev0}
\left(\mathcal{L}^{(0)}-i\lambda^{(0)}J\right)\boldsymbol{\xi}^{(0)} &=& \boldsymbol{0}\\
\label{eqn_ev1}
\left(\mathcal{L}^{(0)}-i\lambda^{(0)}J\right)\boldsymbol{\xi}^{(1)}&=&
\left(i\lambda^{(1)}J-\mathcal{L}^{(1)}\right)\boldsymbol{\xi}^{(0)}.
\end{eqnarray}
Letting $\{\boldsymbol{\eta}_i\}$ be an orthonormal basis for
$E_{-i\ell\Omega}$, upon writing $\boldsymbol{\xi}^{(0)}=\sum
c_i\boldsymbol{\eta}_i$ the solvability condition for (\ref{eqn_ev1}) is the
reduced spectral problem,
\begin{eqnarray}
\label{eqn_ev_reduced}
M \boldsymbol{c} = i\lambda^{(1)}\boldsymbol{c}.
\end{eqnarray}
Here
\[
M_{j,k}=\langle \boldsymbol{\eta}_j, J\mathcal{L}^{(1)} \boldsymbol{\eta}_k  \rangle,
\]
where the inner-product on each component is the standard one for
$L^2(\mathbb{R})$.
Similar calculations have been presented in different examples (chiefly for
single component systems); see for one such example, e.g.,~\cite{todd3}. Note
that if the spectrum of $M$ is purely real, then to leading order the
eigenvalues will be purely imaginary. On the other hand, eigenvalues of $M$
which have nonzero imaginary part lead to an oscillatory instability for the
underlying wave.

%
%
%

For the expansion we need only consider those eigenvalues with
$\lambda^{(0)}=-i\ell\Omega$ for $0<\ell\le n$. The perturbed eigenvalues for
$\ell>n$ will remain purely imaginary. The size of the matrix $M$ will depend
upon the value of $\ell$; in particular, if $0<\ell\le m$, then
$M\in\mathcal{M}_{4\times 4}(\mathbb{R})$, while if $m<\ell\le n$, then
$M\in\mathcal{M}_{3\times 3}(\mathbb{R})$. Defining
\[
D_{p,q,r,s}:=\langle u_pu_q,u_ru_s\rangle,
\]
the explicit expression for $M$ is:
\begin{itemize}

\item [(a)] $\lambda^{(0)}=-i\ell\Omega$ with $0<\ell\leq m$, then $M=M_a$,
    where
    \[
    M_a=\left(
    \begin{array}{cc}
        \tilde{M}_{11} & \tilde{M}_{12}\\
        \tilde{M}_{21} & \tilde{M}_{22}
    \end{array}\right),
    \]
and the individual blocks are defined via
\[
\tilde{M}_{11}=\left(
    \begin{array}{cc}
        2 g_{11} a^2 B_{m, m+\ell}+g_{12}b^2 B_{n,m+\ell}-\mu_1^{(1)} & g_{11}a^2 D_{m,m,m+\ell,m-\ell} \\
        -g_{11}a^2 D_{m,m,m+\ell,m-\ell} & -(2 g_{11} a^2 B_{m, m-\ell}+g_{12}b^2 B_{n,m-\ell})+\mu_1^{(1)}
    \end{array}
\right),
\]
and
\[
\tilde{M}_{12}=g_{12}a b\left(
    \begin{array}{cc}
        D_{m,n,m+\ell,n+\ell} & D_{m,n,m+\ell,n-\ell} \\
        -D_{m,n,m-\ell,n+\ell} & -D_{m,n,m-\ell,n-\ell}
    \end{array}
\right),\quad\tilde{M}_{21}=g_{12}a b\left(
    \begin{array}{cc}
        D_{m,n,m+\ell,n+\ell} & D_{m,n,m-\ell,n+\ell} \\
        -D_{m,n,m+\ell,n-\ell} & -D_{m,n,m-\ell,n-\ell}
    \end{array}
\right),
\]
and
\[
\tilde{M}_{22}=\left(
    \begin{array}{cc}
        2 g_{22} b^2 B_{n, n+\ell}+g_{12}a^2 B_{m,n+\ell}-\mu_2^{(1)} & g_{22}b^2 D_{n,n,n+\ell,n-\ell} \\
        -g_{22}b^2 D_{n,n,n+\ell,n-\ell} & -(2 g_{22} b^2 B_{n, n-\ell}+g_{12}a^2 B_{m,n-\ell})+\mu_2^{(1)}
    \end{array}
\right)
\]
%
%
%
\item [(b)] If $\lambda^{(0)}=-i\ell\Omega$ where $m<\ell\leq n$, then
    $M=M_b$, and $M_b$ is simply the submatrix obtained from $M_a$ after
    removing the second row and second column.
%
%
\end{itemize}

%
Before continuing, we briefly comment on what the above perturbation
calculation says about the spectral stability of one-component solutions. If
$a=0$, then
\[
\tilde{M}_{11}= \left(
    \begin{array}{cc}
        g_{12}b^2 B_{m+\ell,n}-\mu_1^{(1)} & 0 \\
        0 & -g_{12}b^2 B_{m-\ell,n}+\mu_1^{(1)}
    \end{array}
\right),\quad
\tilde{M}_{22}=g_{22} b^2 \left(
    \begin{array}{cc}
        2 B_{n,n+\ell}-C & D_{n,n,n-\ell,n+\ell} \\
        -D_{n,n,n-\ell,n+\ell} & C-2 B_{n-\ell,n}
    \end{array}
\right),
\]
and $\tilde{M}_{12}=\tilde{M}_{21}=\boldsymbol{0}$. Thus, $M_a$ can have
complex eigenvalues only if $|B_{n-\ell,n}+B_{n,n+\ell}-C| < |D_{n,n,n-\ell,n+\ell}| $.
Similarly, when $b=0,\,M_a$ will have a complex spectrum only if
$|B_{m-\ell,m}+B_{m,m+\ell}-A| < |D_{m,m,m-\ell,m+\ell}|$.
%
%
When $a=0$, the condition for $M_b$ to have complex eigenvalues will be the same as that for $M_a$. If $b=0$, $M_b$ will simply become a diagonal matrix and always have a real spectrum.

Therefore, for a one-component solution where $(\psi_1^{(0)},\psi_2^{(0)})=(a
u_m, 0)$, its spectral stability can be examined by checking conditions
$|B_{m-\ell,m}+B_{m,m+\ell}-A_m| < |D_{m,m,m-\ell,m+\ell}|$ for $0<\ell\leq m$. Here we note
that this is the same stability result if we consider a single one-component
equation. For $m\leq 2$ we can directly check the stability conditions for
the one-component solutions to get:
\begin{enumerate}
\item if $m=0$, the solutions continued from $a u_0$ are spectrally stable
    for small $\epsilon$;

\item if $m=1$, it can be checked that
    $|B_{0,1}+B_{1,2}-A_1|=|1/2+7/16-3/4|
    > |\sqrt{2}/8|=|D_{0,1,1,2}|$, so the solutions continued from $a
    u_1$ are spectrally stable for small $\epsilon$;

\item if $m=2$, it can be checked that
    $|B_{1,2}+B_{2,3}-A_2|=|7/16+51/128-41/64|
    > |5\sqrt{6}/64|=|D_{1,2,2,3}|$ but
    $|B_{0,2}+B_{2,4}-A_2|=|3/8+329/1024-41/64| <
    |3\sqrt{6}/128|=|D_{0,2,2,4}|$, so the splitting of eigenvalues at
    $2\Omega$ will enter the complex plane and the solutions continued from
    $a u_2$ are spectrally unstable (these stability features are well
    known, e.g., from the work of~\cite{coles}).
\end{enumerate}
In fact, in Section~\ref{section:summary-stability} we show that, in general,
\[
\frac{|B_{m-1,m}+B_{m,m+1}-A_m|}{|D_{m,m,m-1,m+1}|} = \sqrt{\frac{m}{m+1}}
+\sqrt{\frac{m+1}{m}} >2,\quad m\geq 1.
\]
That is to say, the eigenvalues for one-component solutions near $-i\Omega$ will always stay on the imaginary axis, although this perturbation calculation itself doesn't rule out the possibility for other eigenvalues to enter the complex plane.

\section{Catalogue of Different $(m,n)$ Cases}
\label{section:stability-catologue}

We now use the theory of the previous section to compute the spectral
stability of various two-component solutions. In particular, we will assume
$0\le m\le n\le2$.
{In what follows, the different branches are presented for $\Omega=0.1$, although similar results have been obtained for other values of $\Omega$. In fact, the value of $\Omega$ does not have a significant bearing
on the agreement between analytical predictions and computational results (including in the more physically realistic case of $\Omega\ll 1$).}

\subsection{$(m,n)=(0,0)$}

For $\mu_1^{(0)}=\mu_2^{(0)}=\Omega/2$, we consider  the branches of
solutions continued from $({\psi}_1^{(0)}, {\psi}_2^{(0)})=(a u_{0}, b
u_{0})$, where
\[
a^2=\sqrt{\frac{2\pi}{\Omega}}\,
    \frac{g_{22}\mu_1^{(1)}-g_{12}\mu_2^{(1)}}{g_{11}g_{22}-g_{12}^2},\quad
b^2=\sqrt{\frac{2\pi}{\Omega}}\,
    \frac{g_{11}\mu_2^{(1)}-g_{12}\mu_1^{(1)}}{g_{11}g_{22}-g_{12}^2}.
\]
This state features the fundamental (ground state) waveform in both
components of the system. Since $K_{\mathrm{Ham}}=0$ when $\epsilon=0$, the
wave is spectrally (indeed, orbitally) stable for small $\epsilon$, and it
is not necessary to perform the perturbation calculation.

\subsection{$(m,n)=(0,1)$ (interchange all subscripts to obtain case $(1,0)$)}

If $\mu_1^{(0)}=\Omega/2$ and $\mu_2^{(0)}=3\Omega/2$, we consider the
continuation of $({\psi}_1^{(0)}, {\psi}_2^{(0)})=(a u_{0}, b u_{1})$, where
\[
a^2=\sqrt{\frac{2\pi}{\Omega}}\,\frac{3g_{22}\mu_1^{(1)}-2g_{12}\mu_2^{(1)}}{3g_{11}g_{22}-g_{12}^2},\quad
b^2=2\sqrt{\frac{2\pi}{\Omega}}\,\frac{2g_{11}\mu_2^{(1)}-g_{12}\mu_1^{(1)}}{3 g_{11}g_{22} - g_{12}^2},
\]
which corresponds to a ``dark-bright" configuration. This configuration has been extensively studied in experiments over the past decade, as has been
recently summarized e.g.
in~\cite{revip}.
%
%
%
%
%

Regarding the spectral stability we have $K_{\mathrm{Ham}}=2$, with the
dangerous eigenvalues at $\lambda^{(0)}=\pm i\Omega$. At most one eigenvalue
with positive real part will emerge from $-i\Omega$. For the perturbation
calculation we only need consider case (b), where
$M_b\in\mathcal{M}_{3\times3}(\mathbb{R})$ is
\begin{equation}\label{e:matrix-b-0-1}
M_b=\frac{1}{8}\sqrt{\frac{\Omega}{2\pi}}\left(
    \begin{array}{ccc}
        2 g_{12}b^2 & \sqrt{2}g_{12}a b & 4 g_{12}a b \\
        \sqrt{2} g_{12}a b  &  g_{22}b^2 - g_{12}a^2 & \sqrt{2} g_{22}b^2 \\
        -4 g_{12}a b & -\sqrt{2} g_{22}b^2 & -2 g_{22}b^2 - 4 g_{12}a^2
    \end{array}
    \right).
\end{equation}
Regarding the spectrum of $M_b$, we have the following proposition:
\begin{proposition}\label{proposition-0-1}
$M_b$ for $(m,n)=(0,1)$ (i.e. the matrix in (\ref{e:matrix-b-0-1})) has an eigenvalue zero with associated
eigenvector $(-a/b,-\sqrt{2},1)^\mathrm{T}$, and two other eigenvalues
\[
-\frac{1}{16}\sqrt{\frac{\Omega}{2\pi}}\left(5 a^2 g_{12} - 2 b^2 g_{12} +
    b^2 g_{22}\pm \sqrt{9 a^4 g_{12}^2 + 18 a^2 b^2 g_{12}(g_{22}-2
    g_{12})+b^4(2 g_{12}+g_{22})^2}\right).
\]
The eigenvalues of $M_b$ will have nonzero imaginary parts if and only if
$g_{12}>g_{22}$.
\end{proposition}

The presence of the zero eigenvalue is well-known to be associated with the
invariance of the condensate to dipolar oscillations with the frequency of
the trap $\Omega$, yielding the so-called Kohn mode in the spectrum with the
trap frequency (and hence vanishing perturbations off of the linear
limit)~\cite{stringari}.
%
%
This proposition can be verified via direct calculation and in
Section~\ref{section:summary-stability} we will state more general results.
It is intriguing that the expression for the nonzero eigenvalues for $M_b$
here does not include $g_{11}$. This is due to the fact that
$A_0-2B_{0,1}=0$. As stated in Proposition~\ref{proposition-0-1}, $M_b$ will
have eigenvalues with nonzero imaginary parts --leading to an instability--
if $g_{12}>g_{22}$, i.e., the inter-component interactions have to be
stronger than the interactions within the ``dark'' species.

\begin{figure}[!htbp]
\centering
\begin{tabular}{ccc}
\includegraphics[width=5cm]{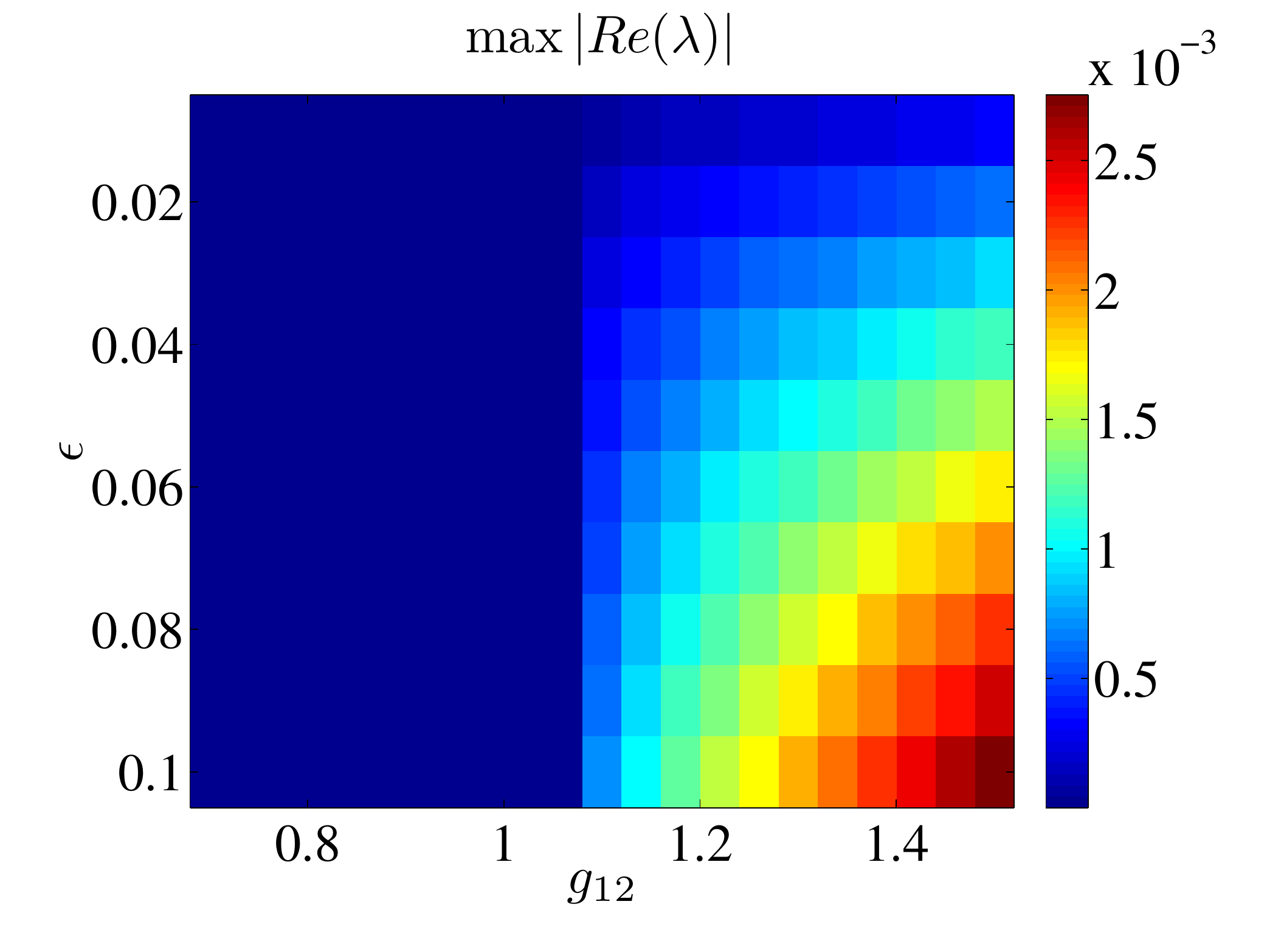}
\includegraphics[width=5cm]{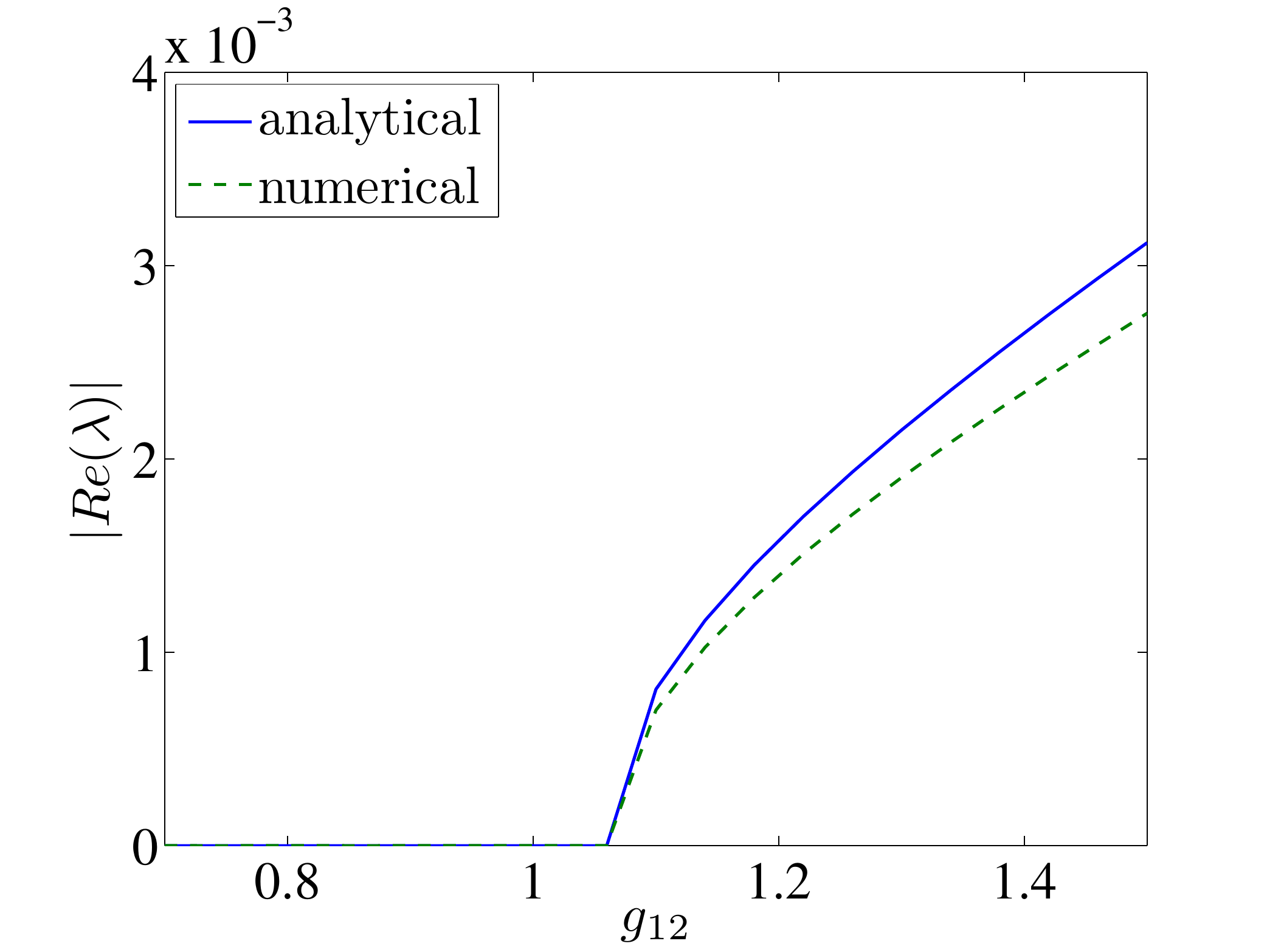}
\includegraphics[width=5cm]{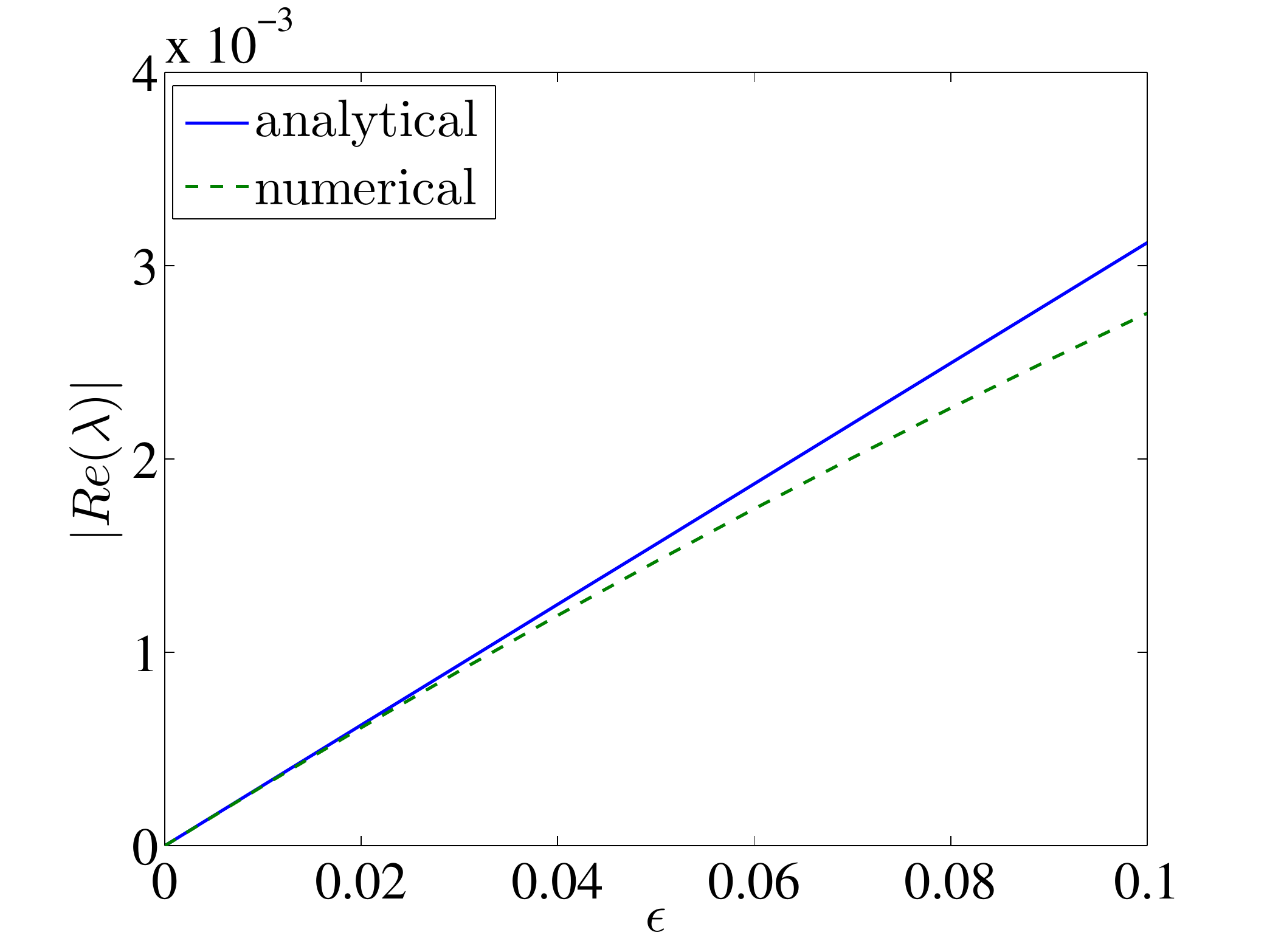}
\end{tabular}
\caption{Stability features of the $(0,1)$ branch:
in the left panel, we set $g_{22}=1.06, a=b=1$ and plot the maximal real part of numerically computed eigenvalues as a function of $g_{12}$ and $\epsilon$. In the middle (right) panel, we fix $\epsilon=0.1$ ($g_{12}=1.5$) for the setups in the left panel and compare the numerical result with analytical prediction $\epsilon|\re(\lambda^{(1)})|$, as a function of
$g_{12}$ ($\epsilon$). We find very good agreement in the dependence
of the relevant eigenvalue.}
\label{fig1_2}
\end{figure}

As an example, when $a=b=1$ the growth rate is
\[
|\re(\lambda^{(1)})|=
\frac{1}{16}\sqrt{\frac{\Omega}{2\pi}(23g_{12}+g_{22})(g_{12}-g_{22})}\,I_{g_{12}>g_{22}}(g_{12}),
\]
where
\[
I_{g_{12}>g_{22}}(g_{12})=\begin{cases}
0,\quad&g_{12}\le g_{22}\\
1,\quad&g_{12}> g_{22}.
\end{cases}
\]
In Figure \ref{fig1_2}, a case  associated with this potential instability
scenario of the $(0,1)$ branch is shown. In particular, the maximal real part
of numerically computed eigenvalues from Eqn.~(\ref{eqn_ev}) is plotted with
respected to $g_{12}$ and $\epsilon$. We see that the numerical result is in
good agreement with with our prediction $\epsilon|\re(\lambda^{(1)})|$. It is
relevant to indicate that in the integrable limit of $g_{ij}=1$, this
instability does not manifest itself, but it should be observable in systems
away from this limit provided that the first excited (dark) state is
initialized in the ``wrong'' component i.e., the one with intra-component
interactions $g_{22}<g_{12}$, while the fundamental state is initialized in
the component with $g_{11}$.

%



\begin{figure}[!htbp]
\begin{tabular}{ccc}
\includegraphics[width=5cm]{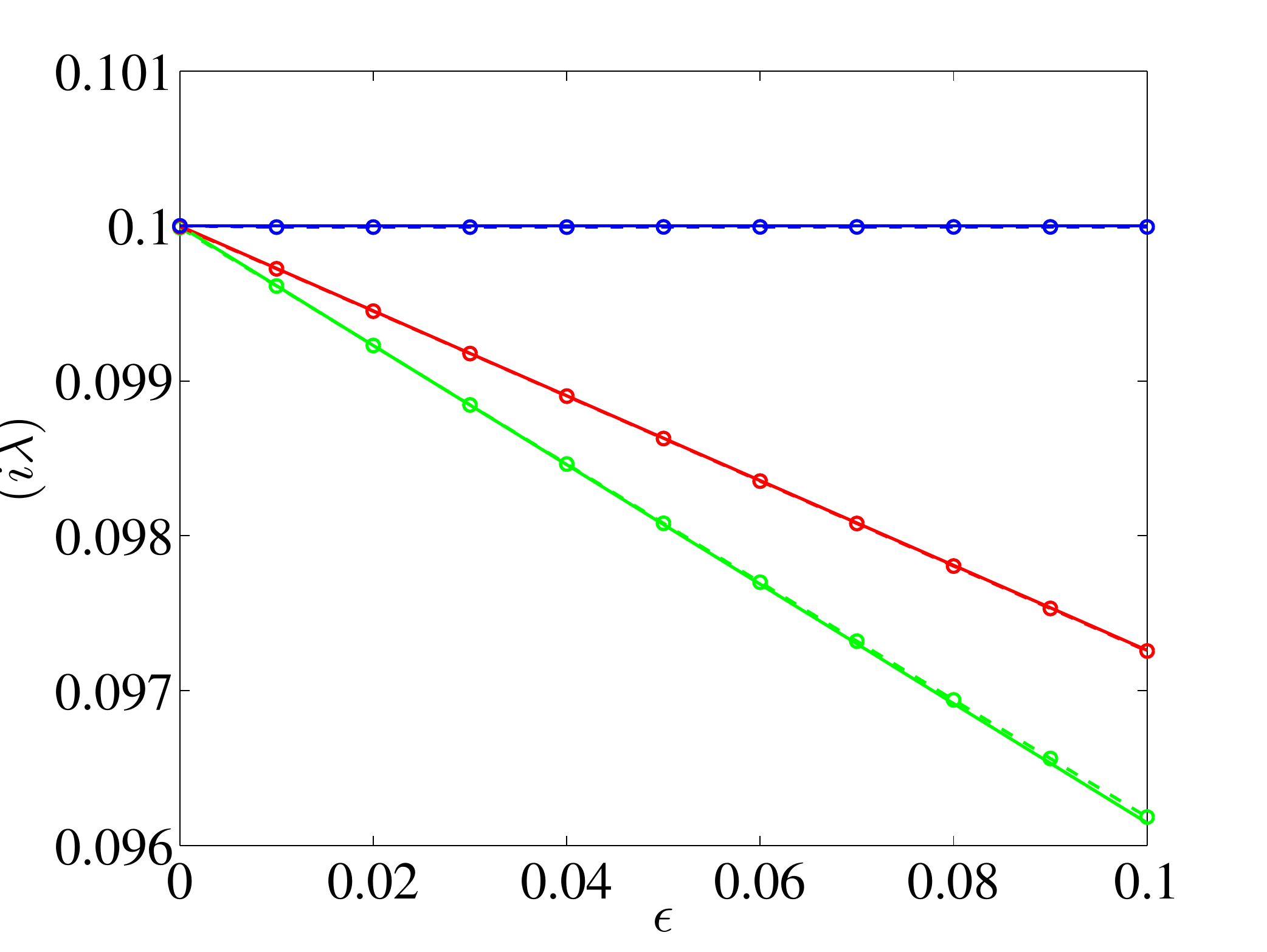}
\includegraphics[width=5cm]{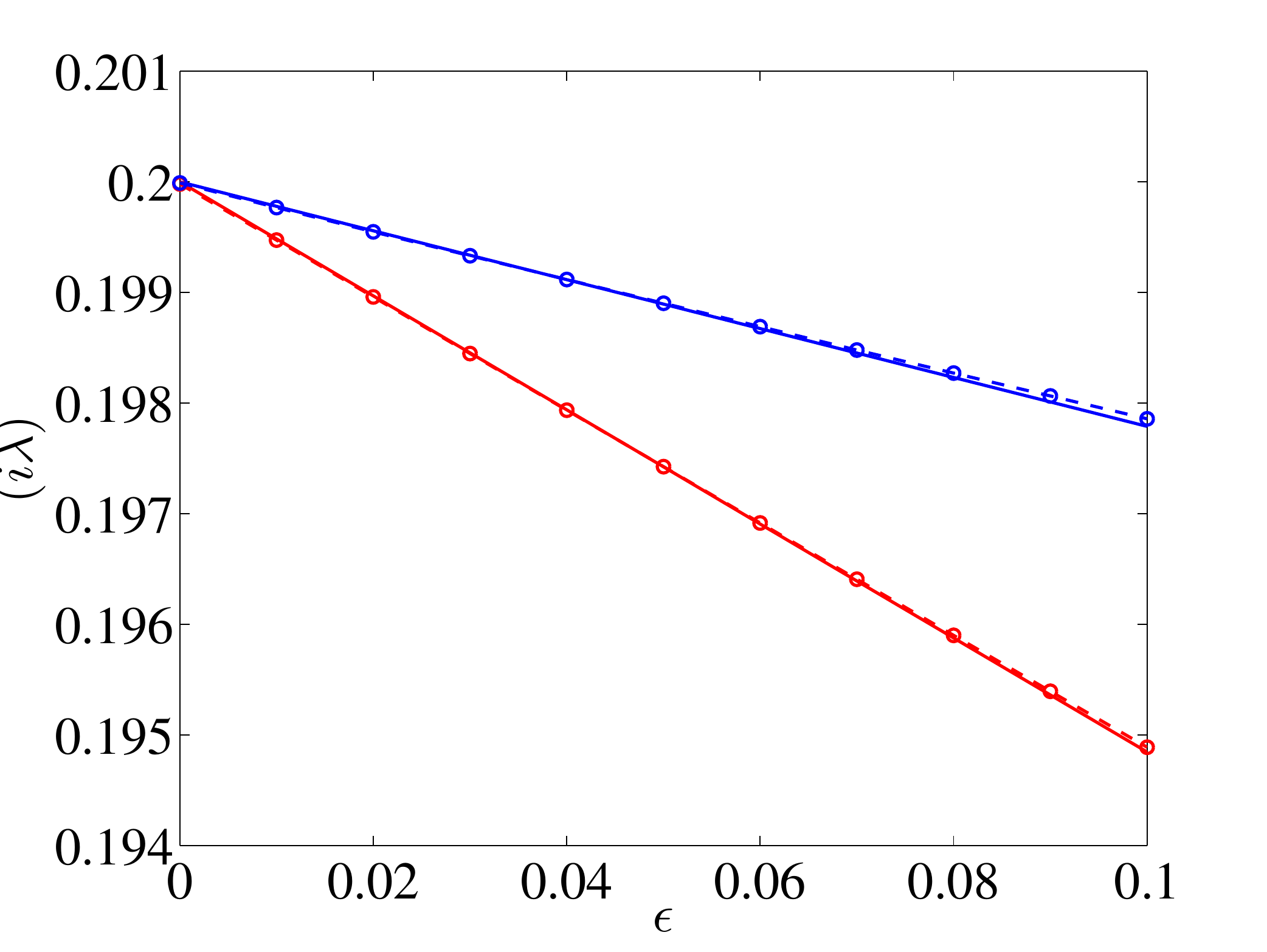}
\includegraphics[width=5cm]{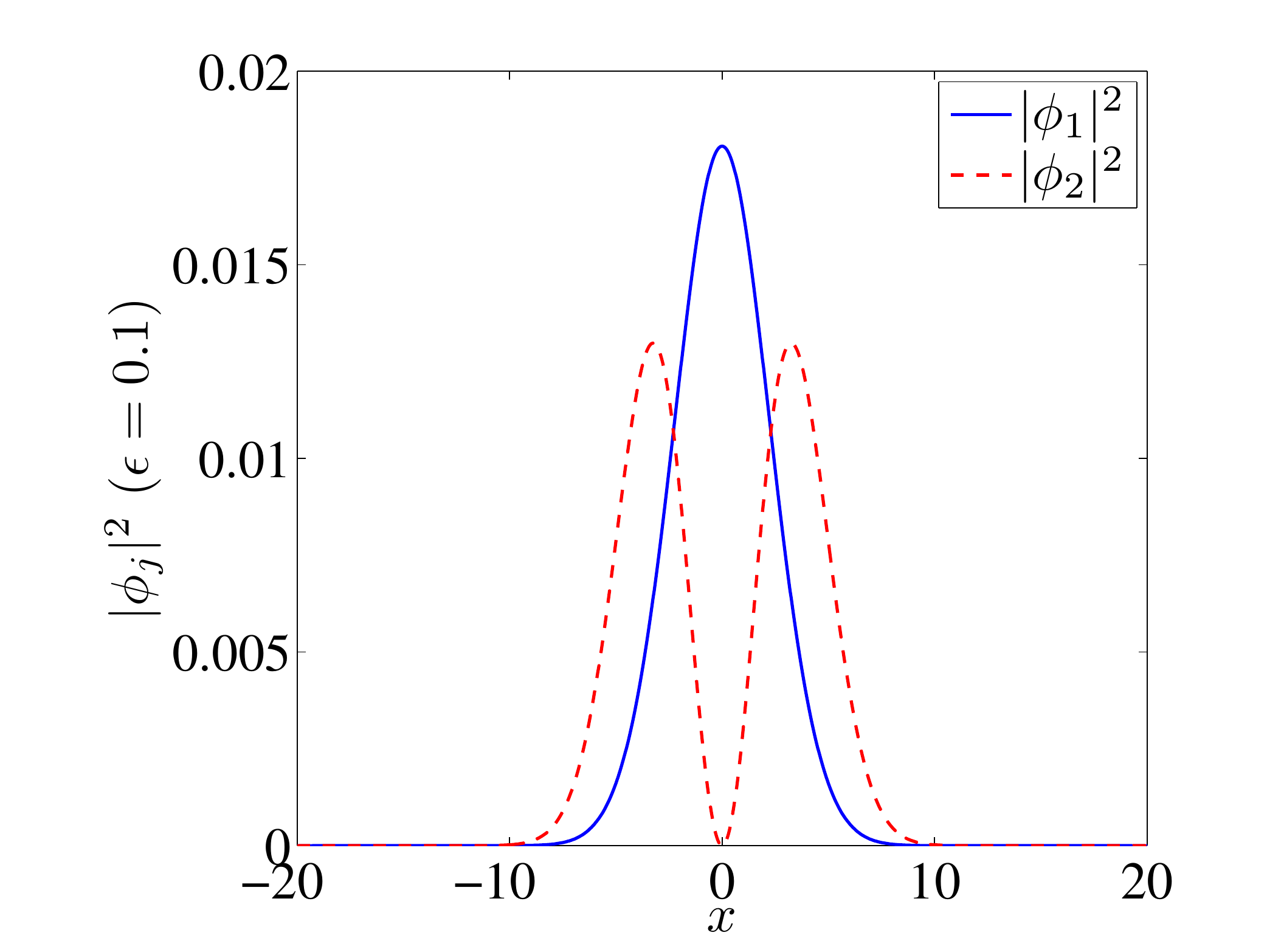}
\end{tabular}
\caption{The left (middle) panel shows the imaginary parts of the eigenvalues
around $-i\Omega$ ($-i2\Omega$) as functions of $\epsilon$ with $O(\epsilon)$ corrections
(solid lines) and corresponding numerical results (dashed lines with circles)
in the case of $(m,n)=(0,1)$, i.e., a prototypical example of a dark state with $n=1$ coupled to
a fundamental state of $m=0$. In the right panel, we show the densities of $\phi_1$ and $\phi_2$ at $\epsilon=0.1$.}
\label{fig1_2_1}
\end{figure}

\begin{figure}[!htbp]
\begin{tabular}{cc}
\includegraphics[width=7cm]{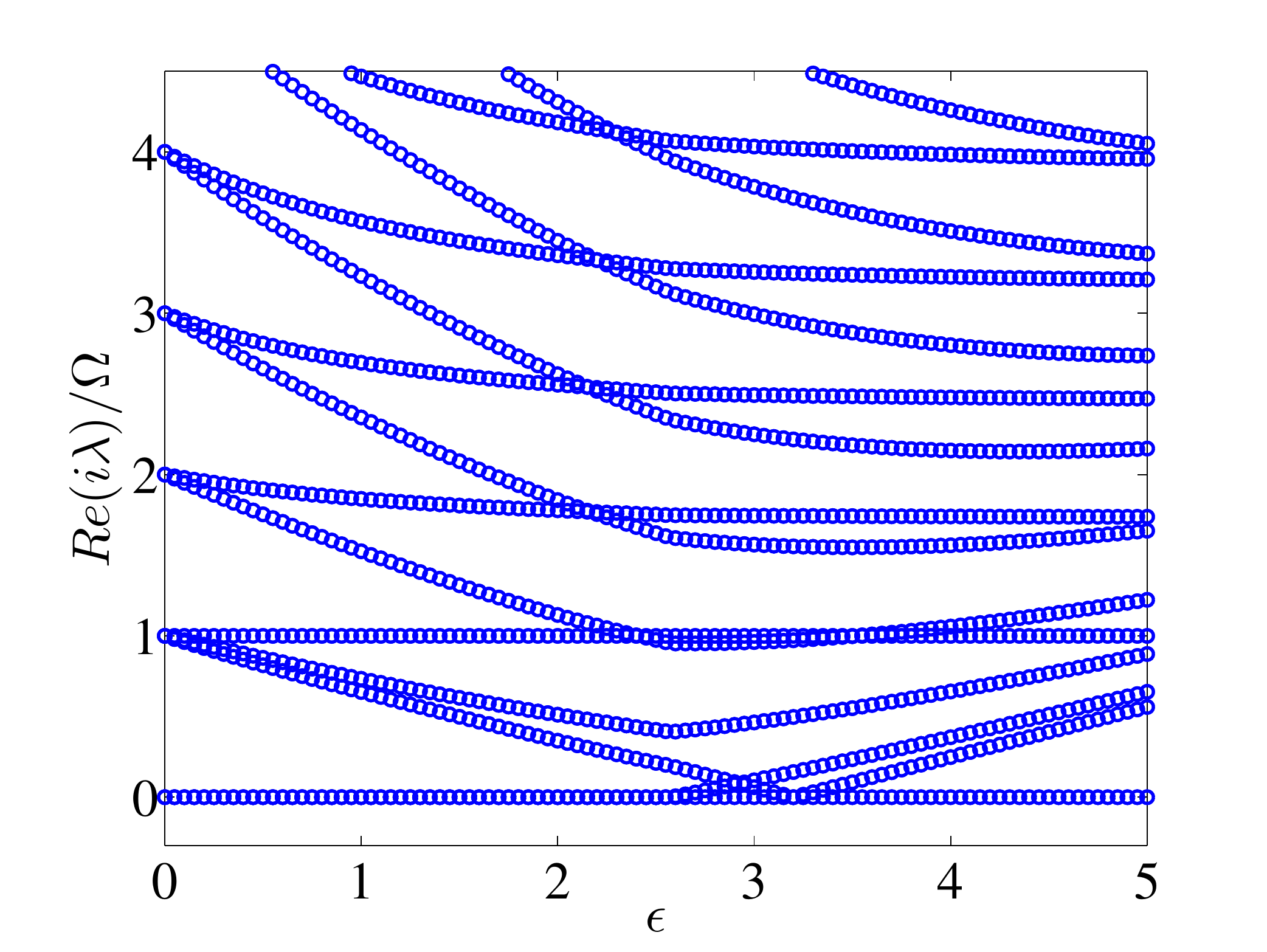}
\includegraphics[width=7cm]{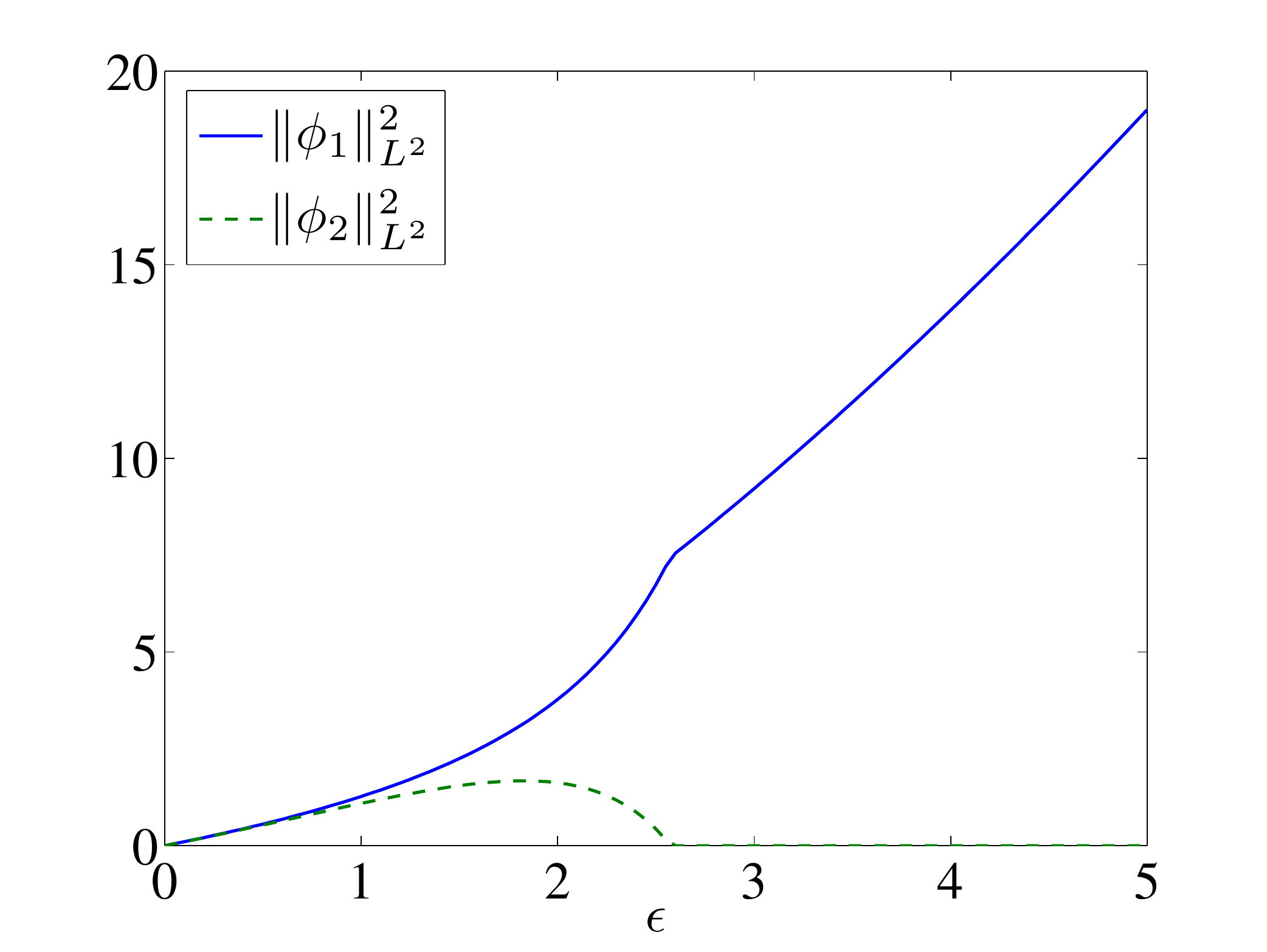}
\end{tabular}
\caption{In the left panel, we plot the imaginary parts of the eigenvalues over a wide interval of parametric variation of $\epsilon$. The right panel shows the $L^2$-norm of the corresponding solution of $\phi_1$ and $\phi_2$ for such interval of $\epsilon$.}
\label{fig1_2_1b}
\end{figure}

In Fig.~\ref{fig1_2_1}, we present the example with $a=b=1, g_{11}=1.03,
g_{12}=1.04, g_{22}=1.06, \Omega=0.1$, comparing eigenvalue predictions with
corrections up to $O(\epsilon)$ with corresponding numerical results.
We find that all of the
eigenvalues in the numerical computation are on the imaginary axis, which
matches our analytical prediction. According to our numerical computation, the spectrum will remain purely imaginary even when $\epsilon$ is large, which is shown in the left panel of Fig.~\ref{fig1_2_1b}. In addition, we find that $\phi_2$ becomes $0$ at
$\epsilon\approx 2.6$ where the branch of solutions meets the branch of
single-component solutions on $\phi_1$.

\begin{figure}[!htbp]
\begin{tabular}{ccc}
\includegraphics[width=5cm]{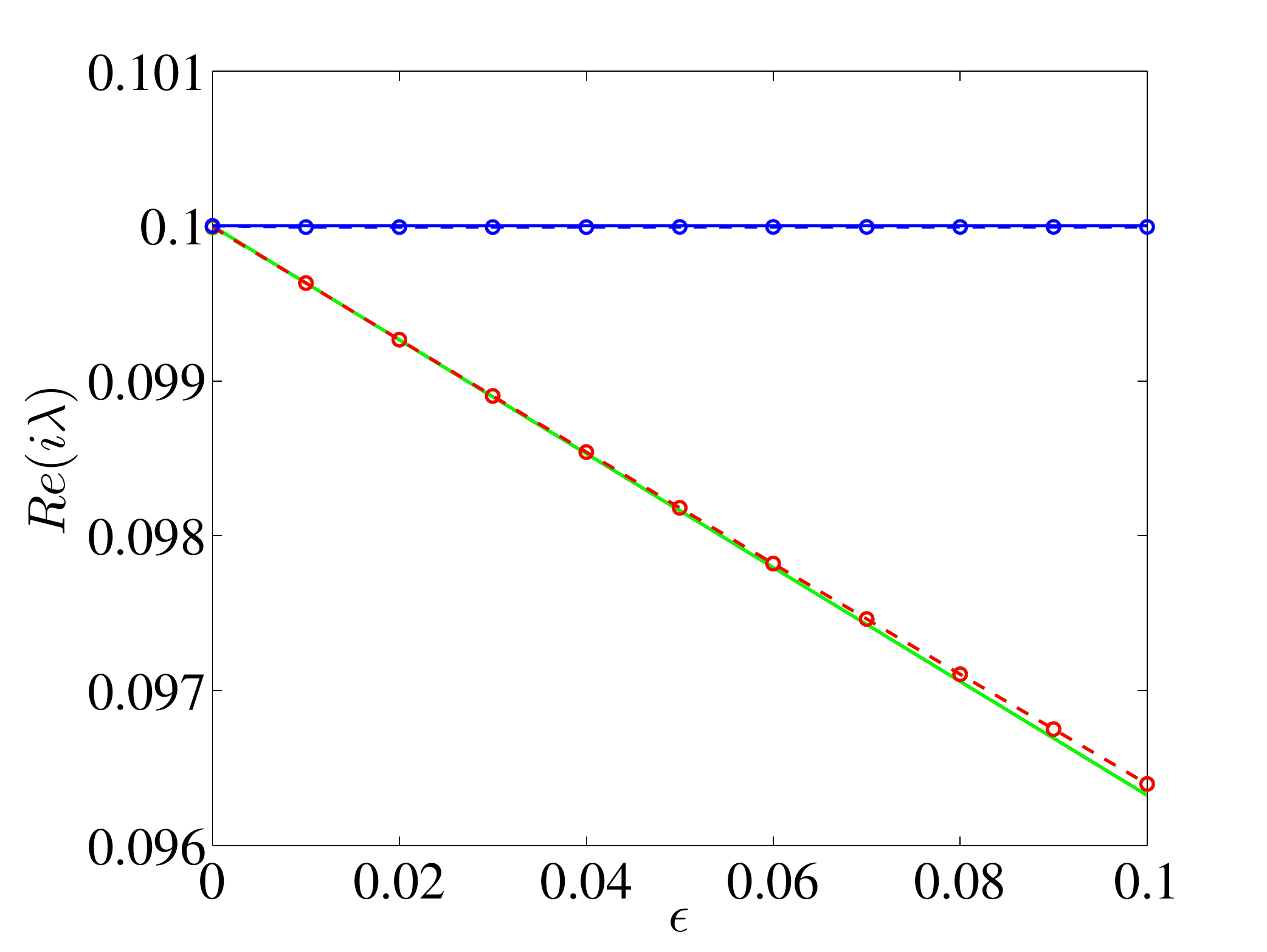}
\includegraphics[width=5cm]{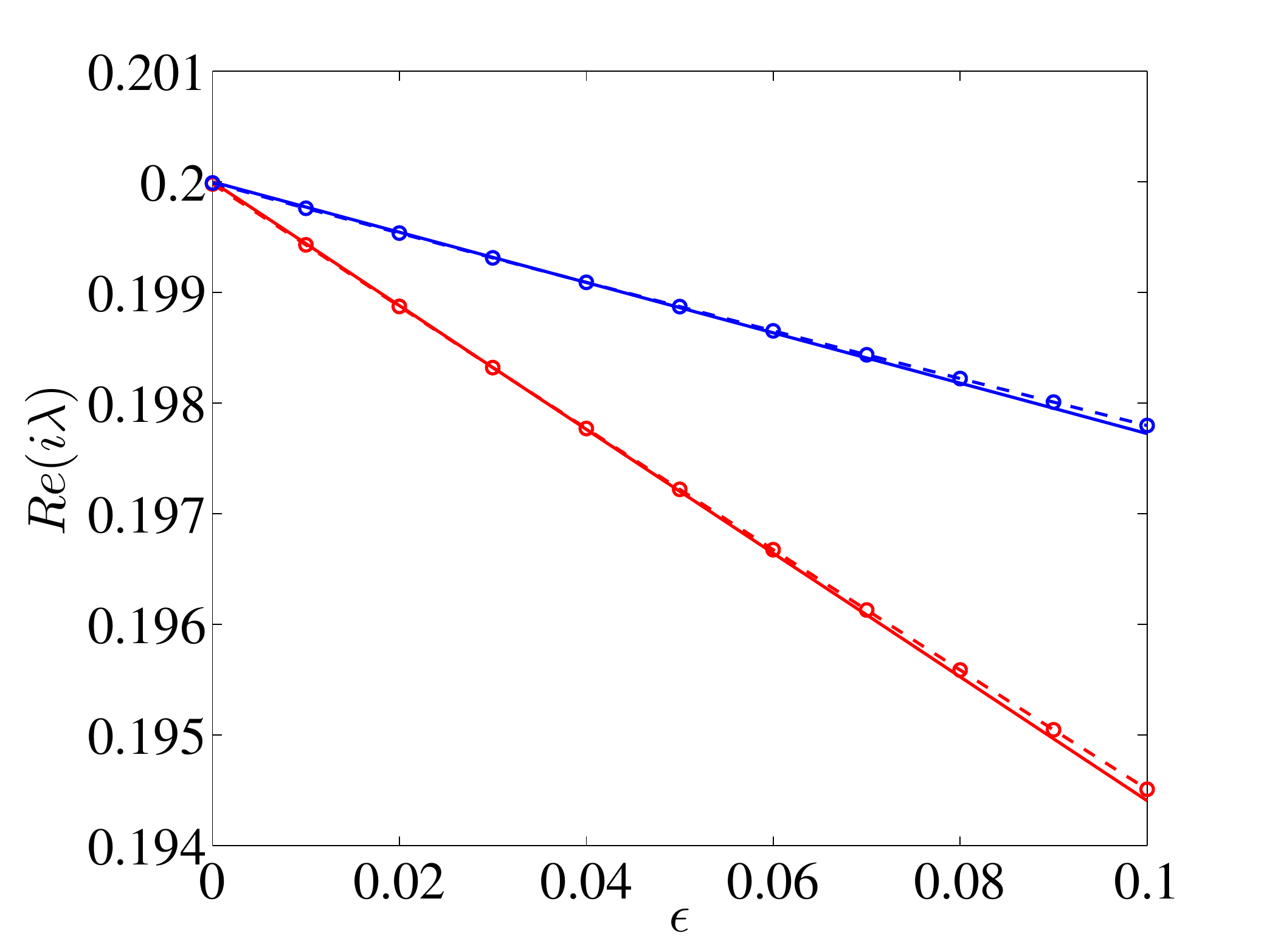}
\includegraphics[width=5cm]{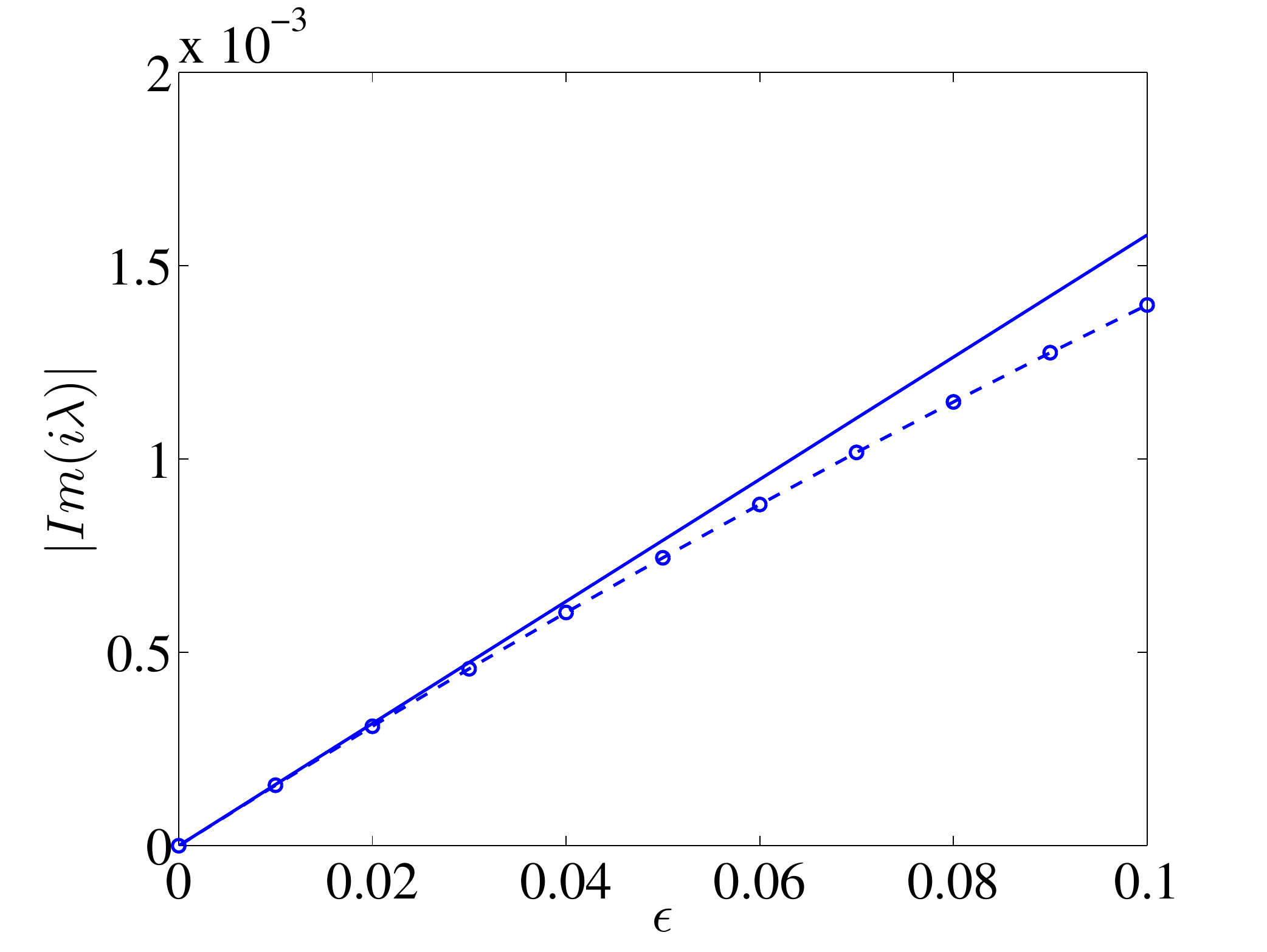}
\end{tabular}
\caption{ The left (middle) panel shows the imaginary parts of the eigenvalues around $-i\Omega$ ($-i2\Omega$) as functions of $\epsilon$ with $O(\epsilon)$ corrections (solid lines) and corresponding numerical results (dashed lines with circles). In the left panel, the red lines and green lines (solid and dashed) are
identical since a pair of eigenvalues of $M_b$ for $\lambda^{(0)}=-i\Omega$ are complex conjugates. Moreover, the nonzero imaginary parts
of $i \lambda$ for this pair imply the instability (this is the only source of the instability) of the solution, as shown in the right panel
(solid line for the $O(\epsilon)$ correction using this pair of complex conjugates and the dashed line for the numerical computation of the real parts of the
relevant eigenvalues).}
\label{fig1_2_2}
\end{figure}

\begin{figure}[!htbp]
\begin{tabular}{ccc}
\includegraphics[width=5cm]{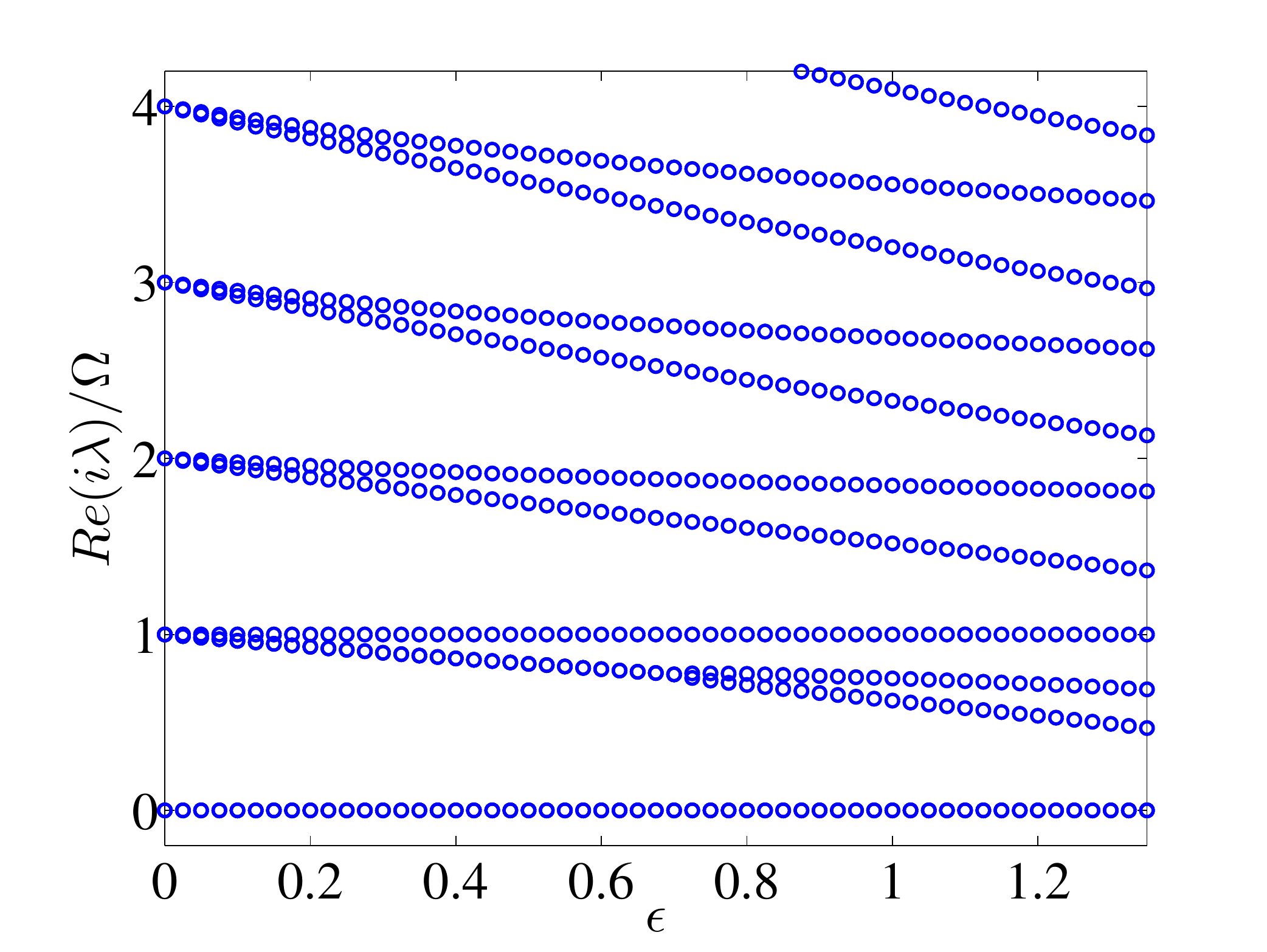}
\includegraphics[width=5cm]{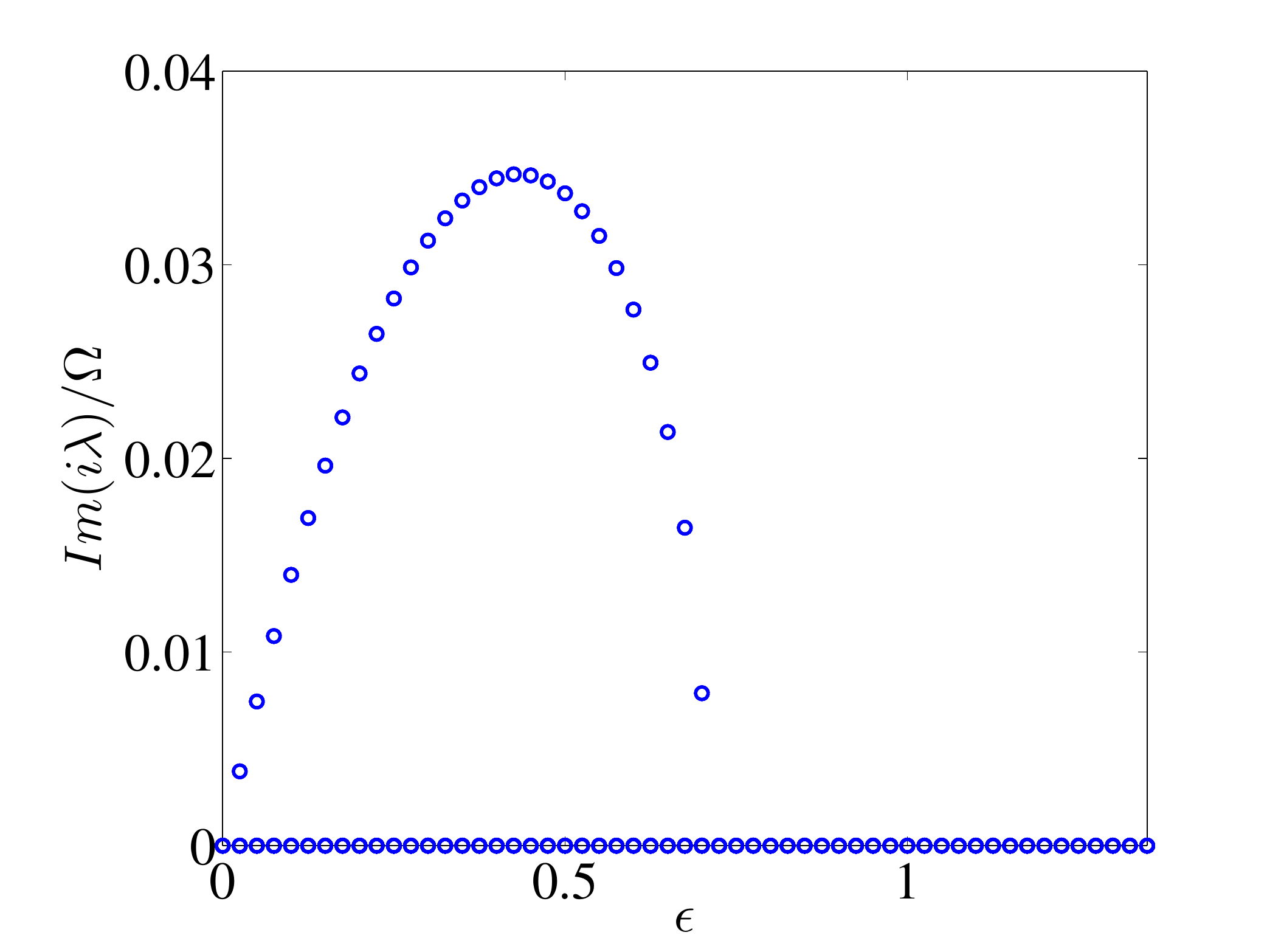}
\includegraphics[width=5cm]{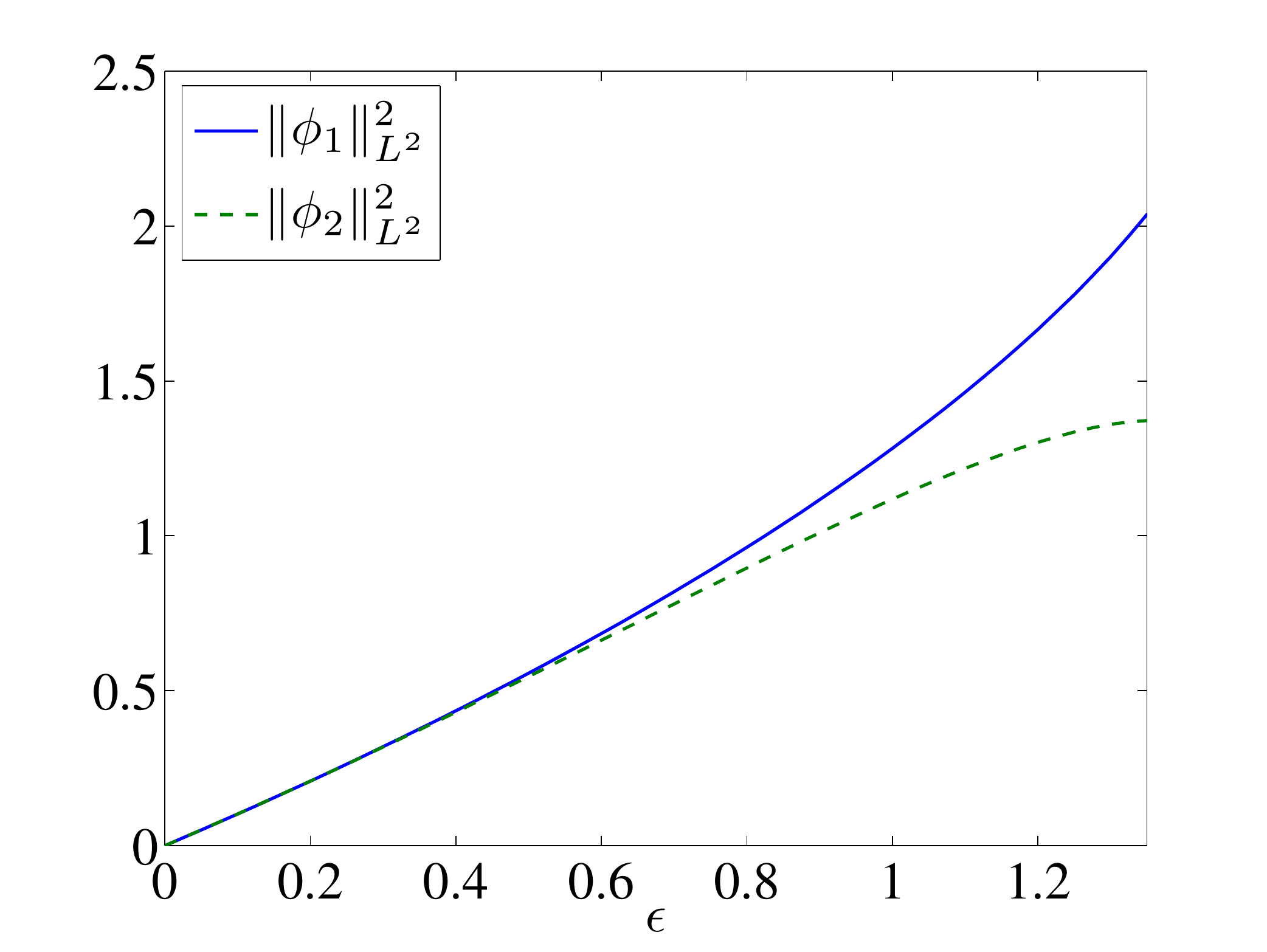}
\end{tabular}
\caption{ The left (middle) panel shows the variation of the
 imaginary (real) parts of the eigenvalues for the branch of solutions with $a=b=1$, $g_{11}=1.03$, $g_{12}=1.2$, $g_{22}=1.06$, $\Omega=0.5$. In the right panel, we plot the $L^2$-norm of the solution of $\phi_j$ over the same interval of $\epsilon$.}
\label{fig1_2_3}
\end{figure}

As another example, we consider a case that is ``immediately unstable''
in the vicinity of the linear limit. In particular,
if $a=b=1, g_{11}=1.03, g_{12}=1.2, g_{22}=1.06, \Omega=0.1$,
the numerical computation shows that all of the eigenvalues except a quartet
(near $\pm i\Omega$) are purely imaginary, as shown in Fig.~\ref{fig1_2_2}.
As $\epsilon$ increases, we notice that the quartet will finally come back to
the real axis at $\epsilon\approx 0.7$ and split along it, which is shown in
Figure \ref{fig1_2_3}.

%

\begin{figure}[!htbp]
\begin{tabular}{cc}
\includegraphics[width=7cm]{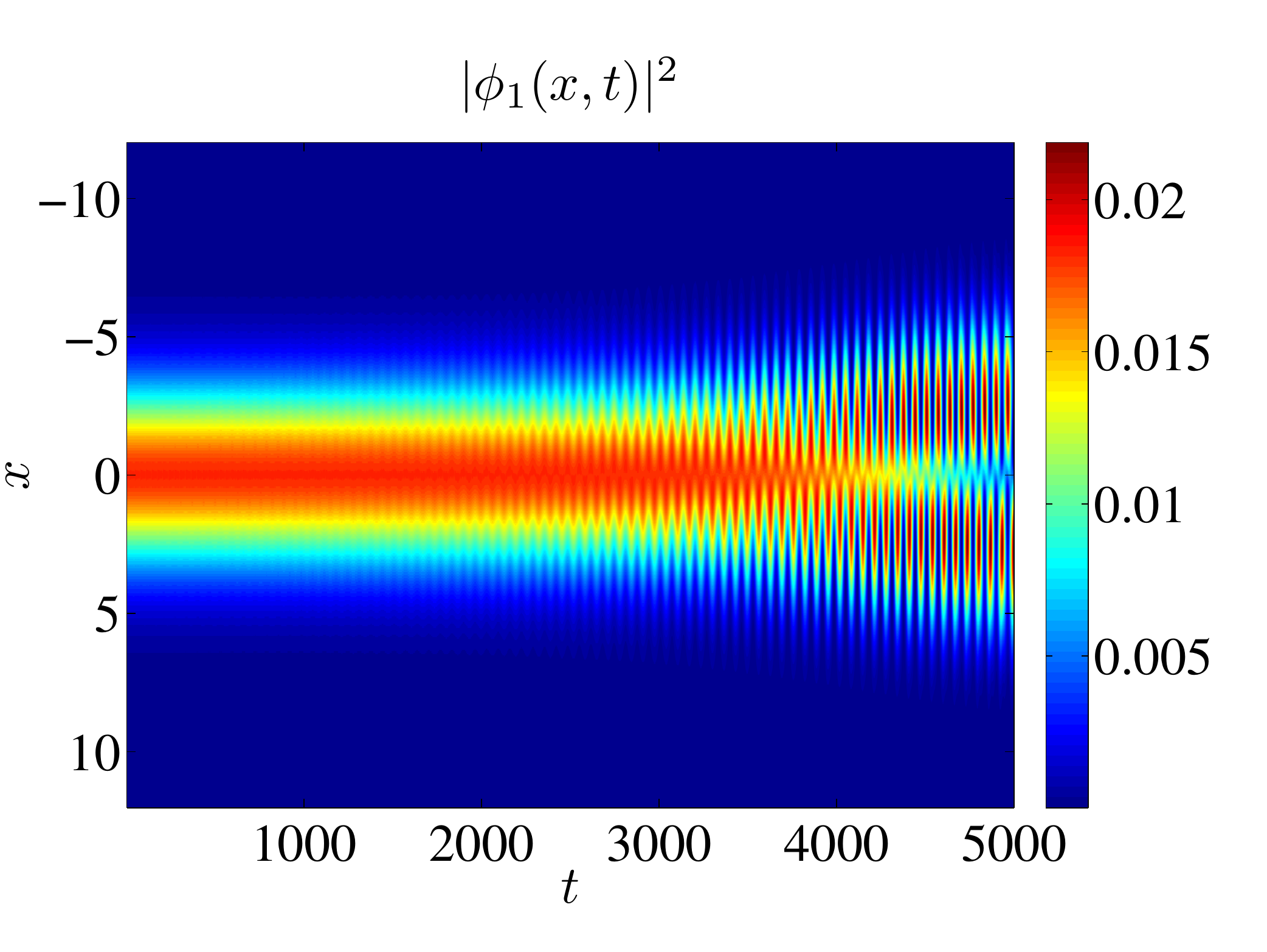}
\includegraphics[width=7cm]{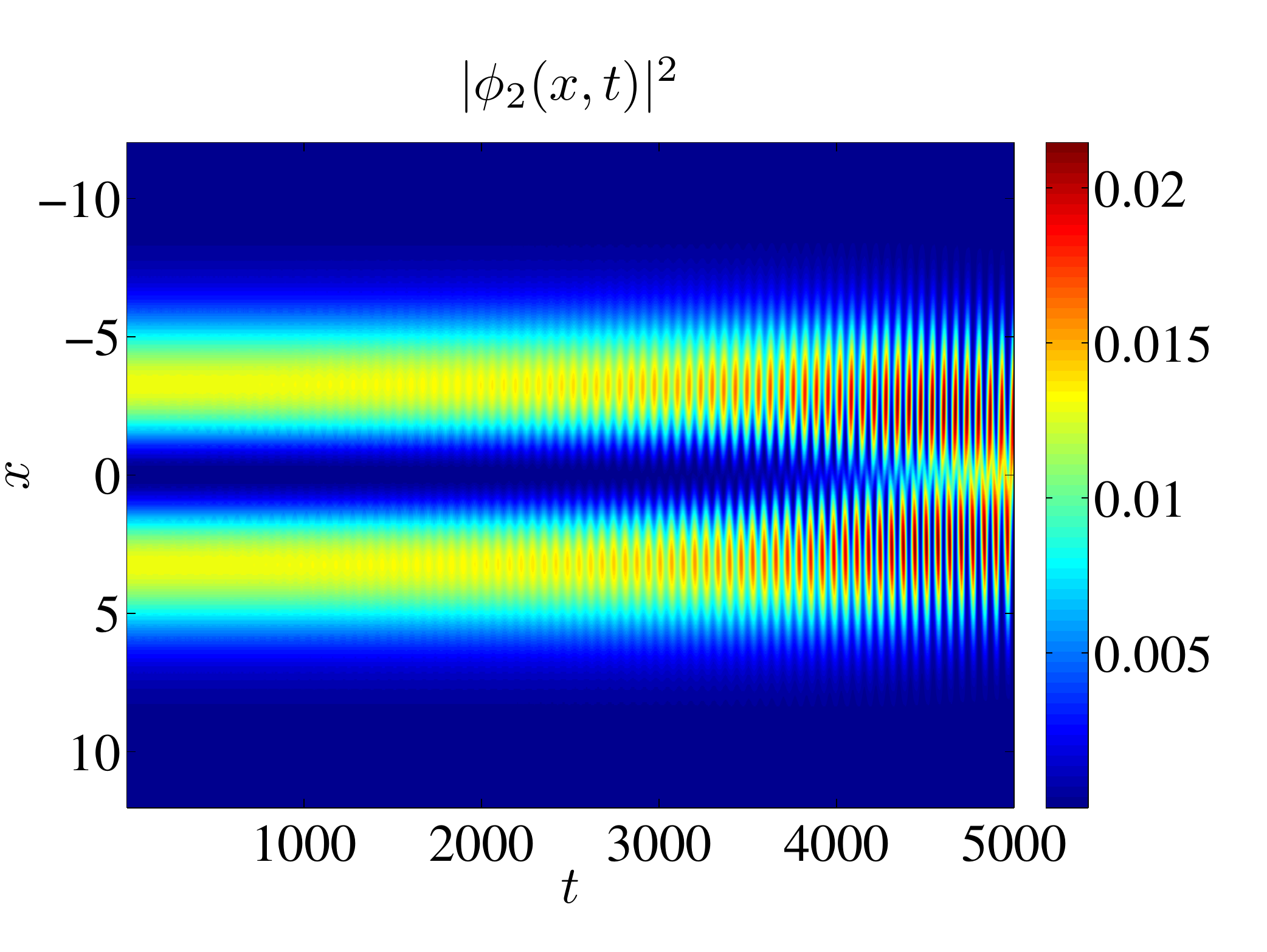}\\
\includegraphics[width=7cm]{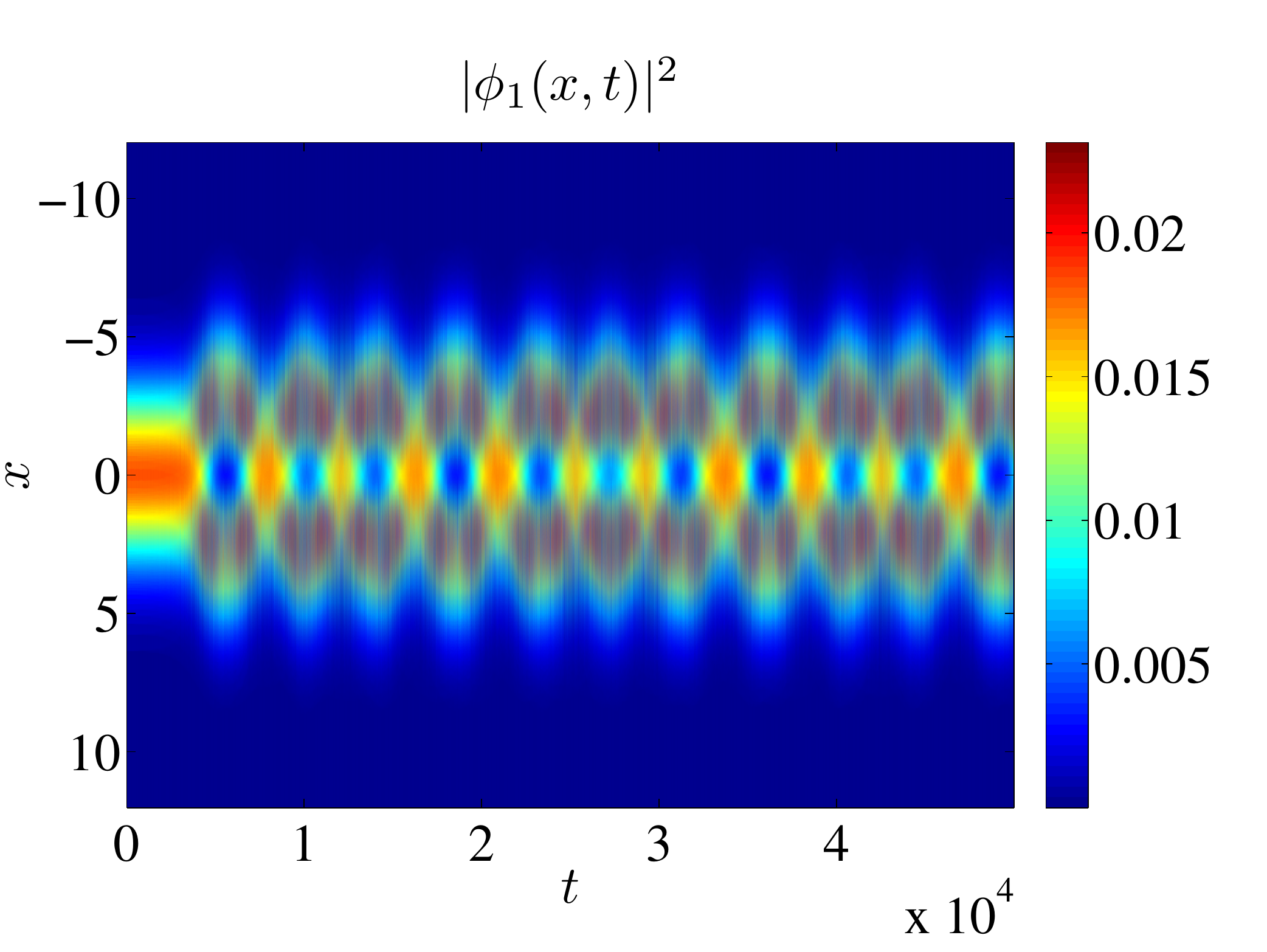}
\includegraphics[width=7cm]{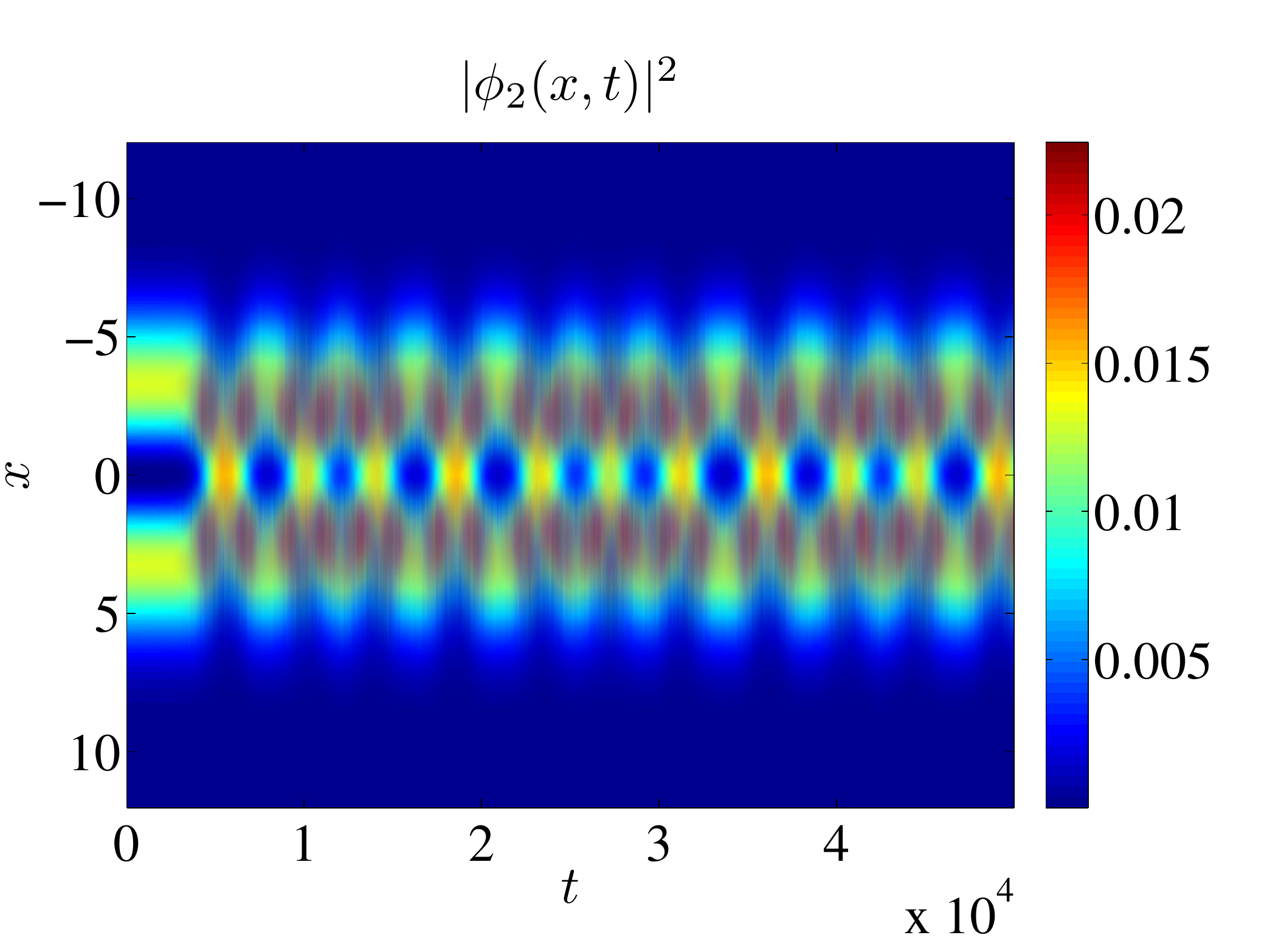}\\
\includegraphics[width=7cm]{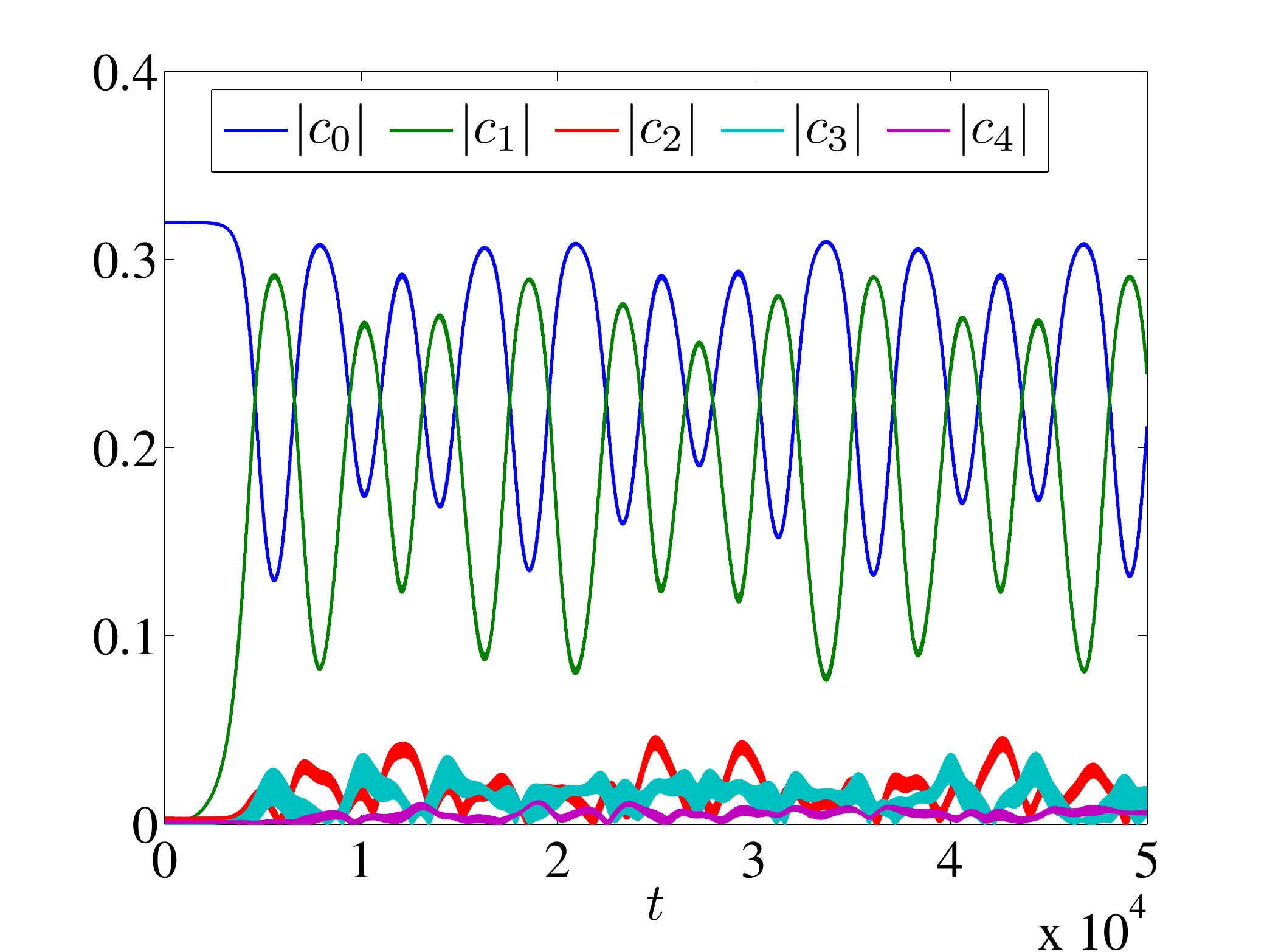}
\includegraphics[width=7cm]{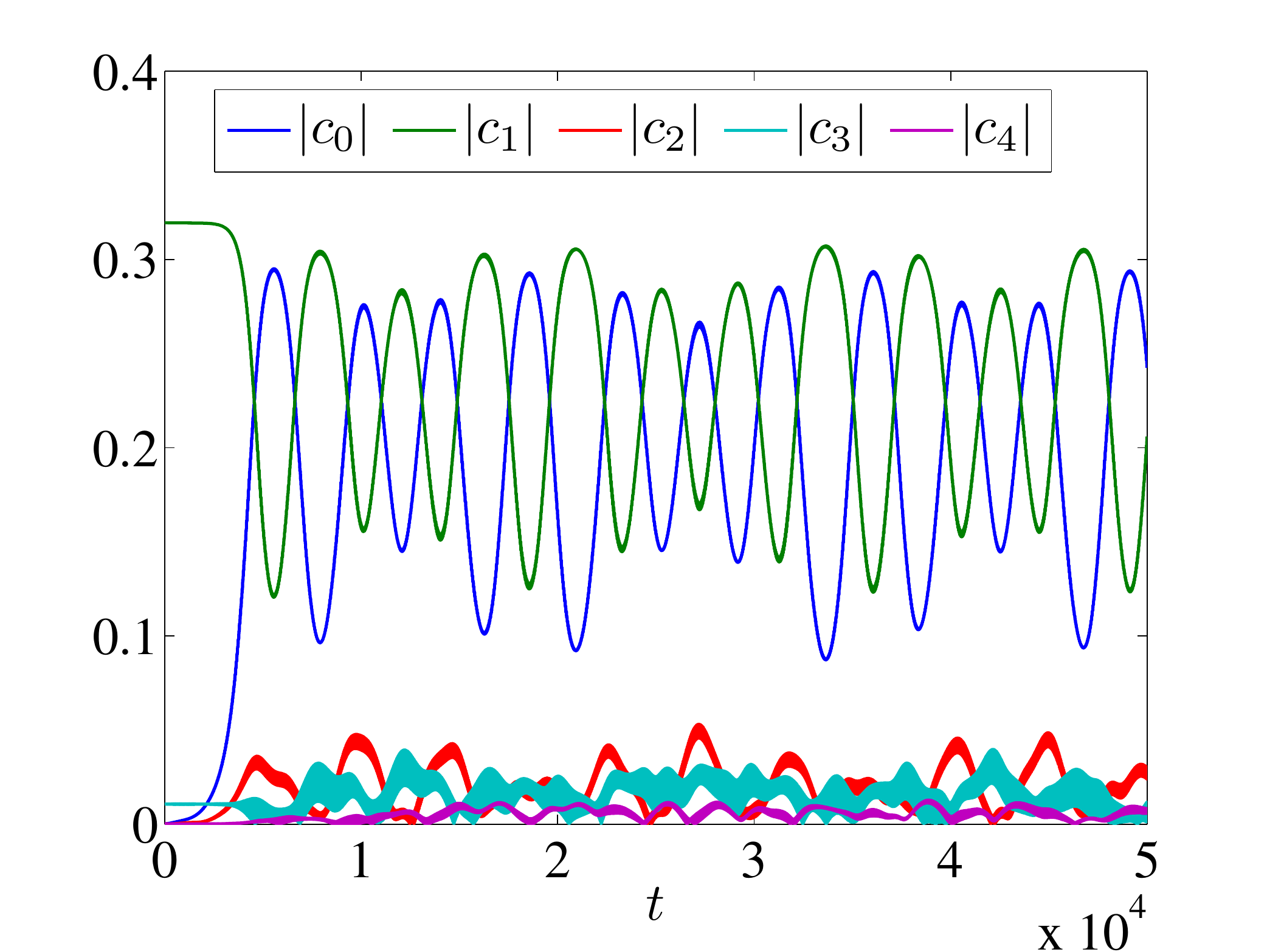}
\end{tabular}
\caption{ The top left (right) panel illustrates an example of the dynamics of $|\phi_1|^2$ ($|\phi_2|^2$) with $a=b=1$, $g_{11}=1.03$, $g_{12}=1.2$, $g_{22}=1.06$, $\Omega=0.1$ and $\epsilon=0.1$ for the $(0,1)$ state. The middle panels show the dynamics for the same setup for a longer time, revealing its oscillatory
nature,  while the bottom panels (left for $\phi_1$ and right for $\phi_2$) show the dynamics upon decomposition to the orthonormal basis $\{u_k\}$, where
the diagnostic $c_k$ used
stands for the coefficient (i.e., prefactor) of $u_k$ in the decomposition.}
\label{fig1_2_4}
\end{figure}

Figure~\ref{fig1_2_4} illustrates the numerical evolution of the unstable configuration shown in Figure \ref{fig1_2_2} with $\epsilon=0.1$. If a small initial perturbation is added to the solution, the development of the instability can be observed over intermediate time scales in Figure~\ref{fig1_2_4}.
To determine the fate of the solution under the action of this
instability, we have performed considerably longer simulations focusing
on the dynamics of the unstable waveform (see the middle panels
of Figure~\ref{fig1_2_4}).
{There we find an oscillatory pattern of the long-term dynamics of the solution,
featuring breathing (yet not genuinely periodic) recurrences over time. To be more specific, the bottom panels in  Figure~\ref{fig1_2_4}
reveal, through a dynamical
decomposition to the lowest order harmonic oscillator
modes, that the system does not stay at a certain state but quantitatively alternates between the states $(0,1)$ and $(1,0)$. We also observe that the instability of the state $(0,1)$ (similar for the state $(1,0)$) is essentially caused by the eigenmodes that are related to the unstable eigenvalues near $\pm i\Omega$, which is verified by the fact that both the time evolution of $|c_1|$ for $\phi_1$ and that of $|c_0|$ for $\phi_2$ bear small oscillations whose frequency is close to $\frac{2\pi}{\Omega}$.}

\subsection{$(m,n)=(1,1)$}

When $\mu_1^{(0)}=\mu_2^{(0)}=3\Omega/2$, we consider $({\psi}_1^{(0)},
{\psi}_2^{(0)})=(a u_{1}, b u_{1})$, where
\[
a^2=\frac{4}{3}\sqrt{\frac{2\pi}{\Omega}}\,
\frac{g_{22}\mu_1^{(1)}-g_{12}\mu_2^{(1)}}{g_{11}g_{22}-g_{12}^2},\quad
b^2=\frac{4}{3}\sqrt{\frac{2\pi}{\Omega}}\,
\frac{g_{11}\mu_2^{(1)}-g_{12}\mu_1^{(1)}}{g_{11}g_{22}-g_{12}^2}.
\]
This state corresponds to a (co-located) dark-dark type configuration
featuring a first excited state in both components.
%
%
%
%
%
Regarding the spectral stability we have $K_{\mathrm{Ham}}=4$, with the
dangerous eigenvalues again at $\lambda^{(0)}=\pm i\Omega$. At most two
eigenvalues with positive real part will emerge from $-i\Omega$. For the
perturbation calculation we need to consider case (a), where
$M_a\in\mathcal{M}_{4\times4}(\mathbb{R})$ is
\[
M_a=\frac{1}{16}\sqrt{\frac{\Omega}{2\pi}}\left(
    \begin{array}{cccc}
        2 g_{11}a^2 -5 g_{12}b^2 & 2\sqrt{2} g_{11} a^2 & 7 g_{12}a b & 2\sqrt{2}g_{12}a b \\
        -2\sqrt{2} g_{11} a^2 & -4 g_{11}a^2 +4 g_{12}b^2 & -2\sqrt{2}g_{12}a b & -8 g_{12}a b\\
        7 g_{12}a b & 2\sqrt{2}g_{12}a b & 2 g_{22}b^2 -5 g_{12}a^2 & 2\sqrt{2} g_{22} b^2\\
        -2\sqrt{2}g_{12}a b & -8 g_{12}a b & -2\sqrt{2} g_{22} b^2 & -4 g_{22}b^2 +4 g_{12}a^2
    \end{array}
\right).
\]
Examining the spectrum of $M_a$, we find that one eigenvalue is zero with
associated eigenvector $(-\sqrt{2}\,a/b,a/b,-\sqrt{2},1)^\mathrm{T}$; once
again, this is associated with the invariance to dipolar oscillations with
the frap frequency. Since the matrix is real-valued, this then implies that
there is at most one pair of eigenvalues with nonzero imaginary part. This is
an important conclusion that is particular to the case of the parabolic trap:
the presence of the well-known symmetry associated with the dipolar
oscillations~\cite{stringari} does not allow in this case the broader
spectrum of two potentially unstable eigendirections to lead to
instabilities; instead, only such instability direction may be realized in
practice.
%
%
%
%

For a particular example, if we fix $a=b=1$ and $g_{11}=g_{22}=1$, then the
remaining three eigenvalues of $M_a$ are
\[
\sqrt{\frac{\Omega}{2\pi}}\,\frac{1+g_{12}}{8},\quad
\sqrt{\frac{\Omega}{2\pi}}\,\frac{-1\pm\sqrt{1-56
g_{12}+136 g_{12}^2}}{16}
\]
Under these specific parameter values, there is one pair of eigenvalues that can
enter the complex plane for
\[
g_{12}\in\left(\frac{14-9\sqrt{2}}{68},\frac{14+9\sqrt{2}}{68}\right) \approx (0.0187,
0.3931).
\]

\begin{figure}[!htbp]
\begin{tabular}{ccc}
\includegraphics[width=5cm]{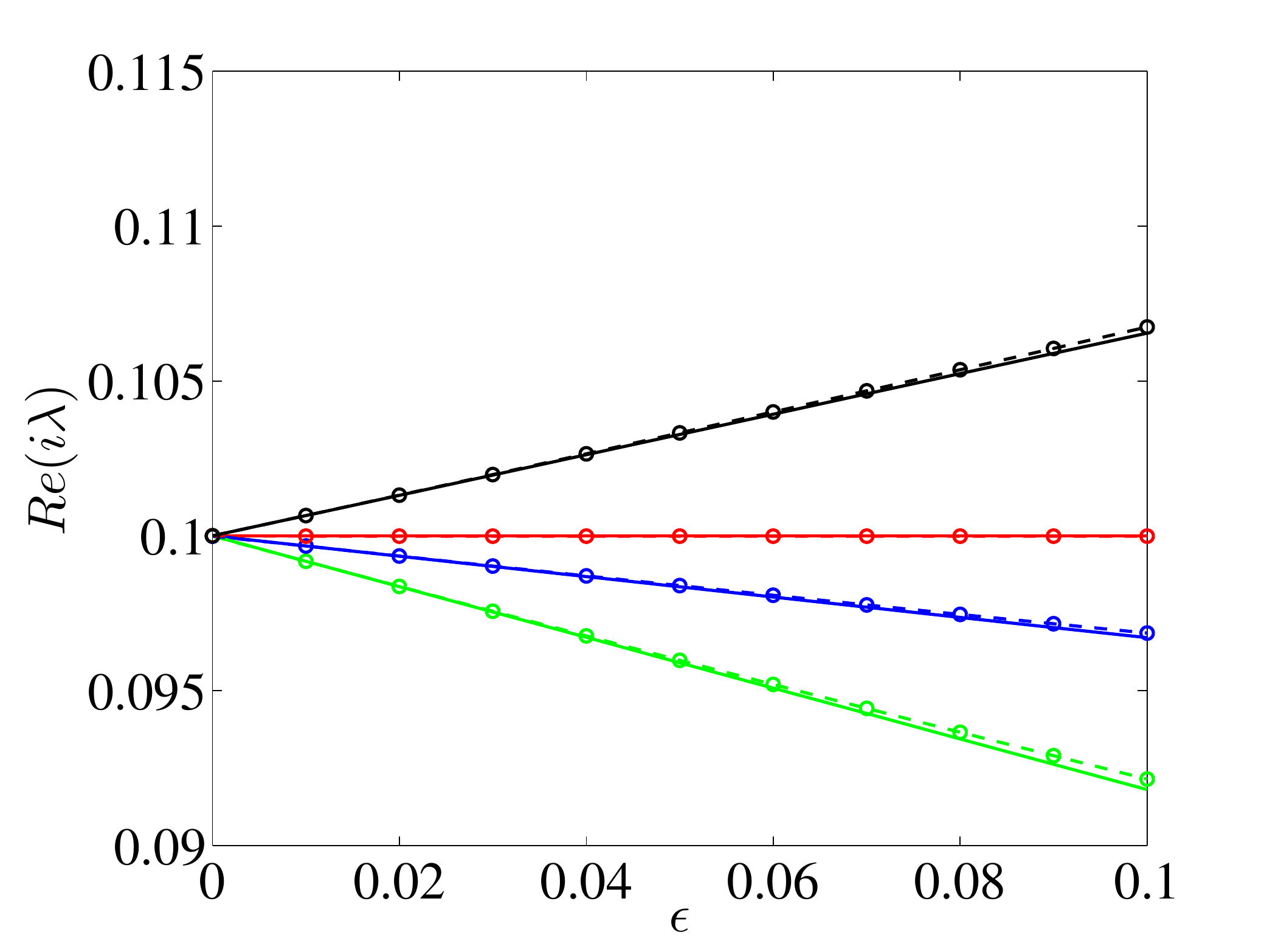}
\includegraphics[width=5cm]{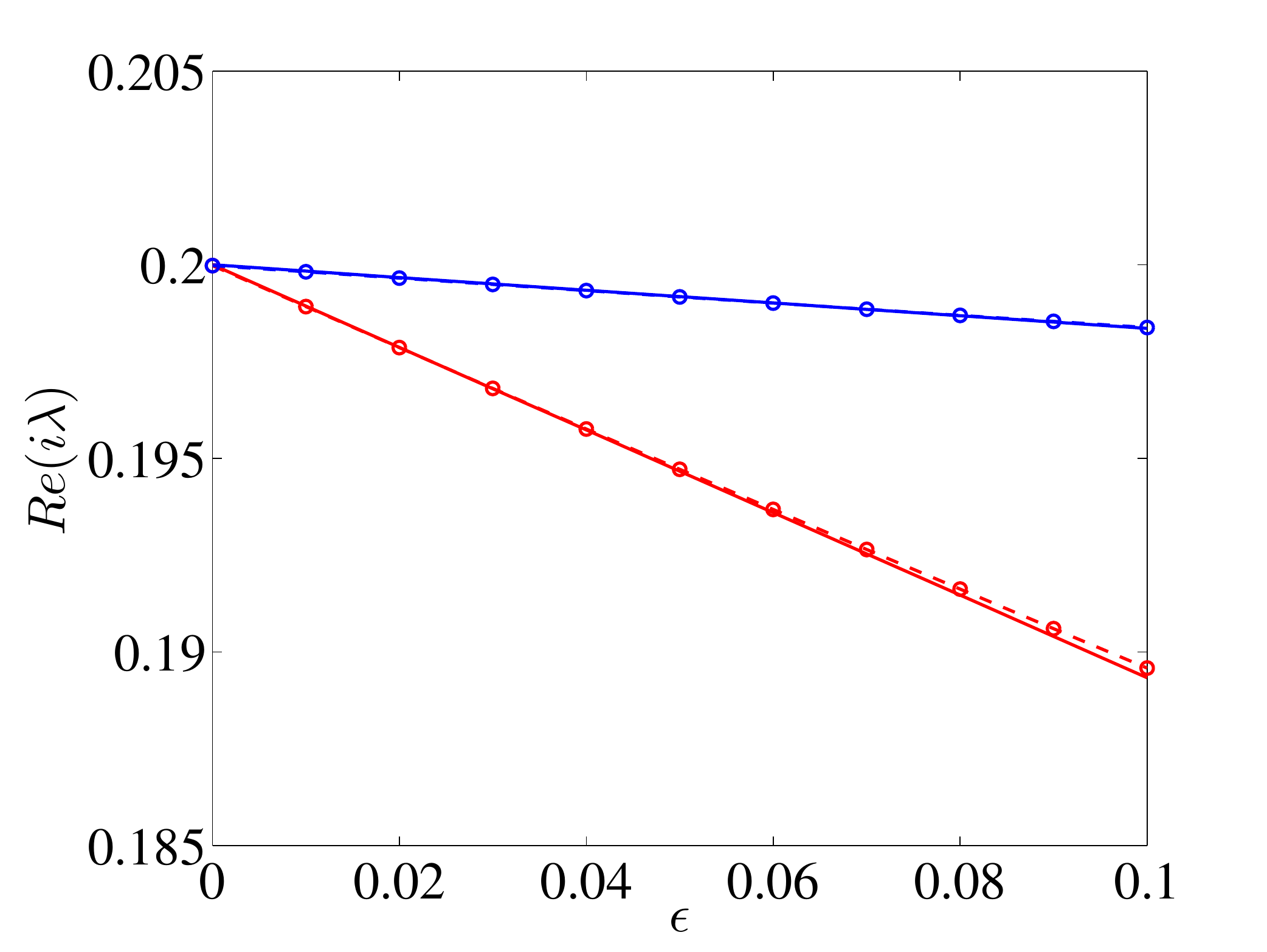}
\includegraphics[width=5cm]{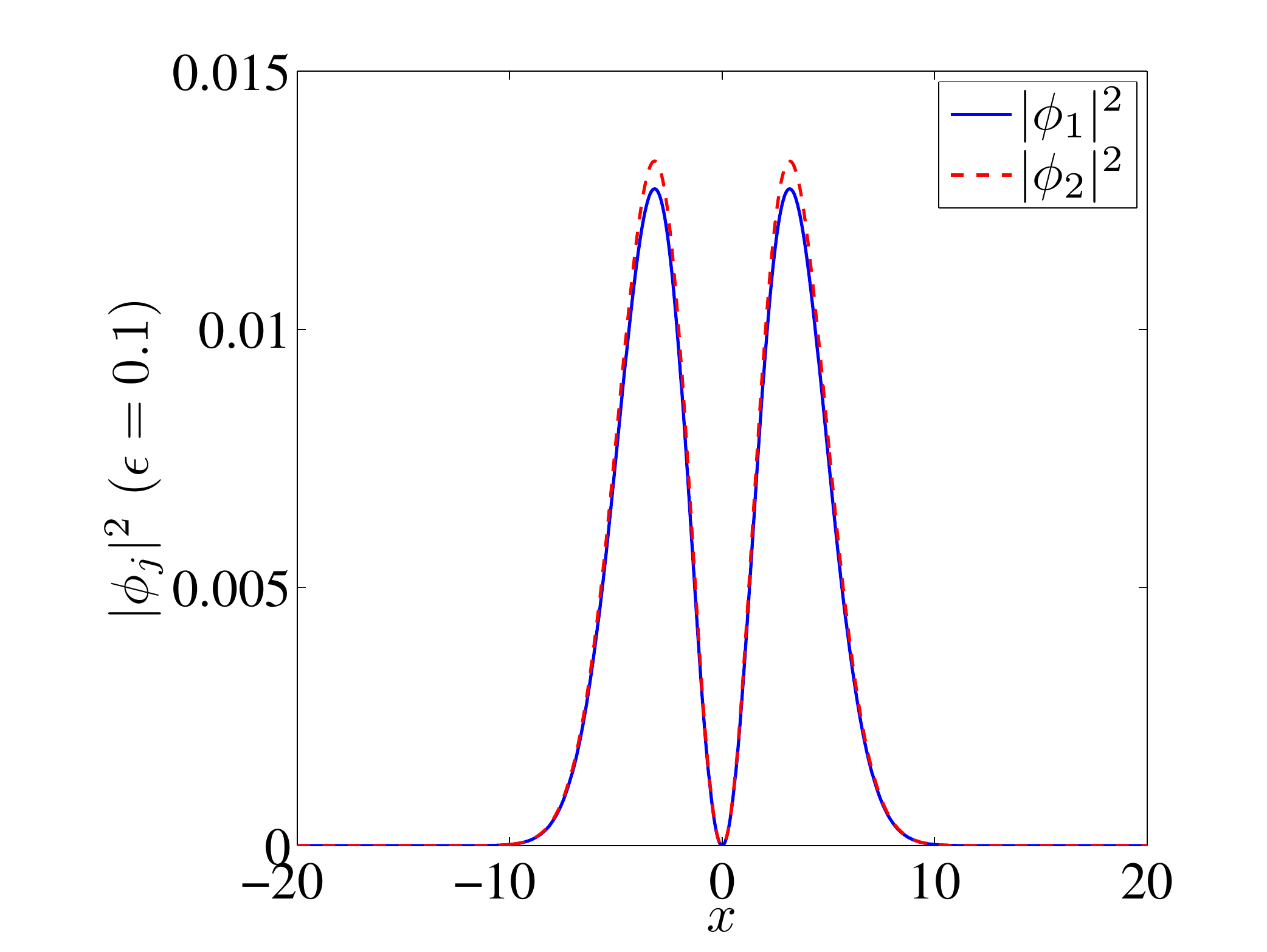}
\end{tabular}
\caption{ Case of $a=b=1, g_{11}=1.03, g_{12}=1.04, g_{22}=1.06, \Omega=0.1$
for the $(1,1)$ branch:
The left (middle) panel shows the imaginary parts of the eigenvalues
around $-i\Omega$ ($- 2i\Omega$) as functions of $\epsilon$ with $O(\epsilon)$ corrections (solid lines) and corresponding numerical results
(dashed lines with circles). The right panel shows the densities of $\phi_1$ and $\phi_2$ at $\epsilon=0.1$.}
\label{fig1_3}
\end{figure}
\begin{figure}[!htbp]
\begin{tabular}{ccc}
\includegraphics[width=5cm]{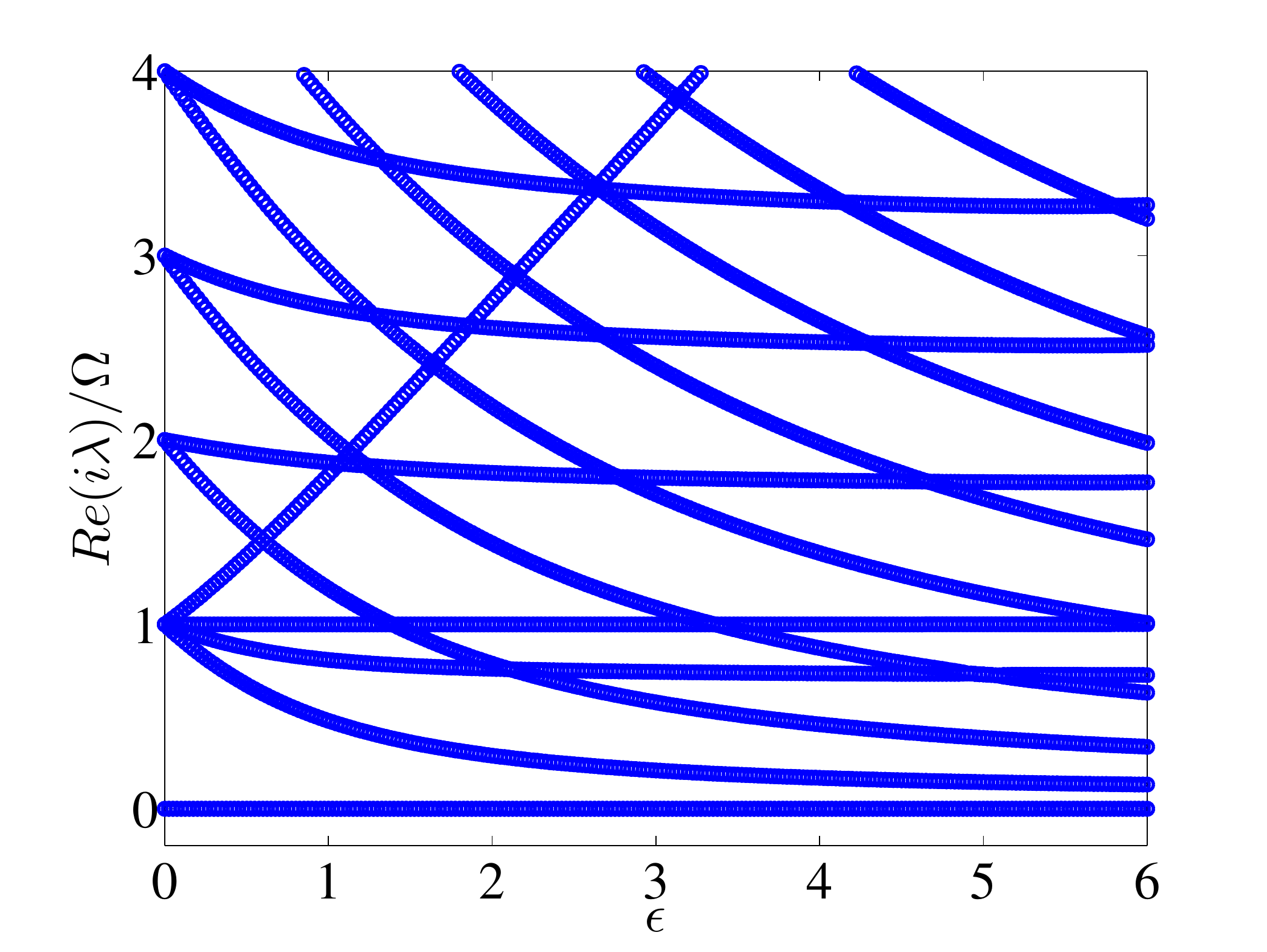}
\includegraphics[width=5cm]{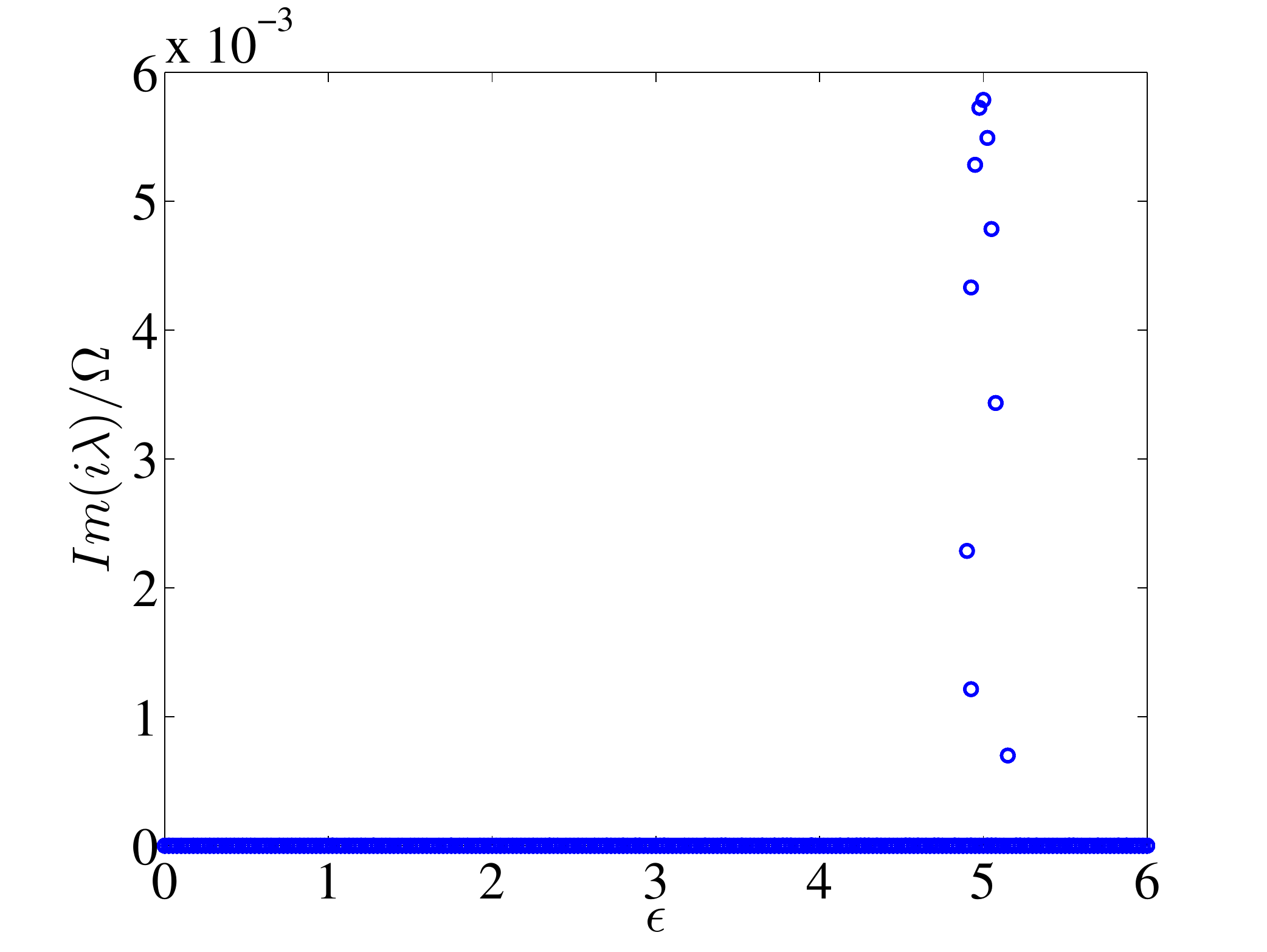}
\includegraphics[width=5cm]{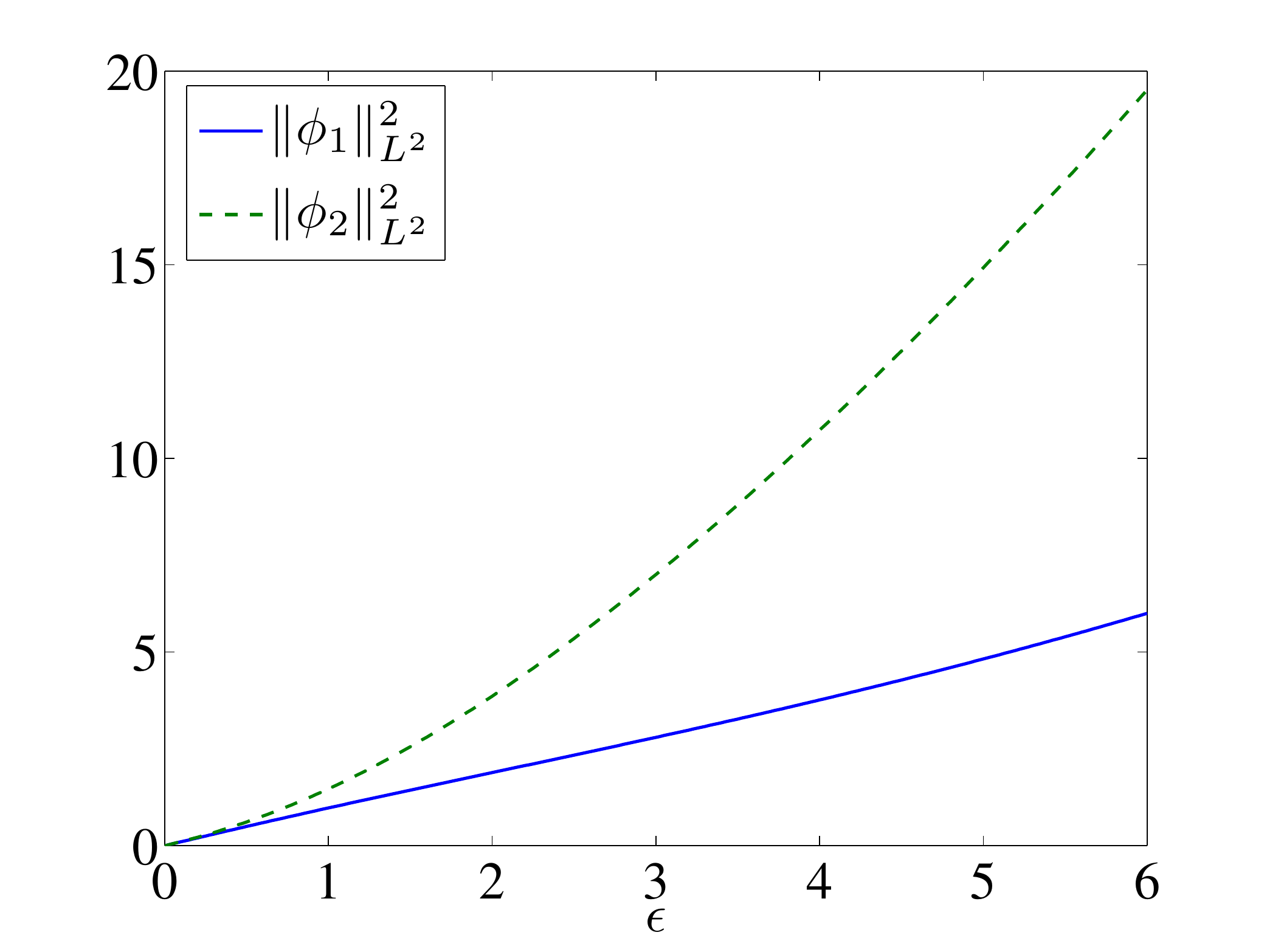}
\end{tabular}
\caption{ The left (middle) panel shows the change of imaginary (real) parts of the eigenvalues for the branch of solutions with $a=b=1, g_{11}=1.03, g_{12}=1.04, g_{22}=1.06, \Omega=0.1$ for the $(1,1)$ branch. In the right panel, we plot the $L^2$-norm of the solution of $\phi_j$ as a function of $\epsilon$.}
\label{fig1_3_2}
\end{figure}

Again, setting $a=b=1, g_{11}=1.03, g_{12}=1.04, g_{22}=1.06, \Omega=0.1$ we
compare some predicted eigenvalues up to $O(\epsilon)$ with corresponding
numerical eigenvalues in Fig.~\ref{fig1_3}. All of the numerically computed
eigenvalues from Eqn.~(\ref{eqn_ev}) for this example are imaginary, which
matches the analytical result of the reduced spectral eigenvalue problem from
Eqn.~(\ref{eqn_ev_reduced}). As $\epsilon$ becomes large, we see a pair of
eigenvalues enter the complex plane near $\pm i\Omega$ at $\epsilon\approx
4.9$ when the eigenvalue from $i \Omega$ collides with the eigenvalue from
$3i\Omega$ (see Figure {\ref{fig1_3_2}}). We also note that this complex pair
will come back to the imaginary axis at $\epsilon\approx 5.2$, i.e., the
parametric interval of instability is fairly narrow in this case.

%

\begin{figure}[!htbp]
\begin{tabular}{ccc}
\includegraphics[width=5cm]{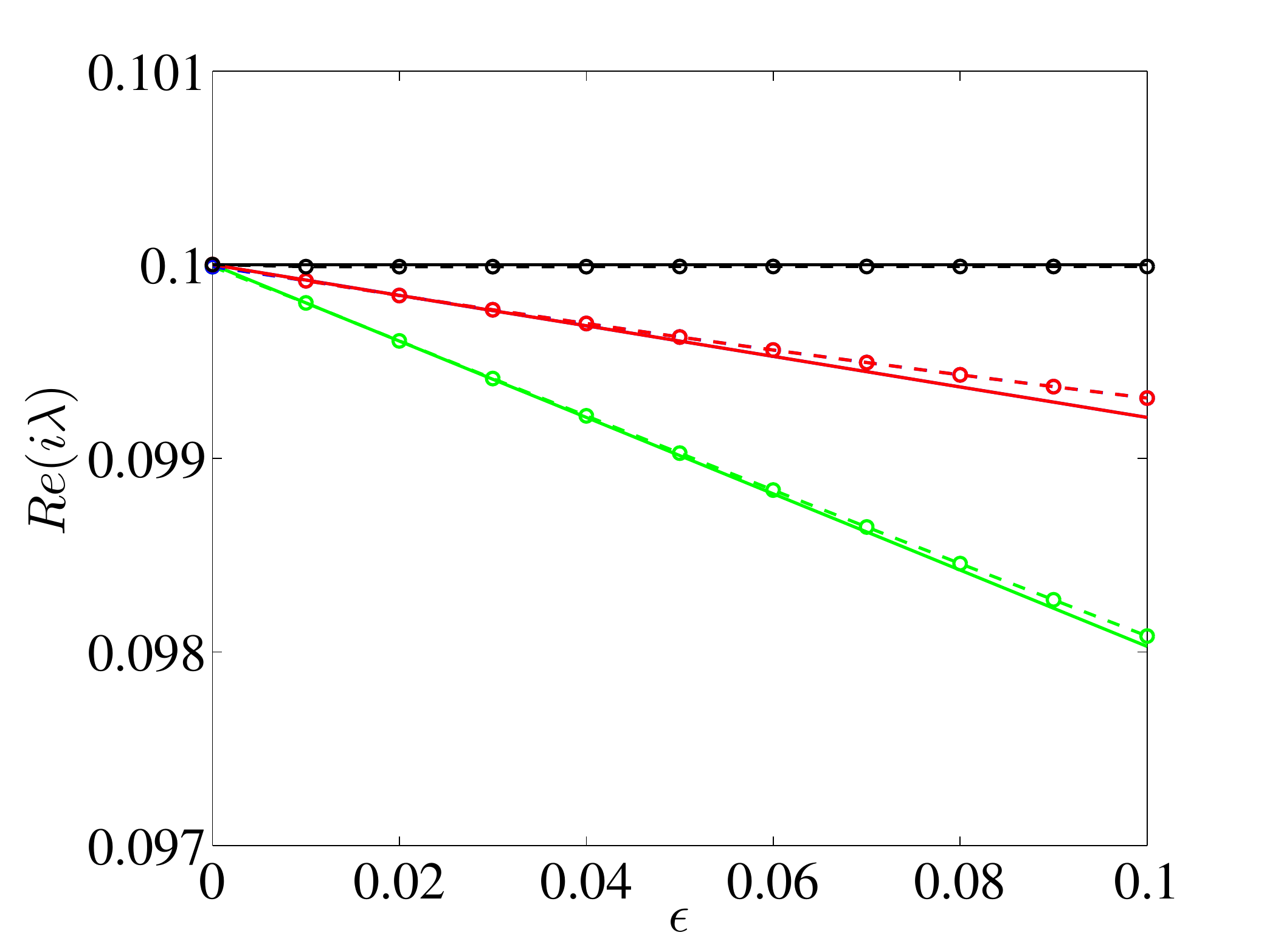}
\includegraphics[width=5cm]{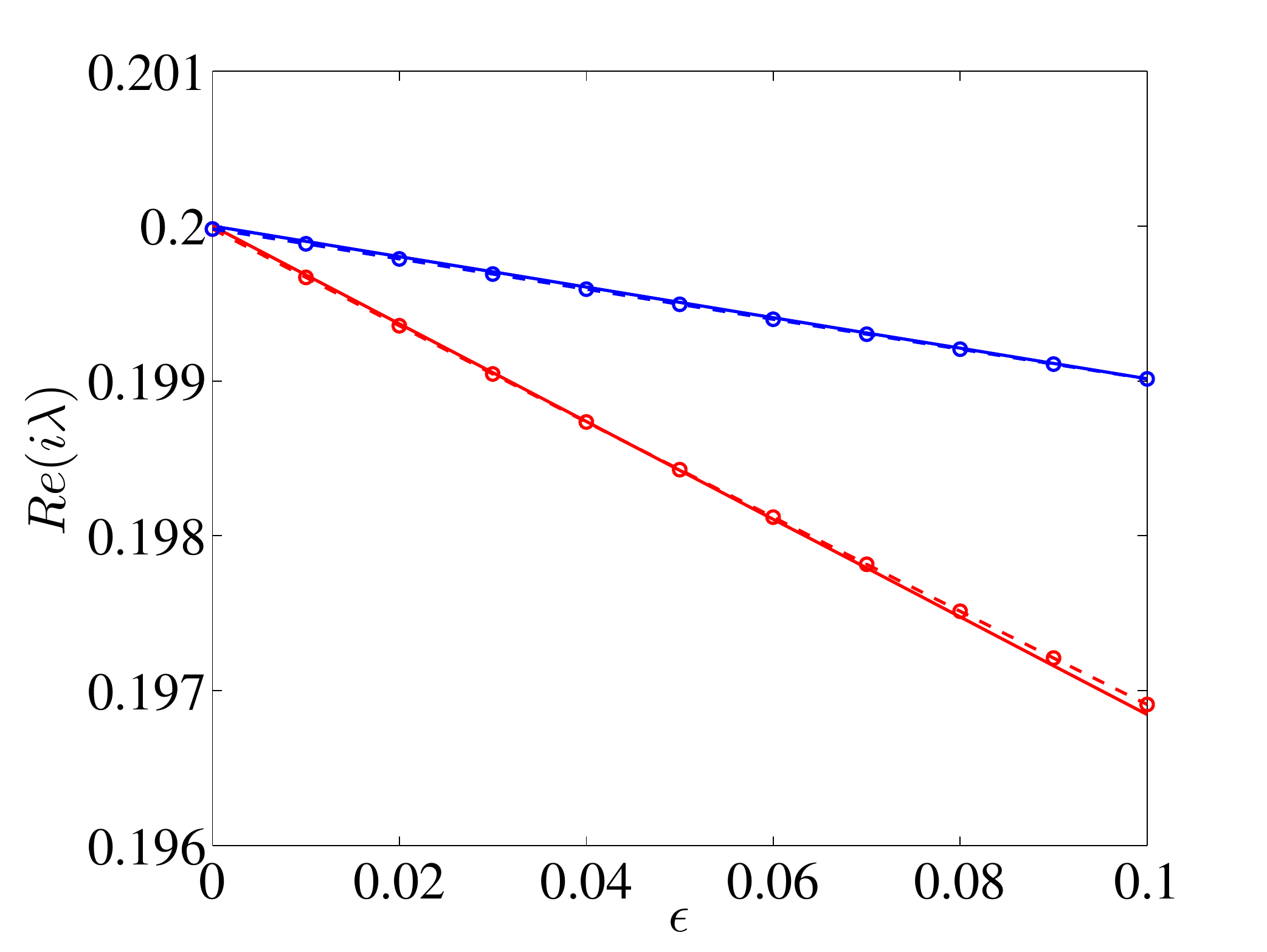}
\includegraphics[width=5cm]{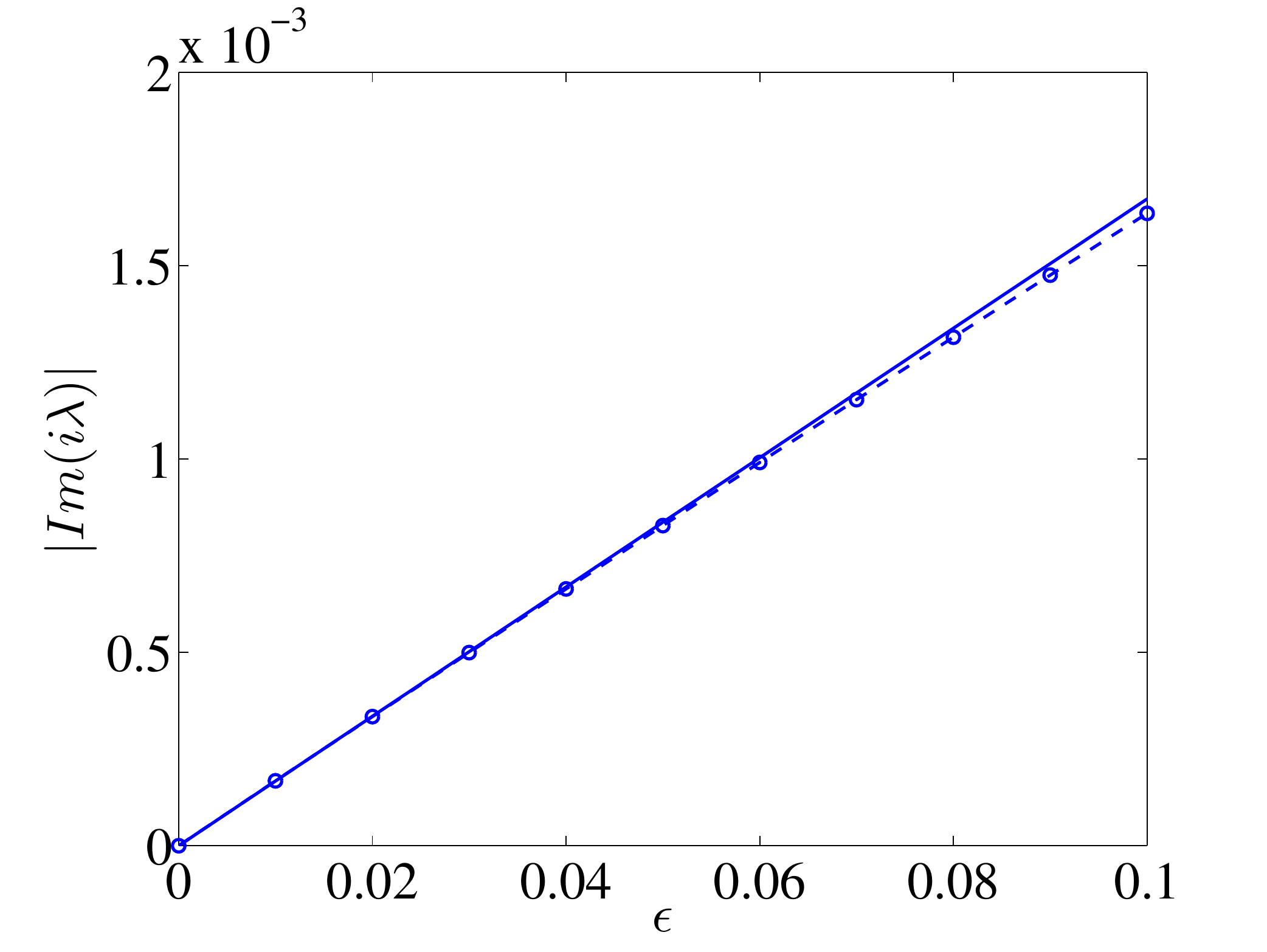}
\end{tabular}
\caption{ Case of the $(1,1)$ branch with
 $a=b=1, g_{11}=1, g_{12}=0.25, g_{22}=1, \Omega=0.1$: The left (middle) panel shows the imaginary parts of the eigenvalues
around $-i\Omega$ ($-i2\Omega$) as functions of $\epsilon$
with $O(\epsilon)$ corrections (solid lines) and corresponding numerical
results (dashed lines with circles). In the left panel, the red lines and
blue lines (solid and dashed) are almost identical since a pair of
eigenvalues of $M_a$ for $\lambda^{(0)}=-i\Omega$ are complex conjugates. Moreover, the nonzero
imaginary parts of this pair imply the instability (this is the only source of the instability) of the solution, as shown in the right panel
(solid line for the $O(\epsilon)$ correction using this pair of complex
conjugates and the dashed line for the numerical computation of the real
parts of the eigenvalues). }
\label{fig1_3_3}
\end{figure}

\begin{figure}[!htbp]
\begin{tabular}{cc}
\includegraphics[width=5cm]{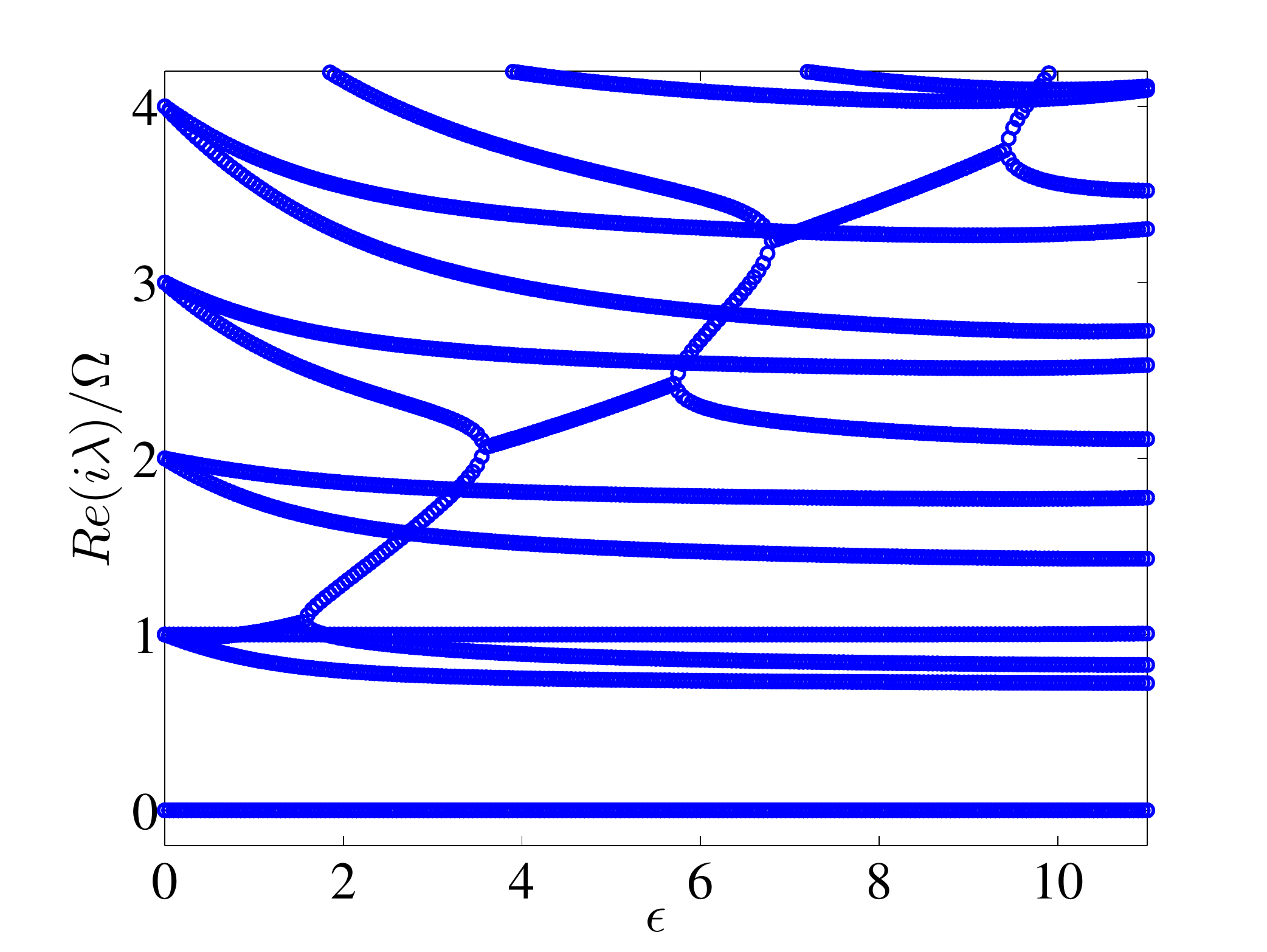}
\includegraphics[width=5cm]{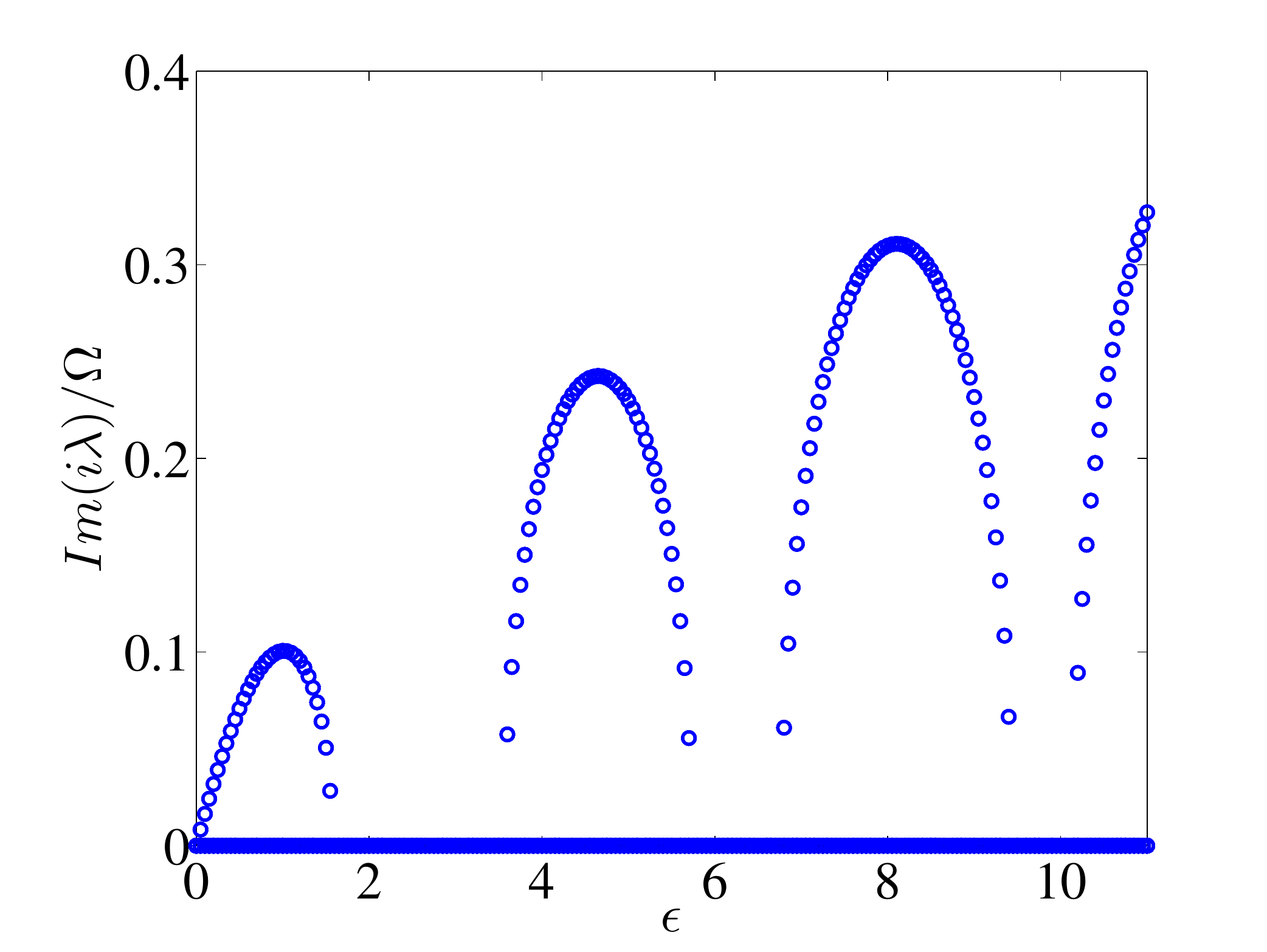}
\includegraphics[width=5cm]{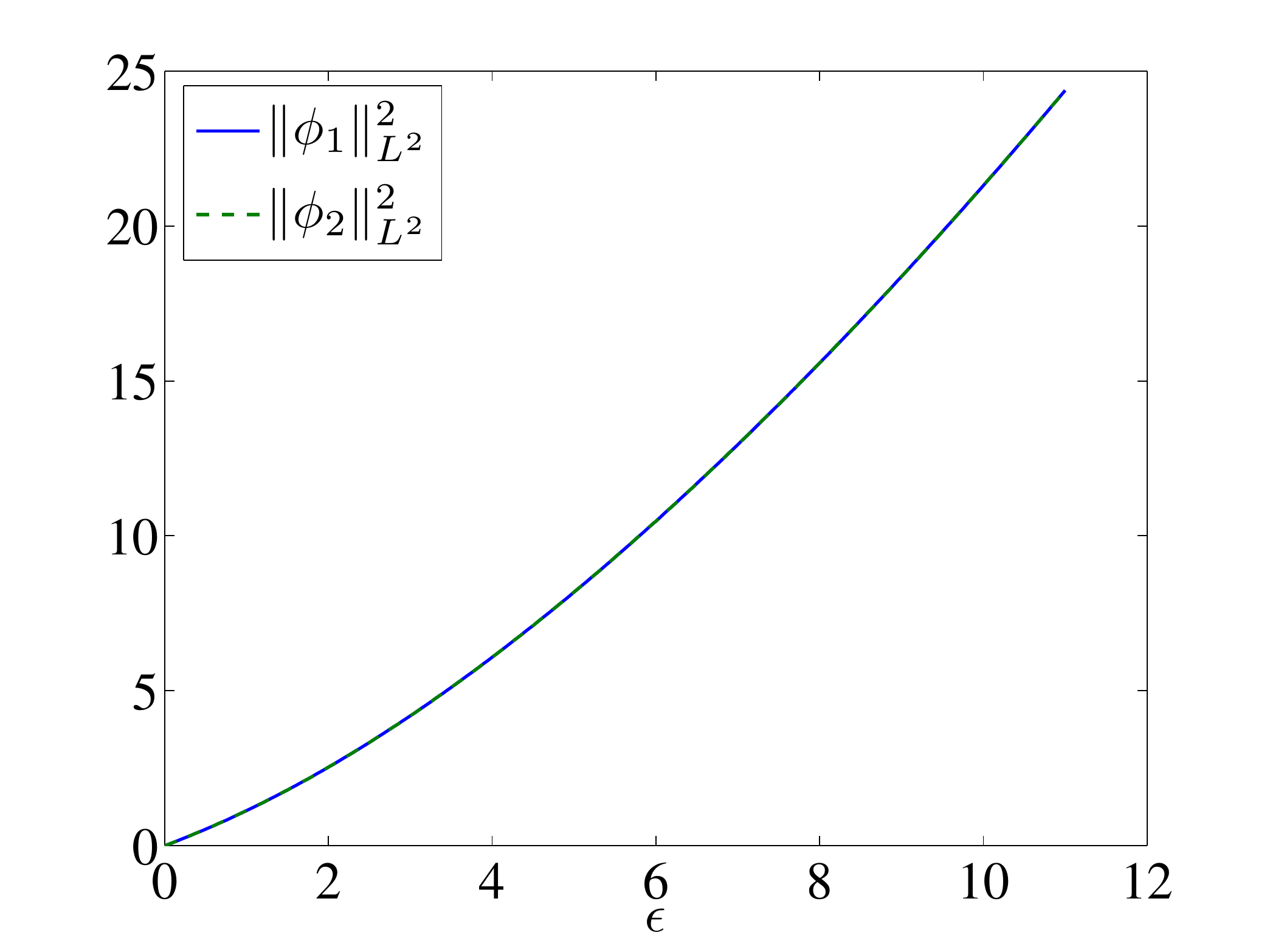}
\end{tabular}
\caption{ The left (middle) panel shows the change of imaginary (real) parts of the eigenvalues for the $(1,1)$
branch of solutions with $a=b=1, g_{11}=1, g_{12}=0.25, g_{22}=1, \Omega=0.1$. This is for a lengthy parametric continuation over
$\epsilon$, featuring not only the original instability near
the linear limit but subsequent splits (stabilizations) and further
collisions (destabilizations) of the relevant solution. Again the right panel shows the $L^2$-norm of the solution of $\phi_j$ as a function of $\epsilon$.
}
\label{fig1_3_4}
\end{figure}

\begin{figure}[!htbp]
\begin{tabular}{cc}
\includegraphics[width=7cm]{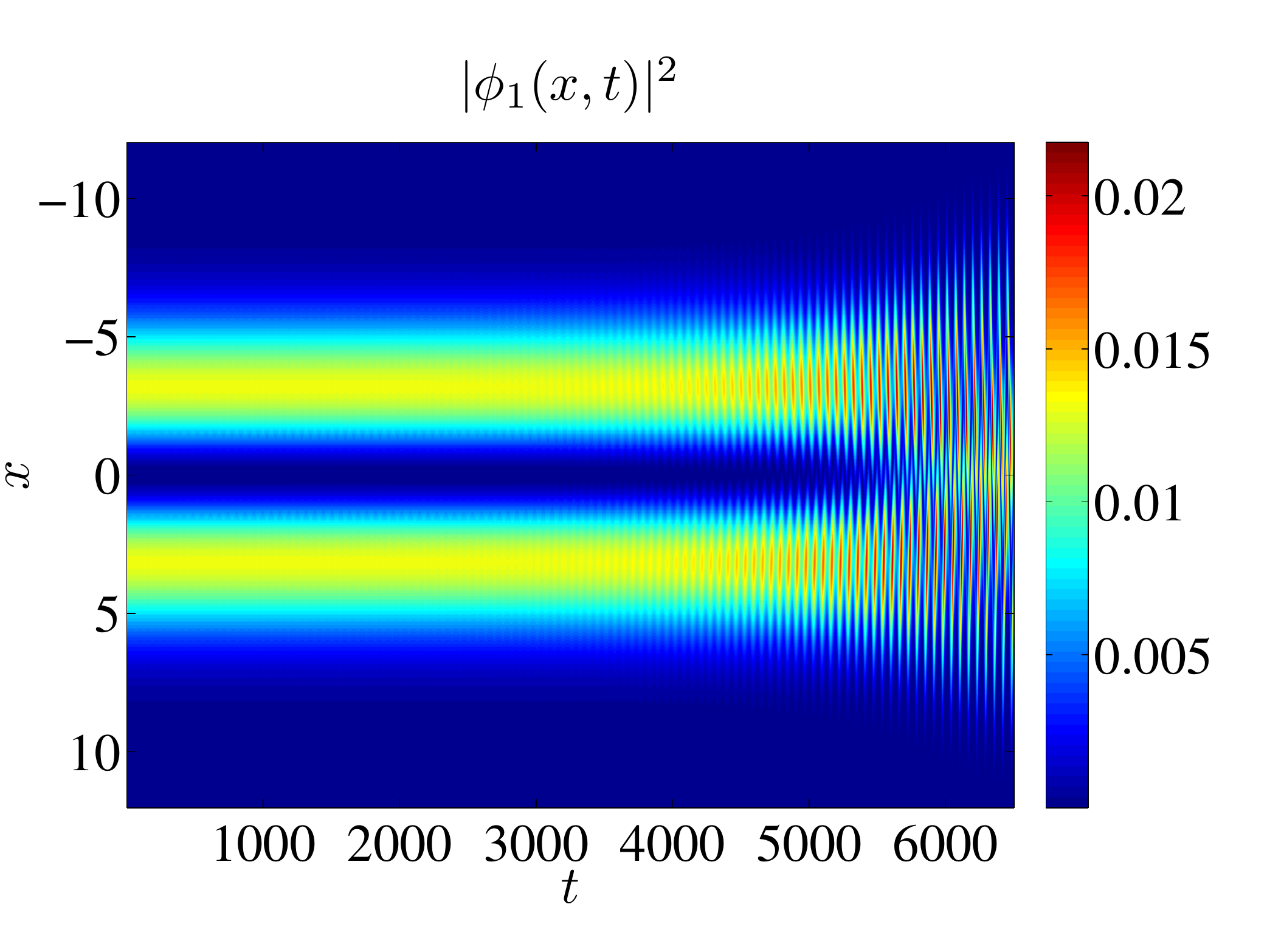}
\includegraphics[width=7cm]{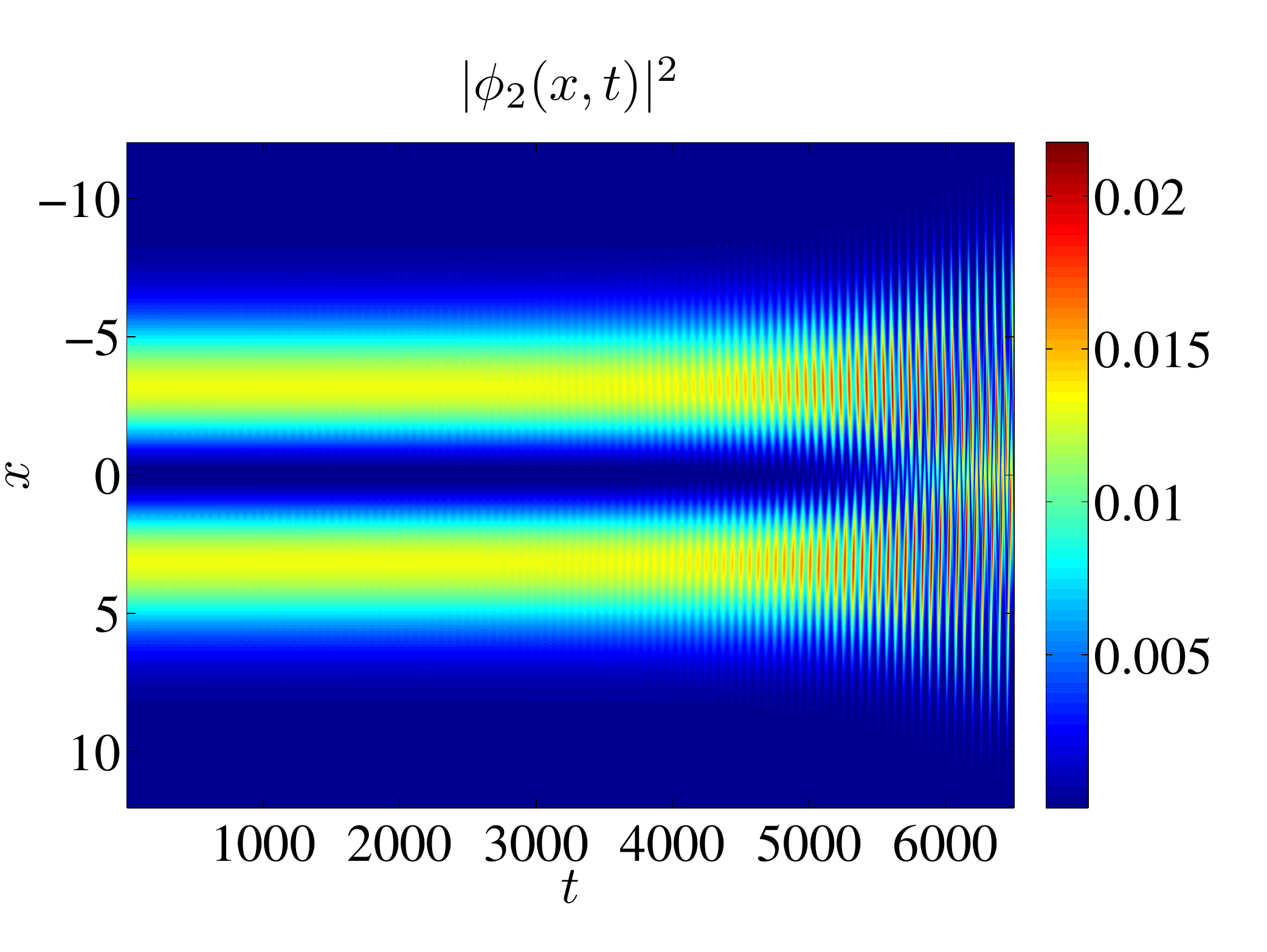}
\end{tabular}
\caption{ The left (right) panel shows an example of the dynamics of $|\phi_1|^2$ ($|\phi_2|^2$) of the $(1,1)$ branch with $a=b=1, g_{11}=1, g_{12}=0.25, g_{22}=1$, $\Omega=0.1$ and $\epsilon=0.1$. }
\label{fig1_3_5}
\end{figure}

If $a=b=1, g_{11}=1, g_{12}=0.25, g_{22}=1, \Omega=0.1$, the numerical
computation shows that all of the eigenvalues except a quartet (near $\pm
i\Omega$) are on the imaginary axis, as shown in Fig.~\ref{fig1_3_3}. As
$\epsilon$ grows, the complex pair of eigenvalues from $-i\Omega$ will return
to the real axis and split into two, as seen in Figure {\ref{fig1_3_4}}. The
split eigenvalue going upward will meet with the eigenvalue coming down from
$-i3\Omega$ and produce another pair (quartet) of complex eigenvalues, which
will go back to the imaginary axis and split again. One of the split
eigenvalue will move upward and collide with the eigenvalue from $-i5\Omega$,
which will again lead to complex eigenvalues, and so on. In
Figure~\ref{fig1_3_5}, we illustrate the numerical evolution of this unstable
configuration for $\epsilon=0.1$. It can be seen that the weak (and clearly
discerned to be oscillatory) nature of the instability only allows it to
manifest over fairly long time scales, resulting in breathing dynamics.

\subsection{$(m,n)=(0,2)$}

For $\mu_1^{(0)}=\Omega/2$ and $\mu_2^{(0)}=5\Omega/2$, we consider the
continuation of $({\psi}_1^{(0)}, {\psi}_2^{(0)})=(a u_{0}, b u_{2})$, where
\[
a^2=64\sqrt{\frac{2\pi}{\Omega}}\,
\frac{41g_{22}\mu_1^{(1)}-24g_{12}\mu_2^{(1)}}{41 g_{11}g_{22}- 9g_{12}^2},\quad
b^2=512\sqrt{\frac{2\pi}{\Omega}}\,
\frac{8 g_{11}\mu_2^{(1)}-3g_{12}\mu_1^{(1)}}{41 g_{11}g_{22}-9g_{12}^2}.
\]
%
%
%
%
%
%
%
%
Regarding spectral stability we have $K_\mathrm{Ham}=4$, except that now the
dangerous eigenvalues are at $\lambda^{(0)}=\pm i\Omega,\pm i2\Omega$. At
most one eigenvalue with positive real part will emerge from each of these
dangerous eigenvalues.

First consider the perturbation calculation associated with
$\lambda^{(0)}=-i\Omega$. We consider case (b), and the matrix
$M_b\in\mathcal{M}_{3\times3}(\mathbb{R})$ is
\begin{equation}\label{E-b-0-2}
M_b=\frac{1}{64}\sqrt{\frac{\Omega}{2\pi}}\left(
    \begin{array}{ccc}
        4 g_{12}b^2 & 4\sqrt{3}\,g_{12}a b & 8\sqrt{2}\, g_{12}a b \\
        4\sqrt{3}\, g_{12}a b  &  10 g_{22}b^2 - 4 g_{12}a^2 & 5\sqrt{6}\, g_{22}b^2 \\
        -8\sqrt{2}\, g_{12}a b & -5\sqrt{6}\, g_{22}b^2 & -15 g_{22}b^2 - 8 g_{12}a^2
    \end{array}
\right).
\end{equation}
\begin{proposition}\label{proposition-0-2}
For $(m,n)=(0,2)$, the eigenvalues of $M_b$ in (\ref{E-b-0-2}) are $0$ and
\[
-\frac{1}{128}\sqrt{\frac{\Omega}{2\pi}}\left(12 a^2 g_{12} - 4 b^2 g_{12}
    + 5 b^2 g_{22}\pm \sqrt{16 a^4 g_{12}^2-8 a^2 b^2 g_{12}(28 g_{12}-25
    g_{22})+b^4(4 g_{12}+5 g_{22})^2 }\right).
\]
The imaginary parts of the eigenvalues for (\ref{E-b-0-2}) will be nonzero if
$g_{12}>5g_{22}/4$.
\end{proposition}
\noindent As is the case for the continuation of $(0,1)$, the parameter
$g_{11}$ does not appear in the expressions of the eigenvalues.

Now consider the perturbation calculation associated with
$\lambda^{(0)}=-i2\Omega$.
In this case, the matrix
$M_b\in\mathcal{M}_{3\times3}(\mathbb{R})$ is
\[
M_b=\frac{1}{512}\sqrt{\frac{\Omega}{2\pi}}\left(
    \begin{array}{ccc}
        -128 g_{11}a^2 + 136 g_{12}b^2 & 12\sqrt{6}g_{12}a b & 192 g_{12}a b \\
        12\sqrt{6}\, g_{12}a b  &  g_{22}b^2 - 52 g_{12}a^2 & 12\sqrt{6}\, g_{22}b^2 \\
        -192 g_{12}a b & -12\sqrt{6}\, g_{22}b^2 & -56 g_{22}b^2 - 320 g_{12}a^2
    \end{array}
\right).
\]
Unfortunately, the expressions of eigenvalues are not as
straightforward/enlightening in an analytical form (although available). As
the numerical computations below show, it is possible for this matrix to have
a pair of eigenvalues with nonzero imaginary part.

\begin{figure}[!htbp]
\begin{tabular}{ccc}
\includegraphics[width=5cm]{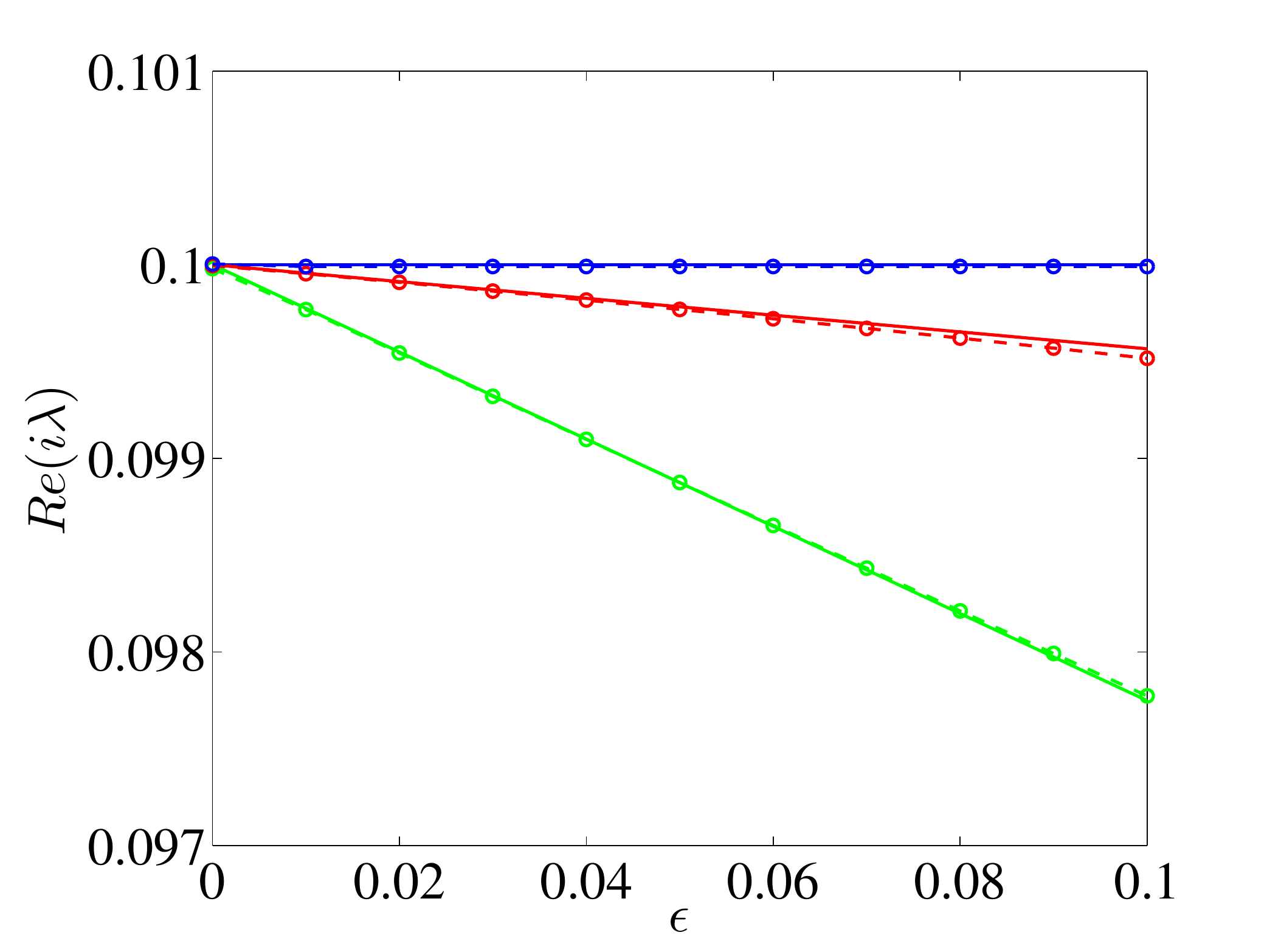}
\includegraphics[width=5cm]{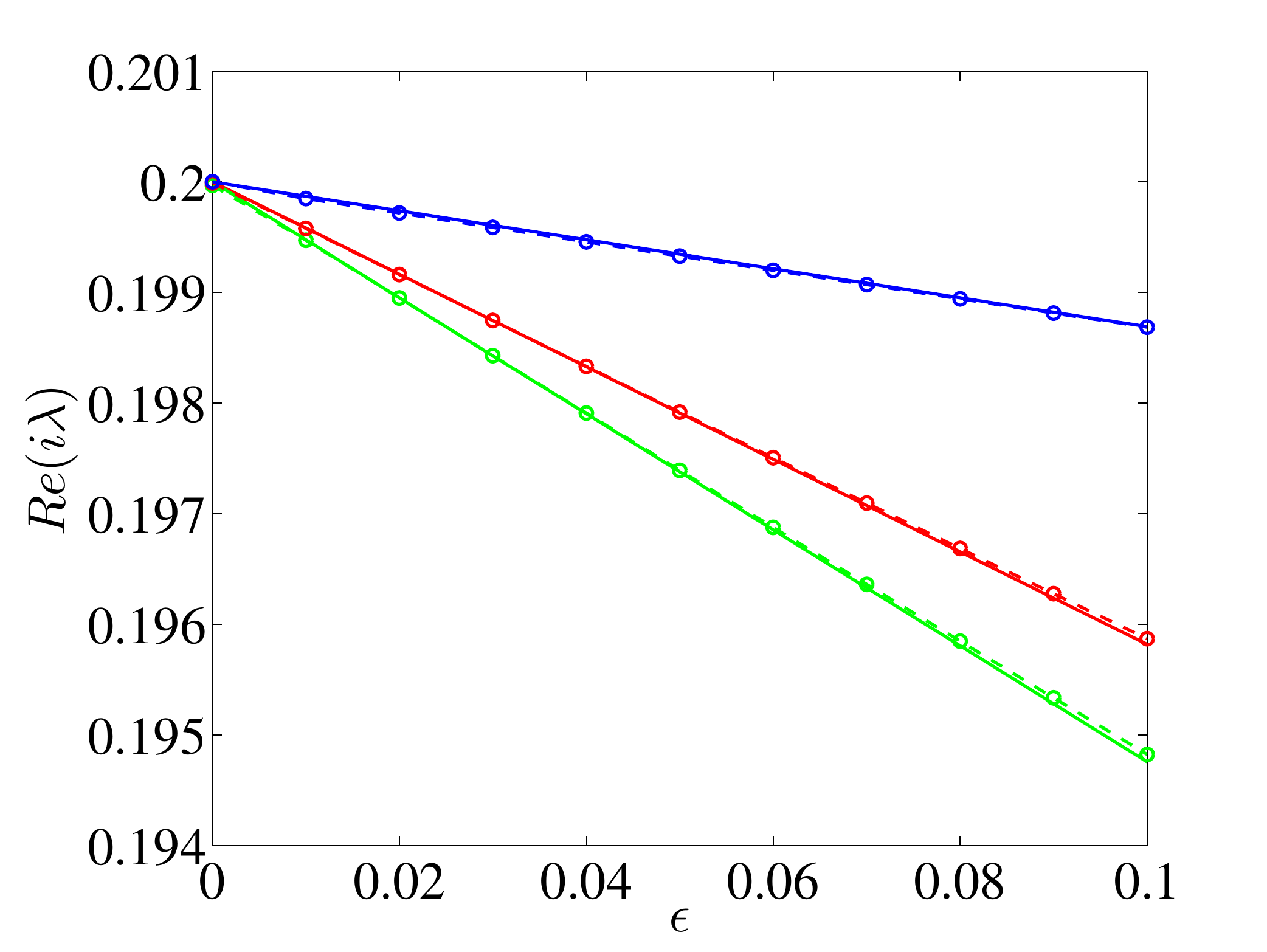}
\includegraphics[width=5cm]{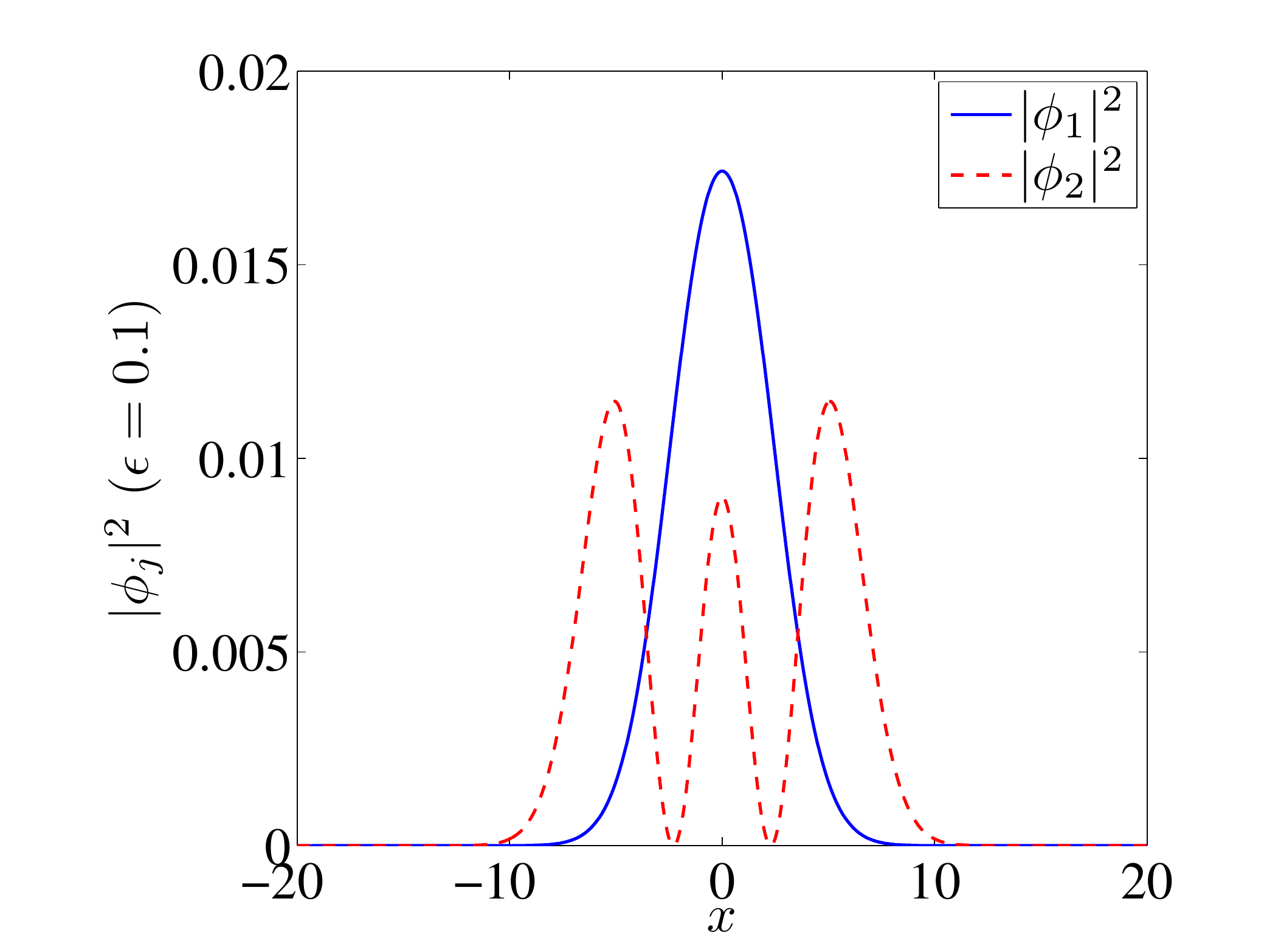}
\end{tabular}
\caption{Case of the $(0,2)$ branch with $a=b=1, g_{11}=1.03,
g_{12}=1.04, g_{22}=1.06, \Omega=0.1$: The left (middle) panel shows the imaginary parts of the eigenvalues
around $-i\Omega$ ($-i2\Omega$) as functions of $\epsilon$ with $O(\epsilon)$ corrections
(solid lines) and corresponding numerical results (dashed lines with circles). The right panel shows the densities of $\phi_1$ and $\phi_2$ at $\epsilon=0.1$ for this $(0,2)$ waveform.}
\label{fig1_4}
\end{figure}

\begin{figure}[!htbp]
\begin{tabular}{cc}
\includegraphics[width=7cm]{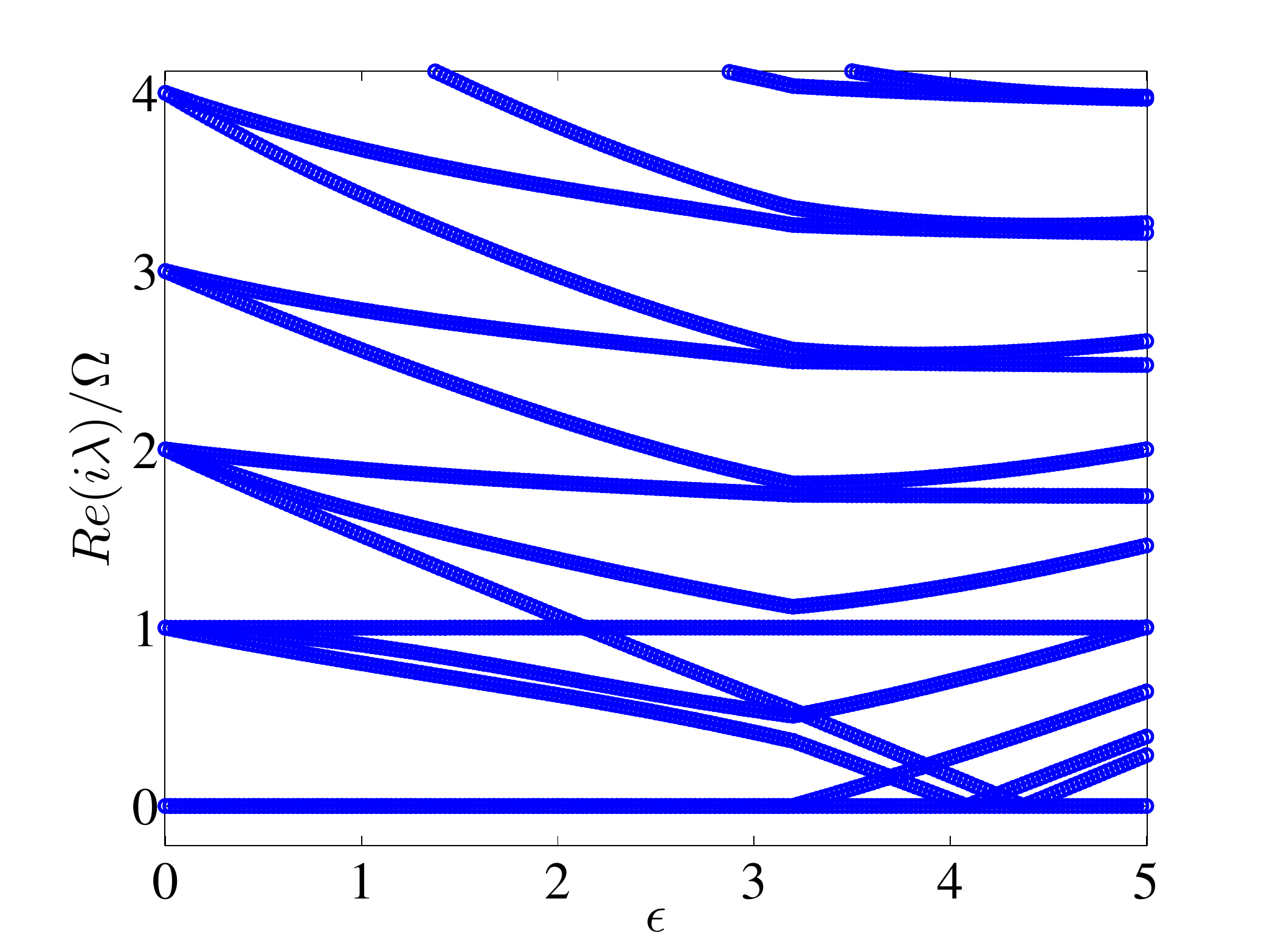}
\includegraphics[width=7cm]{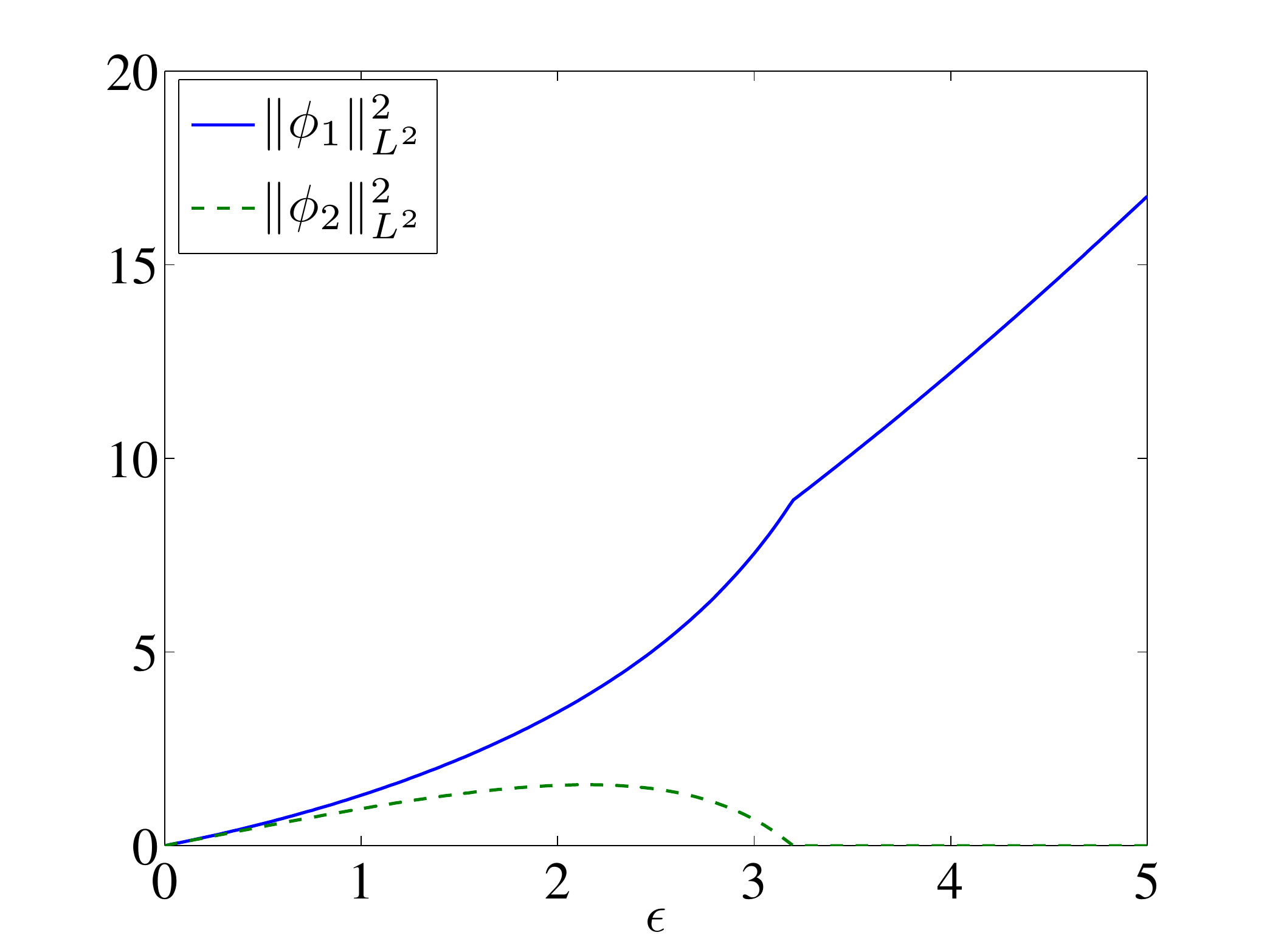}\\
\end{tabular}
\caption{In the left panel, we keep track of the imaginary parts of the eigenvalues for large $\epsilon$ where the imaginary parts remain zero. The right panel shows the $L^2$-norm of the solution of $\phi_j$ for such variation of $\epsilon$.}
\label{fig1_4b}
\end{figure}

For the numerical computations, we again let $a=b=1, g_{11}=1.03,
g_{12}=1.04, g_{22}=1.06, \Omega=0.1$, and compute the continuation of
two-component solutions. We compare
our analytical predictions providing the eigenvalues up to $O(\epsilon)$ with
the corresponding numerical eigenvalues in Fig.~\ref{fig1_4}. As $\epsilon$
grows, all of the numerically computed eigenvalues are on the imaginary axis
and their change with respect to $\epsilon$ is illustrated in the left
panel of Fig.~\ref{fig1_4b}. Additionally, we find that $\phi_2$ becomes zero
at $\epsilon\approx 3.2$ where this branch of solutions meets the branch of
one-component solutions on $\phi_1$. We note that this resembles the first example of case (0,1) very much.

\begin{figure}[!htbp]
\begin{tabular}{ccc}
\includegraphics[width=5cm]{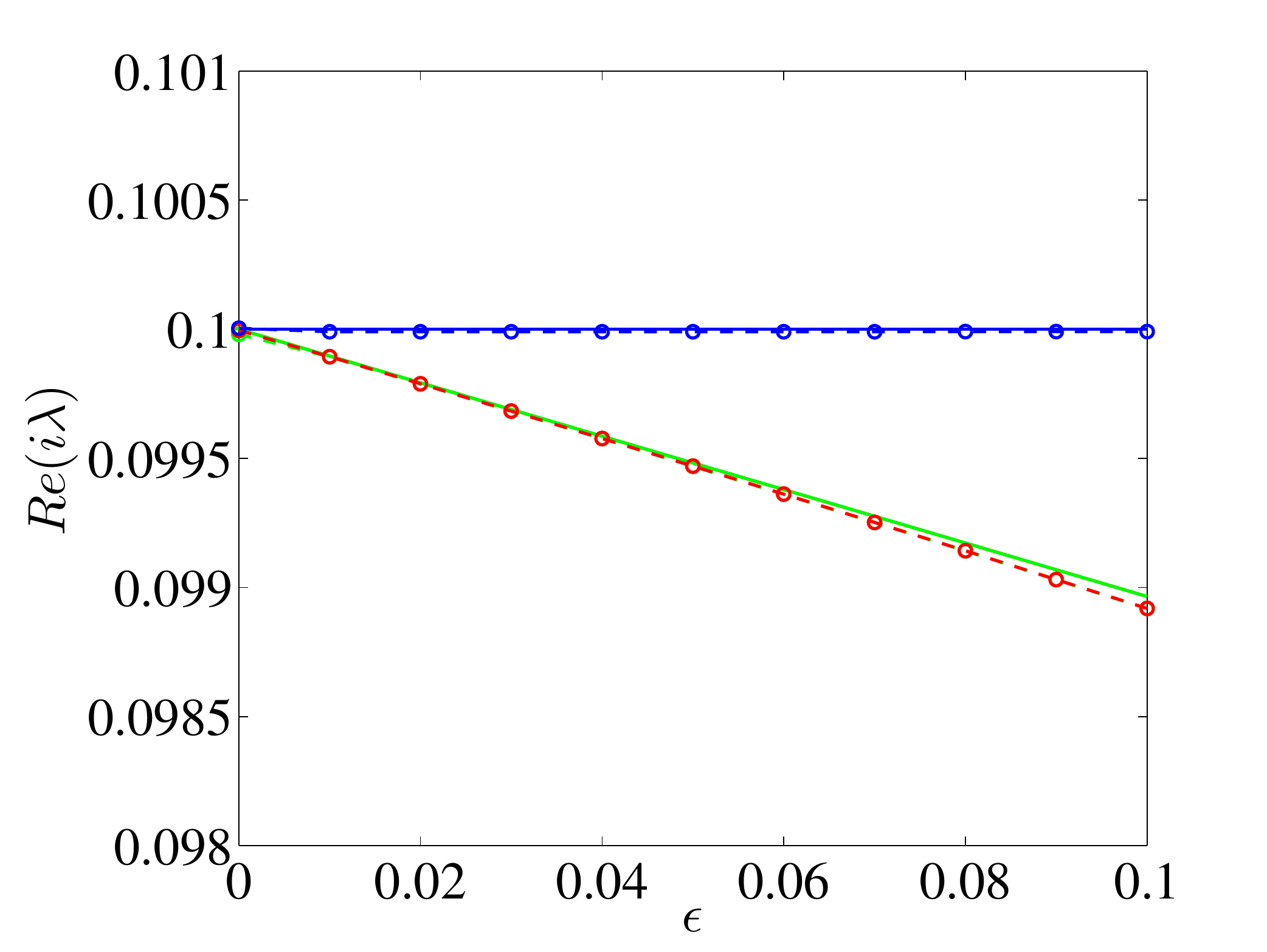}
\includegraphics[width=5cm]{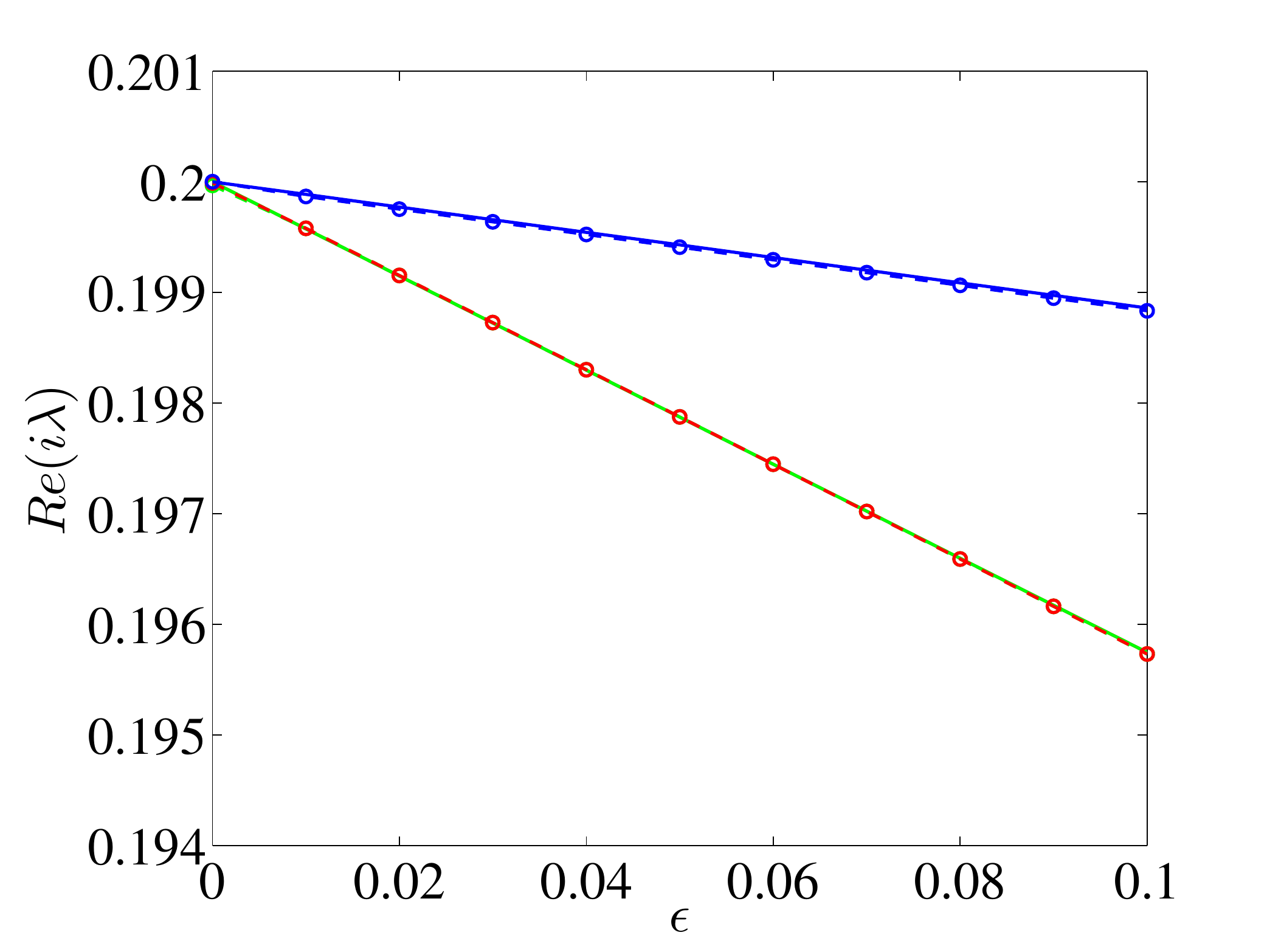}
\includegraphics[width=5cm]{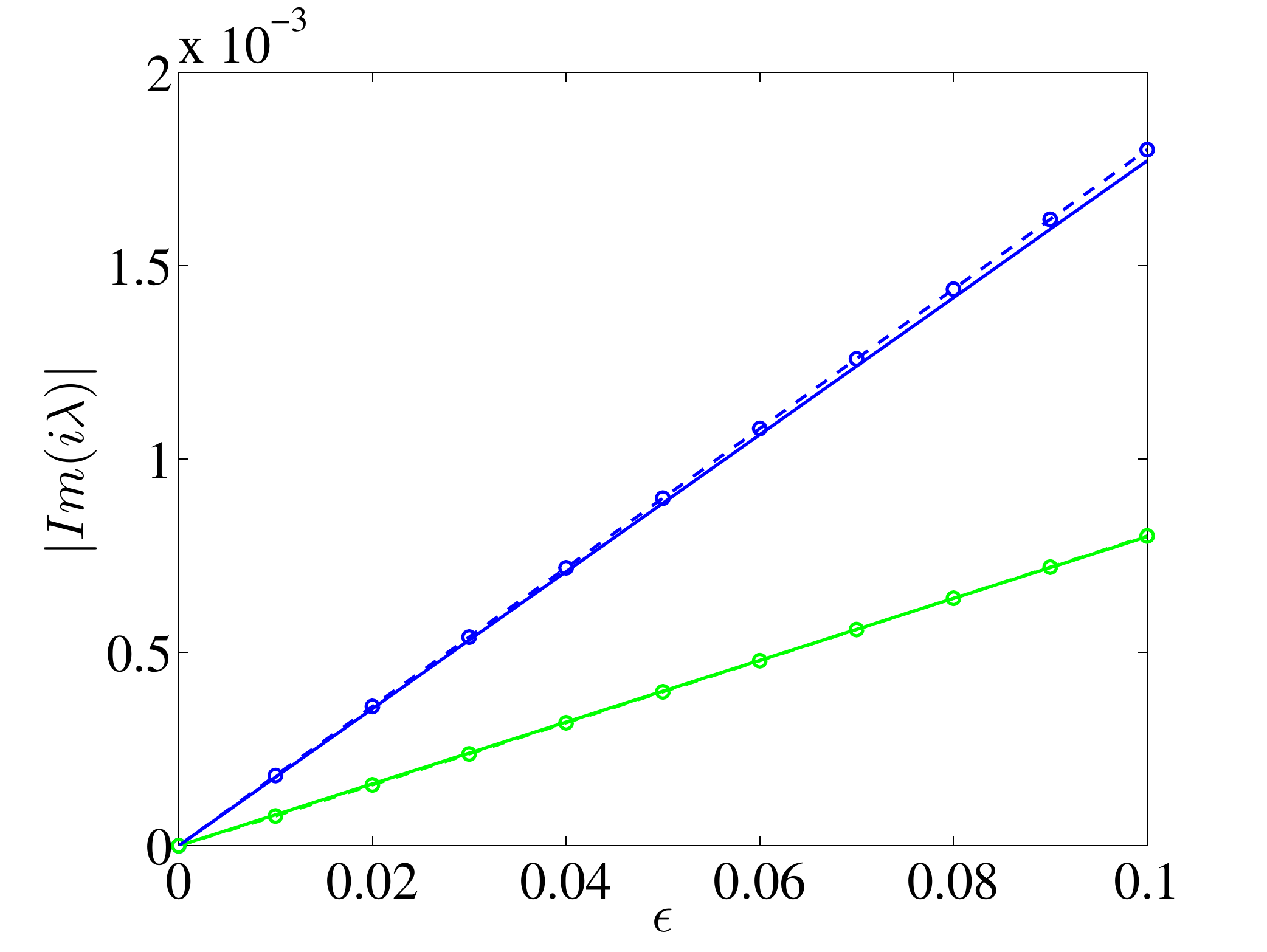}
\end{tabular}
\caption{Case of  $a=b=1, g_{11}=1, g_{12}=1, g_{22}=0.5, \Omega=0.1$
for the $(0,2)$ branch: The left (middle) panel shows the imaginary parts of $\lambda$'s around $\Omega$ ($2\Omega$) as functions of $\epsilon$ with
$O(\epsilon)$ corrections (solid lines) and corresponding numerical
results (dashed lines with circles). In these two panels, the red lines and green lines
(solid and dashed) are almost identical since both $M_b$ for $\lambda^{(0)}=-i\Omega$ and
$M_b$ for $\lambda^{(0)}=-i2\Omega$ have a pair of eigenvalues that are complex conjugates. Moreover, the
nonzero imaginary parts of these pairs imply the instability of the solution, as shown in the right panel (solid lines for the $O(\epsilon)$ corrections using these pairs of complex conjugates and the dashed lines for the numerical computation of the real parts of the eigenvalues, blue for the ones
near $-i\Omega$ and green for the ones near $-i2\Omega$). }
\label{fig1_4_3}
\end{figure}

\begin{figure}[!htbp]
\begin{tabular}{ccc}
\includegraphics[width=5cm]{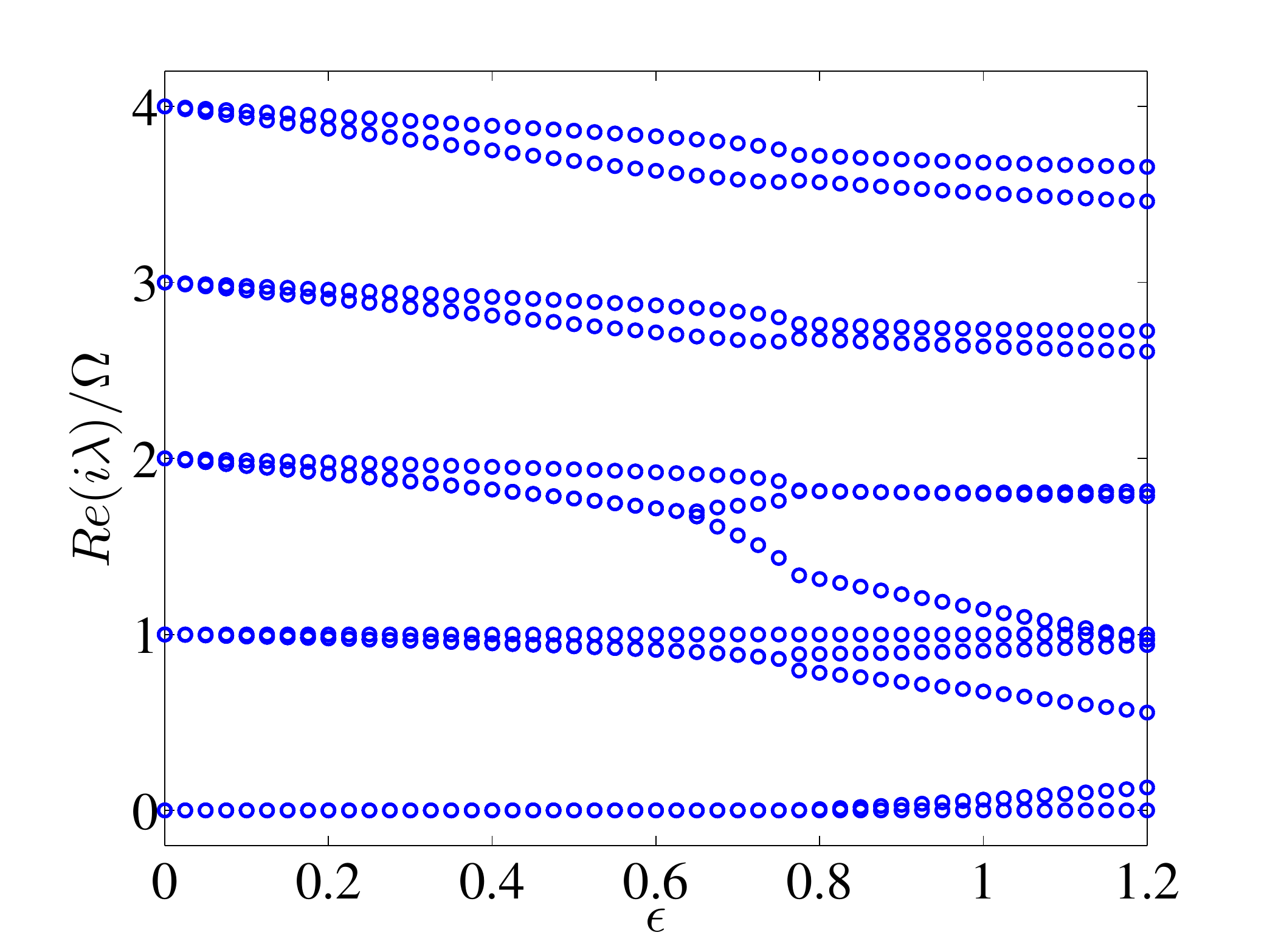}
\includegraphics[width=5cm]{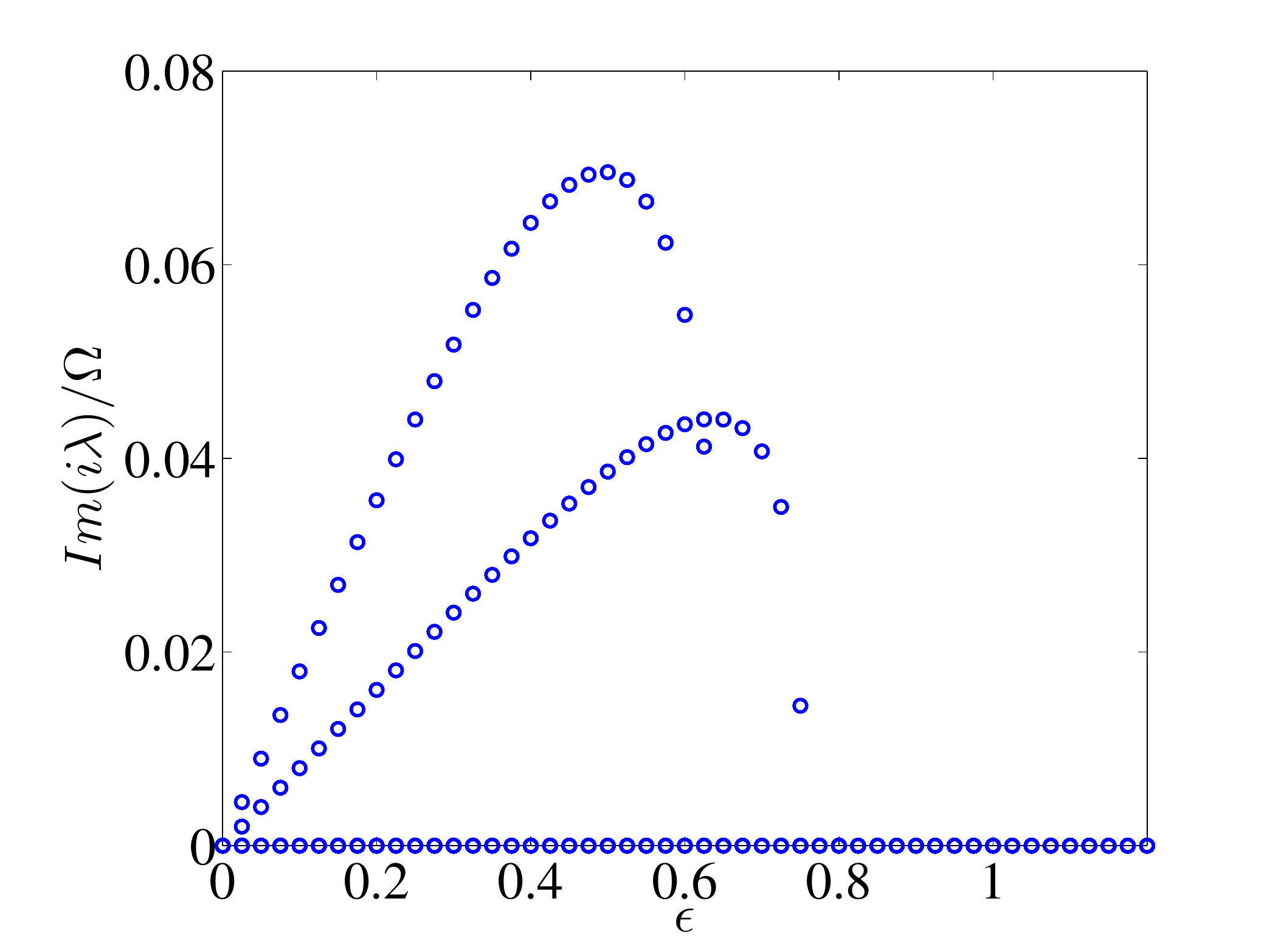}
\includegraphics[width=5cm]{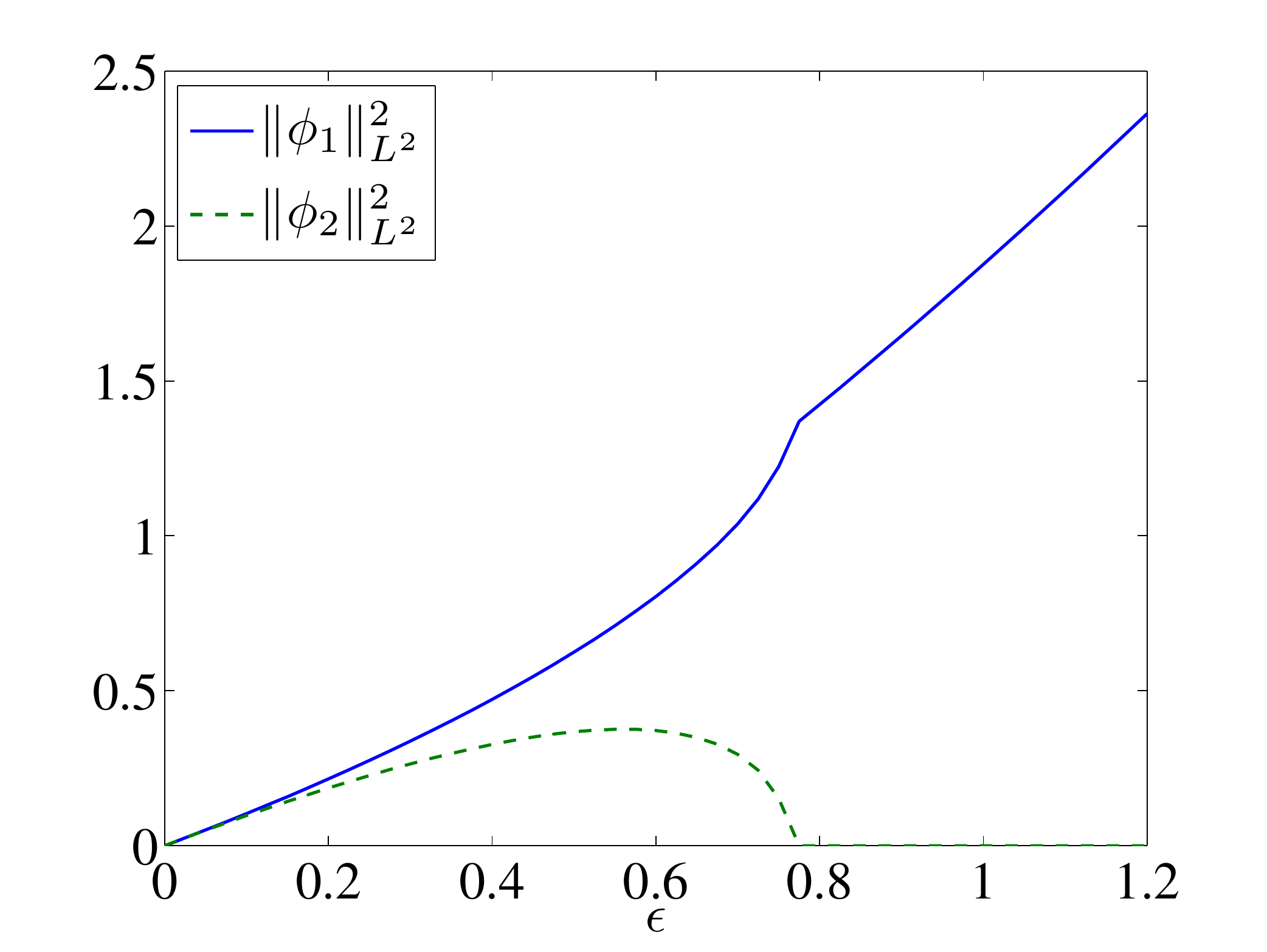}
\end{tabular}
\caption{ The left (middle) panel shows the change of
imaginary (real) parts of $i \lambda$ for the branch of solutions with
$a=b=1, g_{11}=1, g_{12}=1, g_{22}=0.5, \Omega=0.1$. The presence of the two associated instabilities is
evident in the middle panel. In the right panel, we monitor the change of the $L^2$-norm of the solution of $\phi_j$ over $\epsilon$.
}
\label{fig1_4_4}
\end{figure}

\begin{figure}[!htbp]
\begin{tabular}{cc}
\includegraphics[width=7cm]{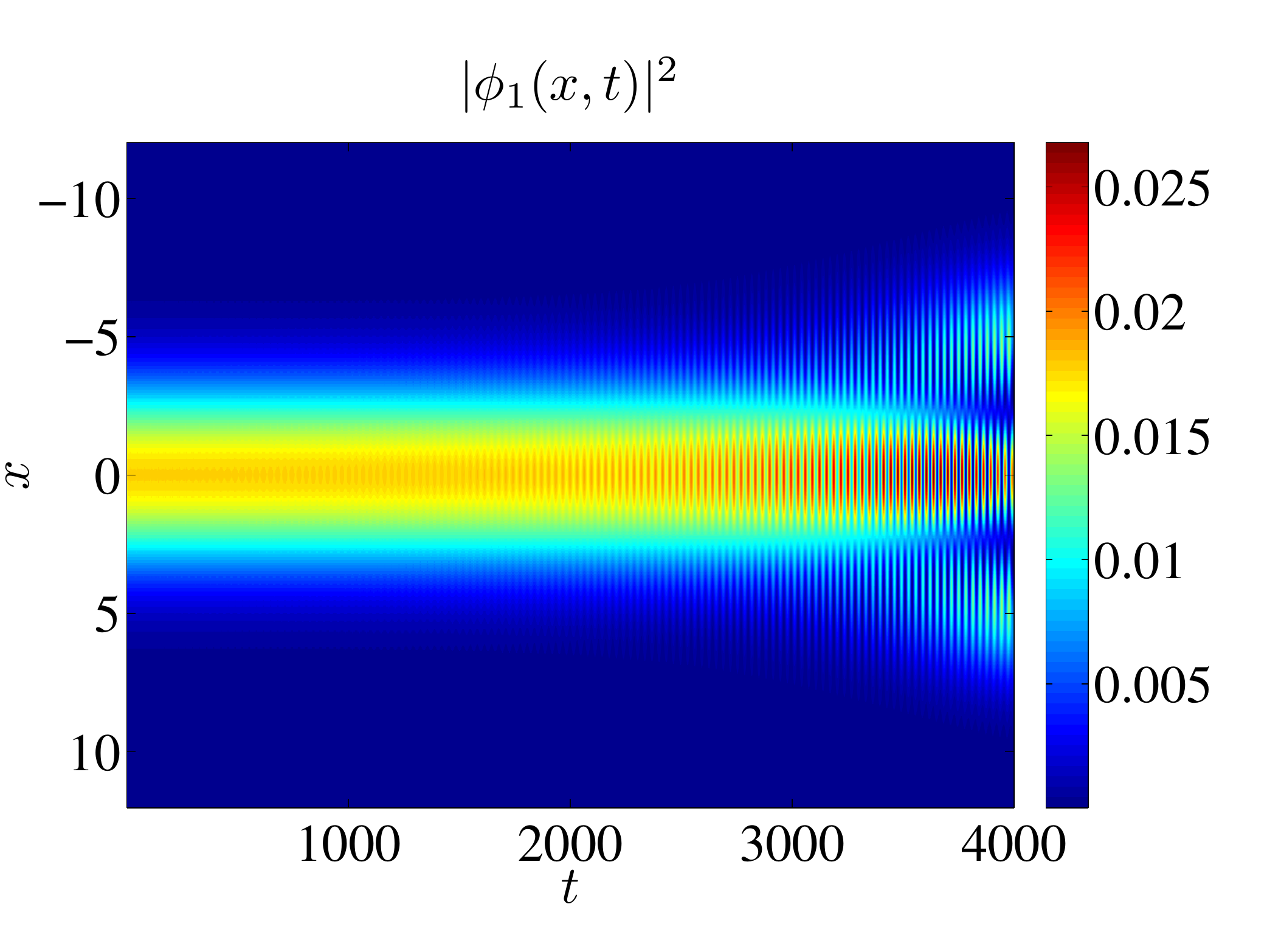}
\includegraphics[width=7cm]{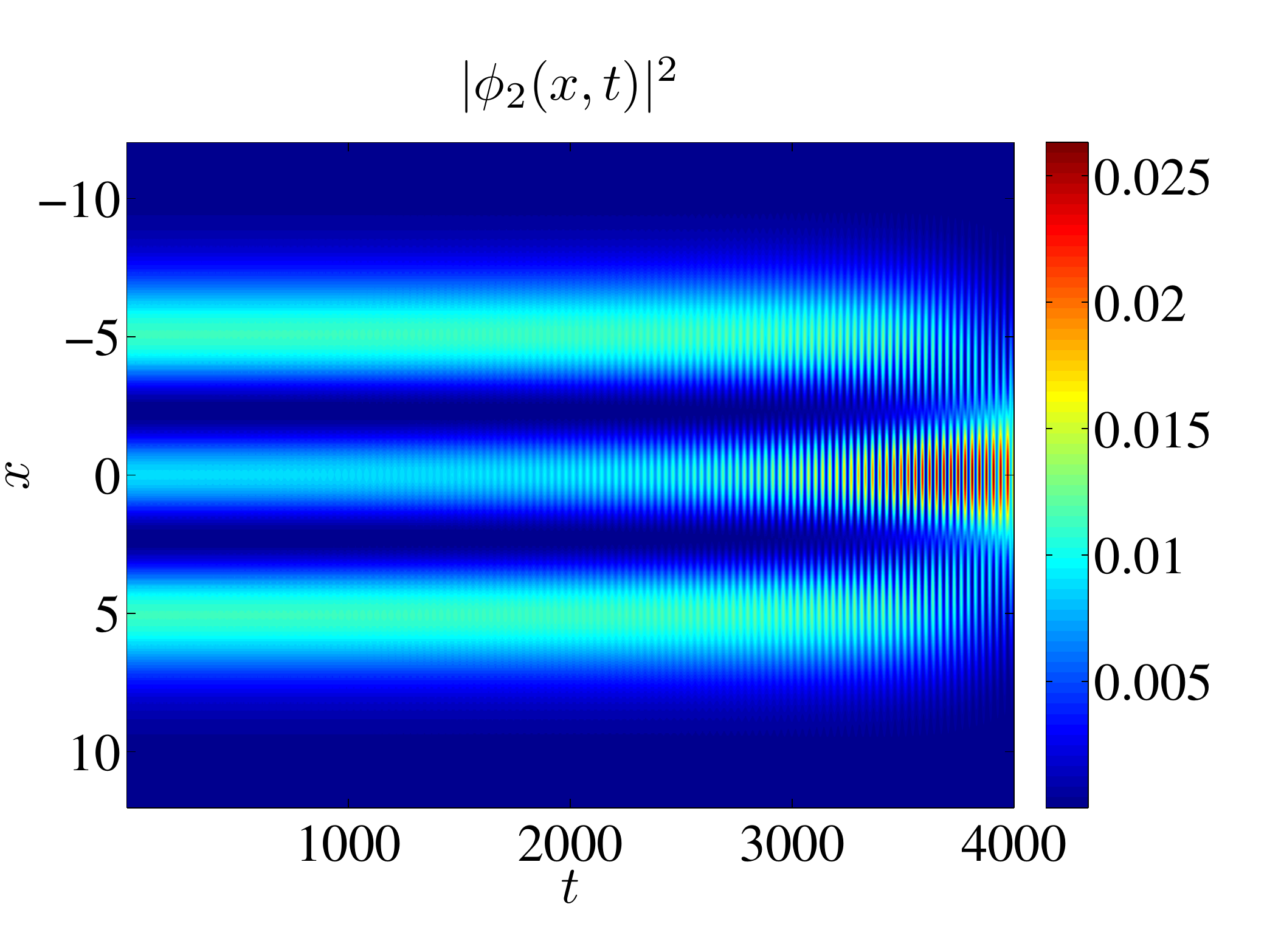}\\
\includegraphics[width=7cm]{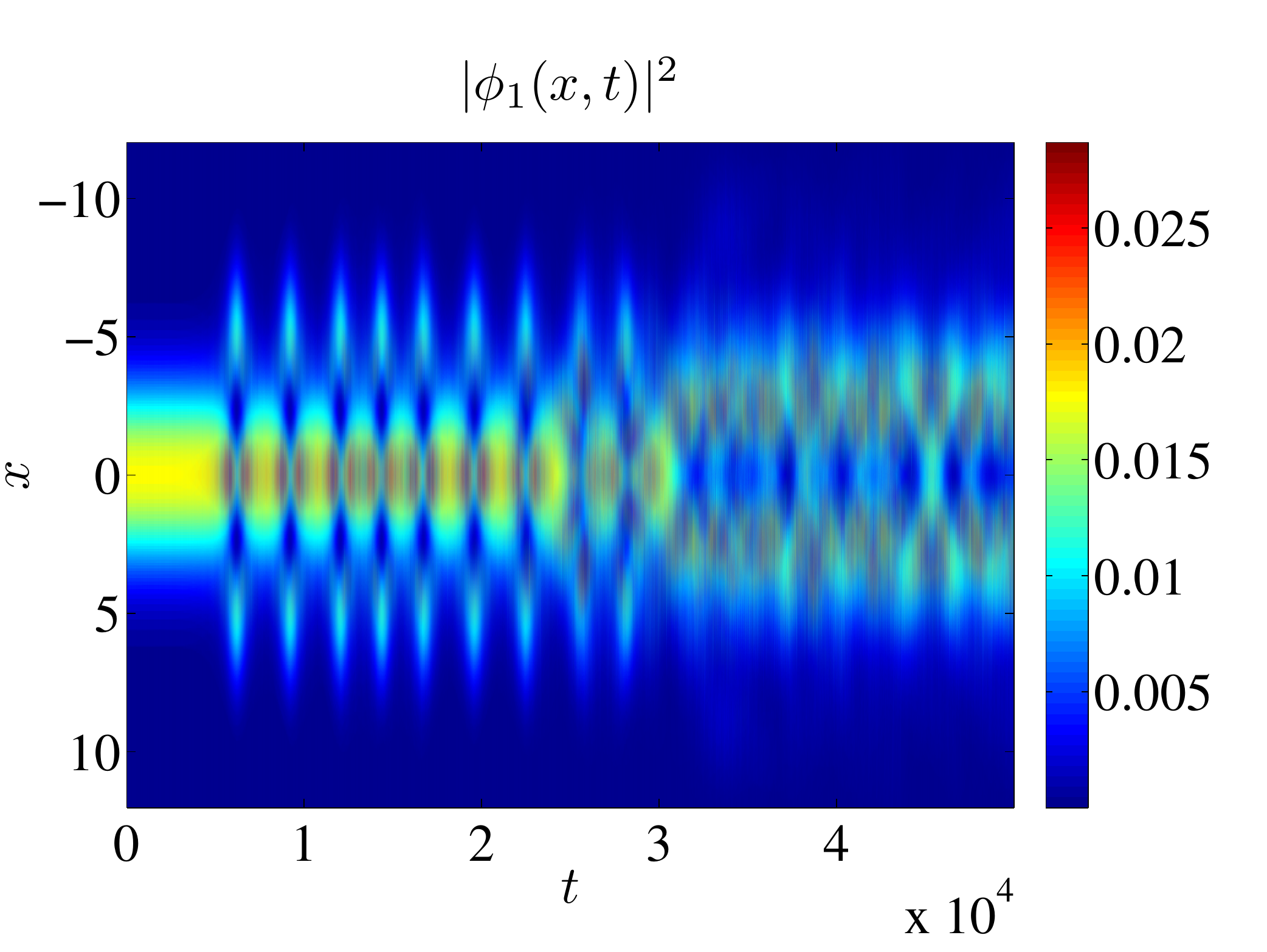}
\includegraphics[width=7cm]{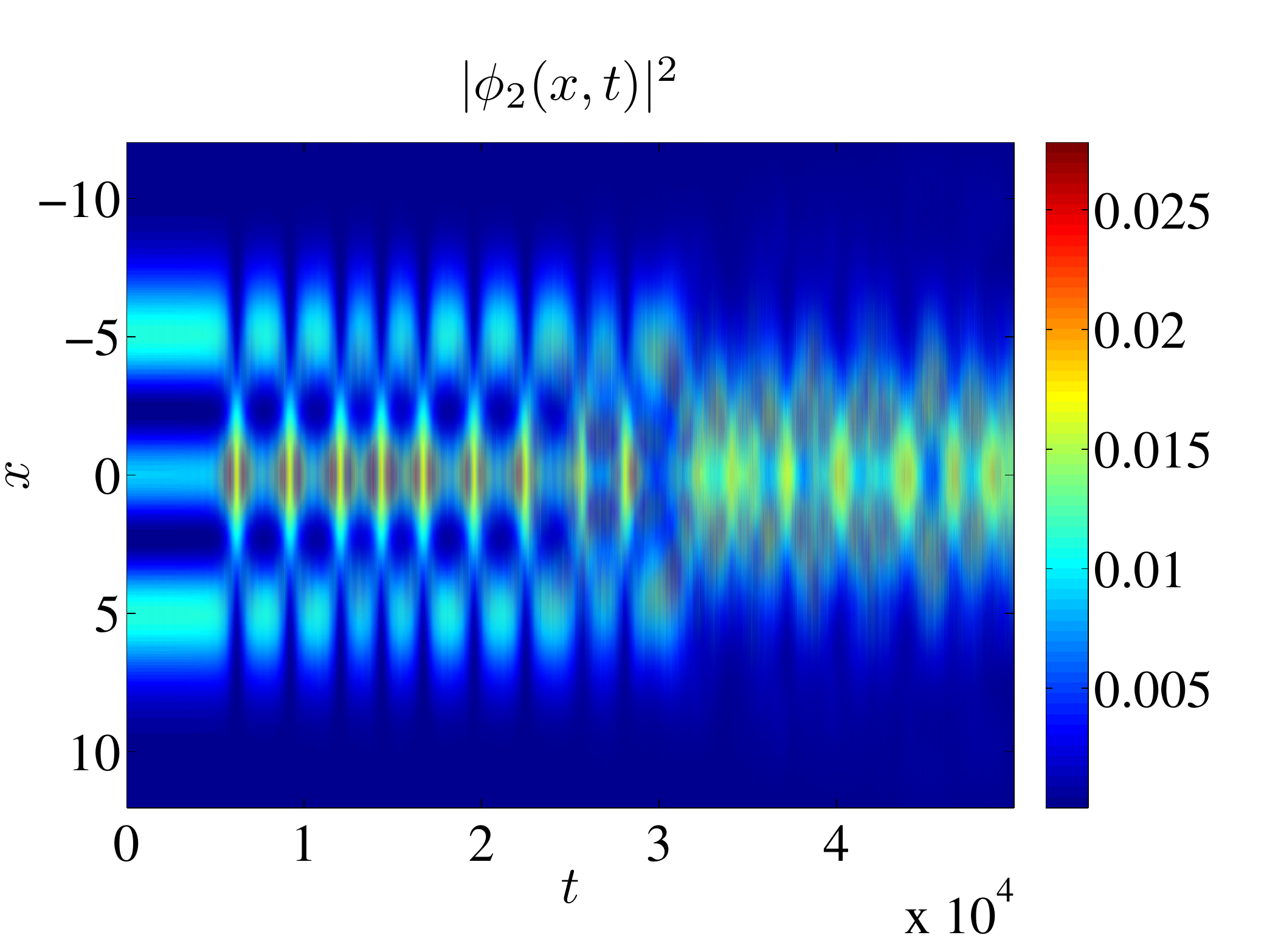}\\
\includegraphics[width=7cm]{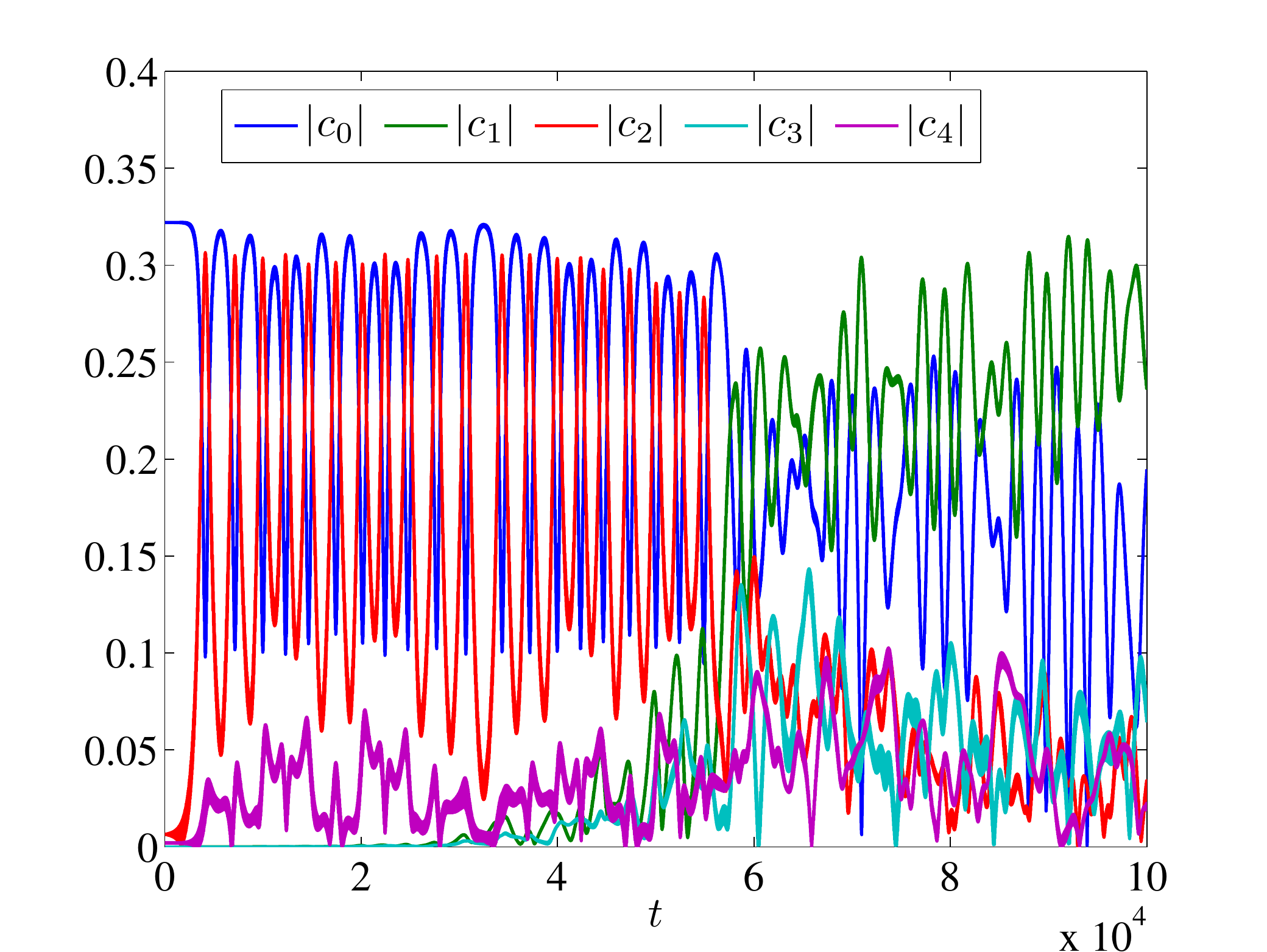}
\includegraphics[width=7cm]{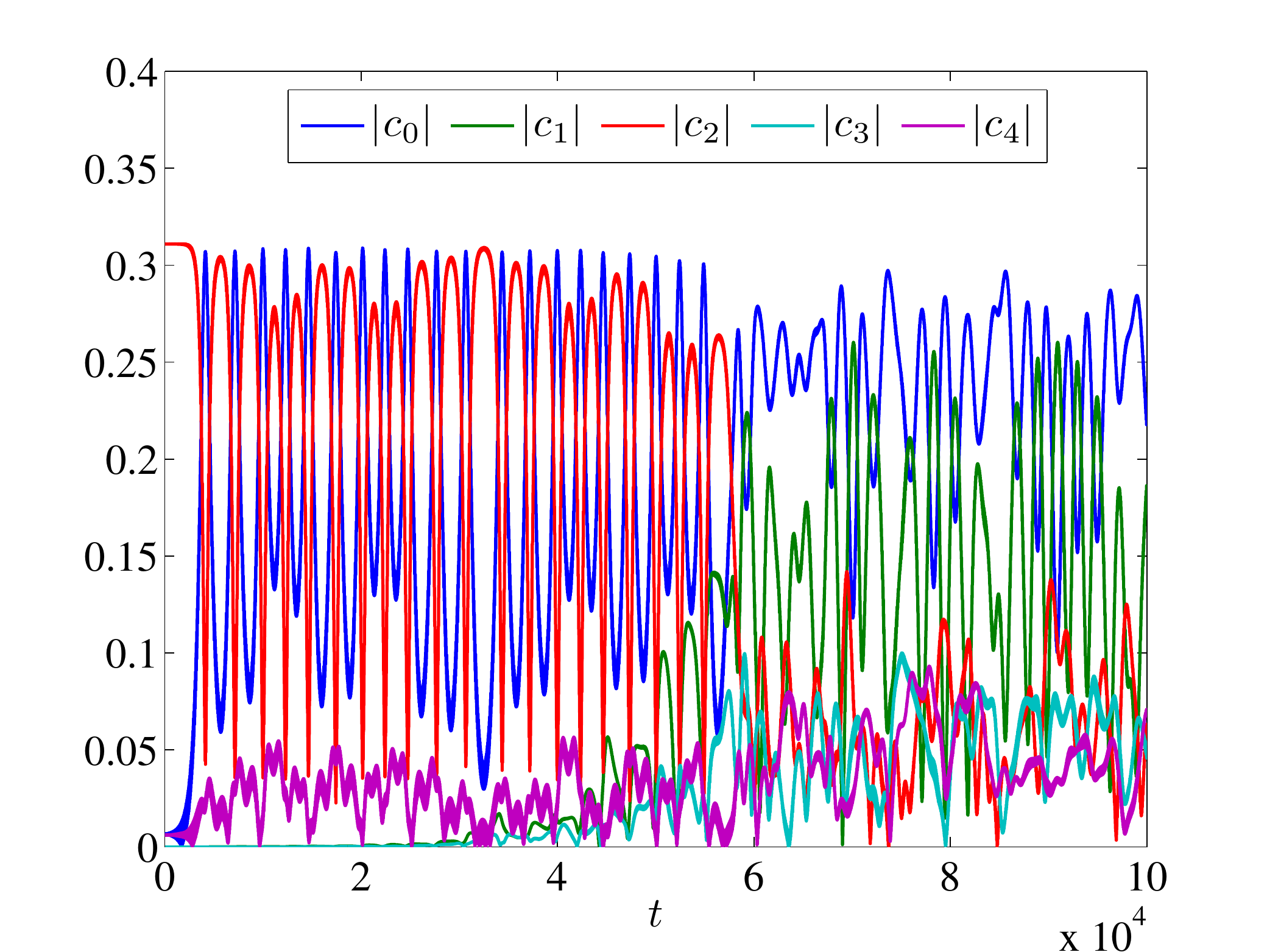}
\end{tabular}
\caption{ The top left (right) panel shows an example of the dynamics
of $|\phi_1|^2$ ($|\phi_2|^2$) for the $(0,2)$ branch
with $a=b=1, g_{11}=1, g_{12}=1, g_{22}=0.5$,
$\Omega=0.1$ and $\epsilon=0.1$. In the middle panels, we monitor the dynamics of the same solution for a longer time, revealing the two stage
nature of the evolution of the instability (see the relevant
discussion in the text). The bottom panels show the same dynamics (left for $\phi_1$ and right for $\phi_2$) using the dynamical decomposition to the
orthonormal basis $\{u_k\}$, where $c_k$ stands for the coefficient for $u_k$
in the decomposition. The transition between modes $(0,2)$ and $(2,0)$
originally to
 $(0,1)$ and $(1,0)$ eventually is evident in the mode dynamics.}
\label{fig1_4_5}
\end{figure}

However, in this case too, we can explore realistic scenarios where the
instability manifests itself immediately in the vicinity of the linear limit.
In particular, if $a=b=1, g_{11}=1, g_{12}=1, g_{22}=0.5, \Omega=0.1$, the
numerical computation shows that there exist two quartets of eigenvalues
(near $\pm i\Omega$ and $\pm i2\Omega$) that do not lie on the imaginary
axis, as shown in Fig.~\ref{fig1_4_3}. I.e., in this case, {\it both}
unstable eigendirections of the system are realized and, in fact, potentially
concurrently (contrary, e.g., to the case of $(1,1)$ waves). As $\epsilon$
grows, we observe that the complex pairs near $-i\Omega$ and $-i2\Omega$ tend
to come back to the imaginary axis and split along the axis, as shown in
Fig.~\ref{fig1_4_4}. We observe that $\phi_2$ vanishes at $\epsilon\approx
0.8$ where the branch of solutions meets the one-component branch of
solutions on $\phi_1$ there.
In Figure~\ref{fig1_4_5}, the
numerically-monitored dynamics of the steady-state solution with a small
initial perturbation verifies its instability. Here too, the instability
manifests its oscillatory character and weak growth rate over longer time
scales.
{In particular, the middle and bottom panels of Figure~\ref{fig1_4_5} show that the system first quantitatively alternates between unstable states $(0,2)$ and $(2,0)$ and then transits to the states that are close to $(0,1)$ and $(1,0)$. It can be checked that the dynamics of $|c_2|$ and $|c_0|$ comes with small oscillations with frequency close to $\frac{2\pi}{2\Omega}$, which implies that the instability in the first phase is related to the unstable eigenvalues near $\pm 2i\Omega$. Similarly, the time evolution of $|c_1|$ oscillates at the frequency of approximately $\frac{2\pi}{\Omega}$, which is connected to the unstable eigenvalues near $\pm i\Omega$.
We note that the two-phase time evolution shown in Figure~\ref{fig1_4_5} is typical for our setup.}


\subsection{$(m,n)=(1,2)$ (interchange all subscripts to obtain $(m,n)=(2,1)$)}

For $\mu_1^{(0)}=3\Omega/2$ and $\mu_2^{(0)}=5\Omega/2$ we consider the
continuation of $({\psi}_1^{(0)}, {\psi}_2^{(0)})=(a u_{1}, b u_{2})$, where
\[
a^2=4\sqrt{\frac{2\pi}{\Omega}}\,
\frac{41g_{22}\mu_1^{(1)}-28g_{12}\mu_2^{(1)}}{123 g_{11}g_{22}- 49g_{12}^2},\quad
b^2=16\sqrt{\frac{2\pi}{\Omega}}\,
\frac{12 g_{11}\mu_2^{(1)}-7g_{12}\mu_1^{(1)}}{123 g_{11}g_{22}-49g_{12}^2}.
\]
Regarding spectral stability we have $K_\mathrm{Ham}=6$, and the dangerous
eigenvalues are $\lambda^{(0)}=\pm i\Omega,\pm i2\Omega$. It is possible for
a pair of eigenvalues with nonzero real part to emerge from $\pm i\Omega$,
while at most one eigenvalue with positive real part can emerge from $\pm
i2\Omega$.

First consider the perturbation calculation with $\lambda^{(0)}=-i\Omega$. We
consider case (a), and $M_a\in\mathcal{M}_{4\times4}(\mathbb{R})$ is
\[
M_a=\frac{1}{64}\sqrt{\frac{\Omega}{2\pi}}\left(
    \begin{array}{cccc}
        8 g_{11} a^2 + 13 g_{12}b^2 & 8\sqrt{2}\,g_{11}a^2  & 5\sqrt{6}\,g_{12}a b & 28 g_{12}a b \\
        -8\sqrt{2}\,g_{11}a^2 & -16 g_{11} a^2+4 g_{12}b^2 & -4\sqrt{3}\,g_{12}a b & -8\sqrt{2}\, g_{12}a b \\
        5\sqrt{6}\, g_{12}a b  & 4\sqrt{3}\,g_{12}a b & 10 g_{22}b^2 - 6 g_{12}a^2 & 5\sqrt{6}\, g_{22} b^2 \\
        -28 g_{12}a b & -8\sqrt{2}\, g_{12}a b & -5\sqrt{6}\, g_{22} b^2 & -15 g_{22}b^2 - 20 g_{12}a^2
    \end{array}
\right).
\]
As per the dipolar mode that we discussed previously,
one eigenvalue of $M_a$ is zero, with associated eigenvector
$(-a/b,-a/(\sqrt{2}\,b,-\sqrt{6}/2,1)^\mathrm{T}$. Consequently, $M_a$ can
have at most one pair of eigenvalues with nonzero imaginary part, which
implies that at most one pair of eigenvalues with nonzero real part can
emerge under the perturbation. If we particularly set $g_{11}=g_{22}=1$ and
$a=b=1$, numerical results suggest a pair of complex conjugate eigenvalues
will arise for $g_{12}>1$, as illustrated in the example below.

\begin{figure}[!htbp]
\begin{tabular}{cc}
\includegraphics[width=8cm]{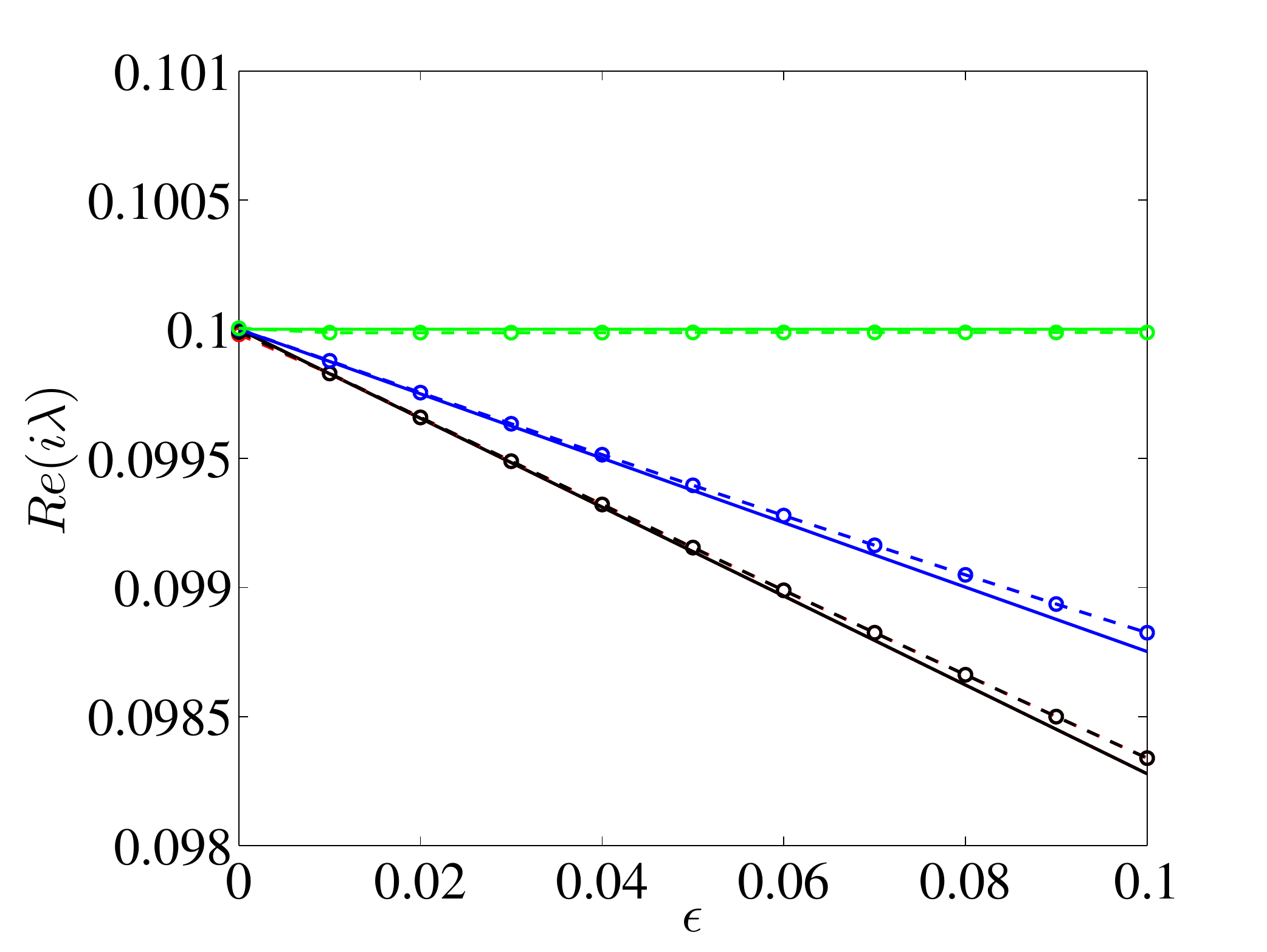}
\includegraphics[width=8cm]{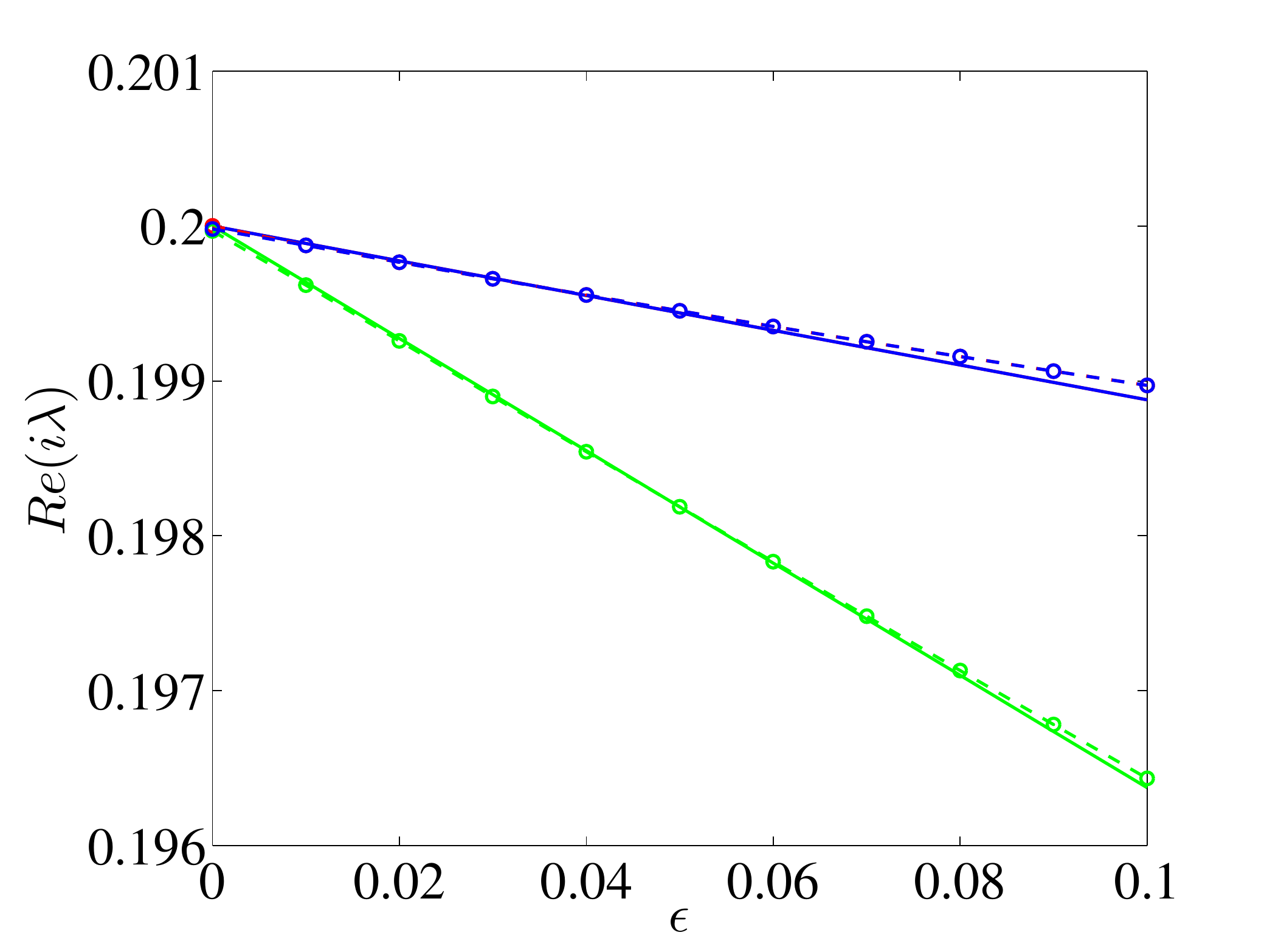}\\
\includegraphics[width=8cm]{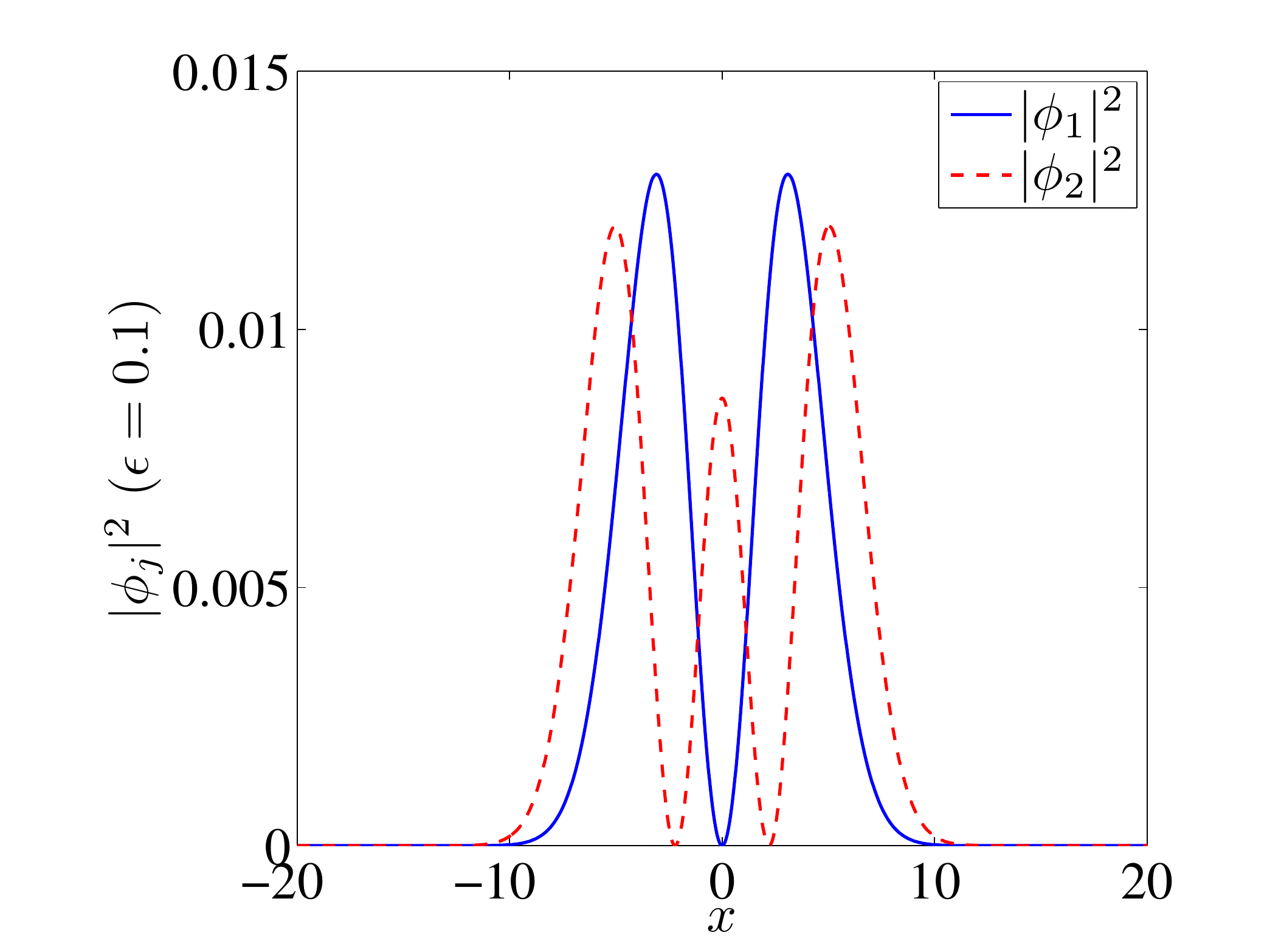}
\includegraphics[width=8cm]{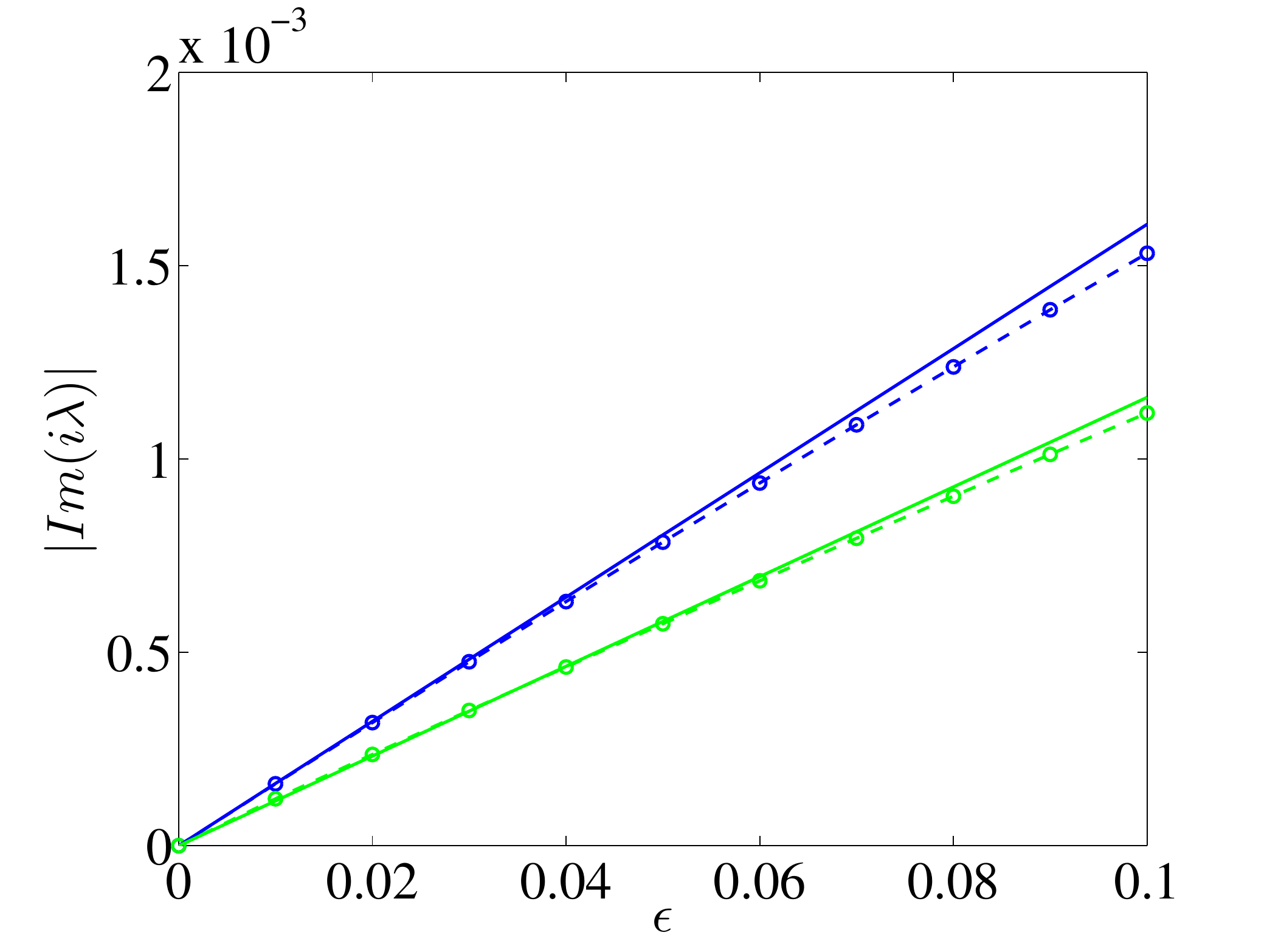}
\end{tabular}
\caption{ Case of $a=b=1, g_{11}=1, g_{12}=1.2, g_{22}=1,
\Omega=0.1$ for the $(1,2)$ branch:  The top left panel shows the real parts of $i\lambda$ around $\Omega$ as functions
of $\epsilon$ with $O(\epsilon)$ corrections (solid lines) and
corresponding numerical results (dashed lines with circles). The top right panel shows the real parts of $i \lambda$ around $2\Omega$. In both panels,
it should be noticed that the red lines and green lines (solid and dashed)
are almost identical since a pair of eigenvalues of $M_a$ and $M_b$ are complex conjugates. Moreover,
the nonzero imaginary parts
of these two pairs imply the instability of the solution, as shown in the bottom right panel
(the growth rates are shown as
solid lines for the predicted $O(\epsilon)$ corrections and as dashed lines for the
numerical computation of the real parts of the eigenvalues; green for $M_b$ with $\lambda^{(0)}=-i2\Omega$
and blue for $M_a$ with $\lambda^{(0)}=-i\Omega$). The densities of $\phi_1$ and $\phi_2$ at $\epsilon=0.1$
are given in the bottom left panel.}
\label{fig1_5}
\end{figure}

Now we consider the perturbation calculation with $\lambda^{(0)}=-i2\Omega$.
We consider case (b), and $M_b\in\mathcal{M}_{3\times3}(\mathbb{R})$ is
\[
\frac{1}{512}\sqrt{\frac{\Omega}{2\pi}}\left(
    \begin{array}{ccc}
        -32 g_{11} a^2-20 g_{12}b^2 & 46\sqrt{2}\,g_{12}a b & 32\sqrt{3}\, g_{12}a b \\
        46\sqrt{2}\, g_{12}a b  &  g_{22}b^2 - 74 g_{12}a^2 & 12\sqrt{6}\, g_{22}b^2 \\
        -32\sqrt{3}\, g_{12}a b & -12\sqrt{6}\, g_{22}b^2 & -56 g_{22}b^2 - 32 g_{12}a^2
    \end{array}
\right).
\]
This matrix can have at most one pair of eigenvalues with nonzero imaginary
part, an example of which can be obtained for $g_{11}=g_{22}=1$, $a=b=1$ and
$g_{12}>0$ (see Fig.~\ref{fig1_5}).

\begin{figure}[!htbp]
\begin{tabular}{ccc}
\includegraphics[width=5cm]{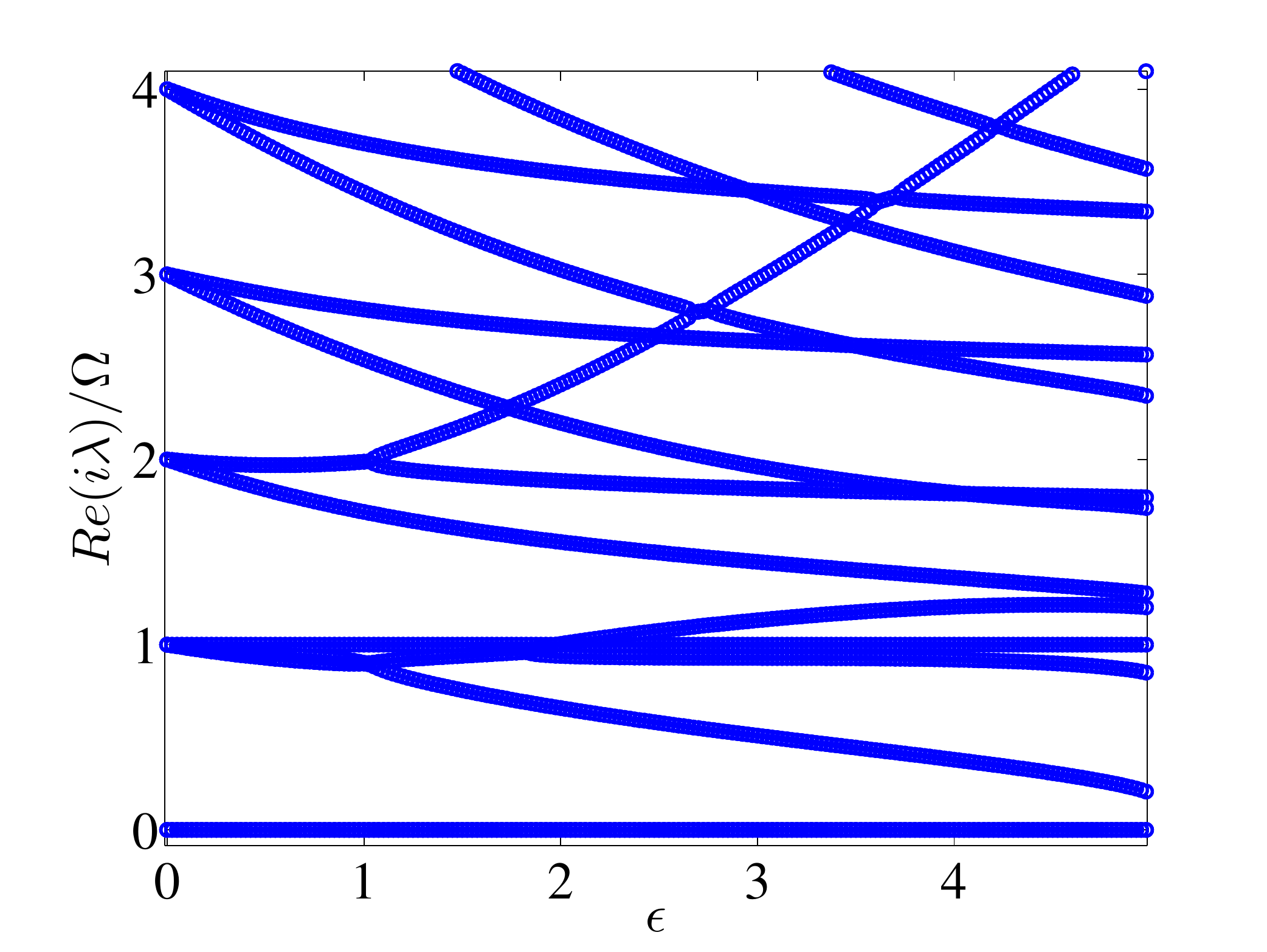}
\includegraphics[width=5cm]{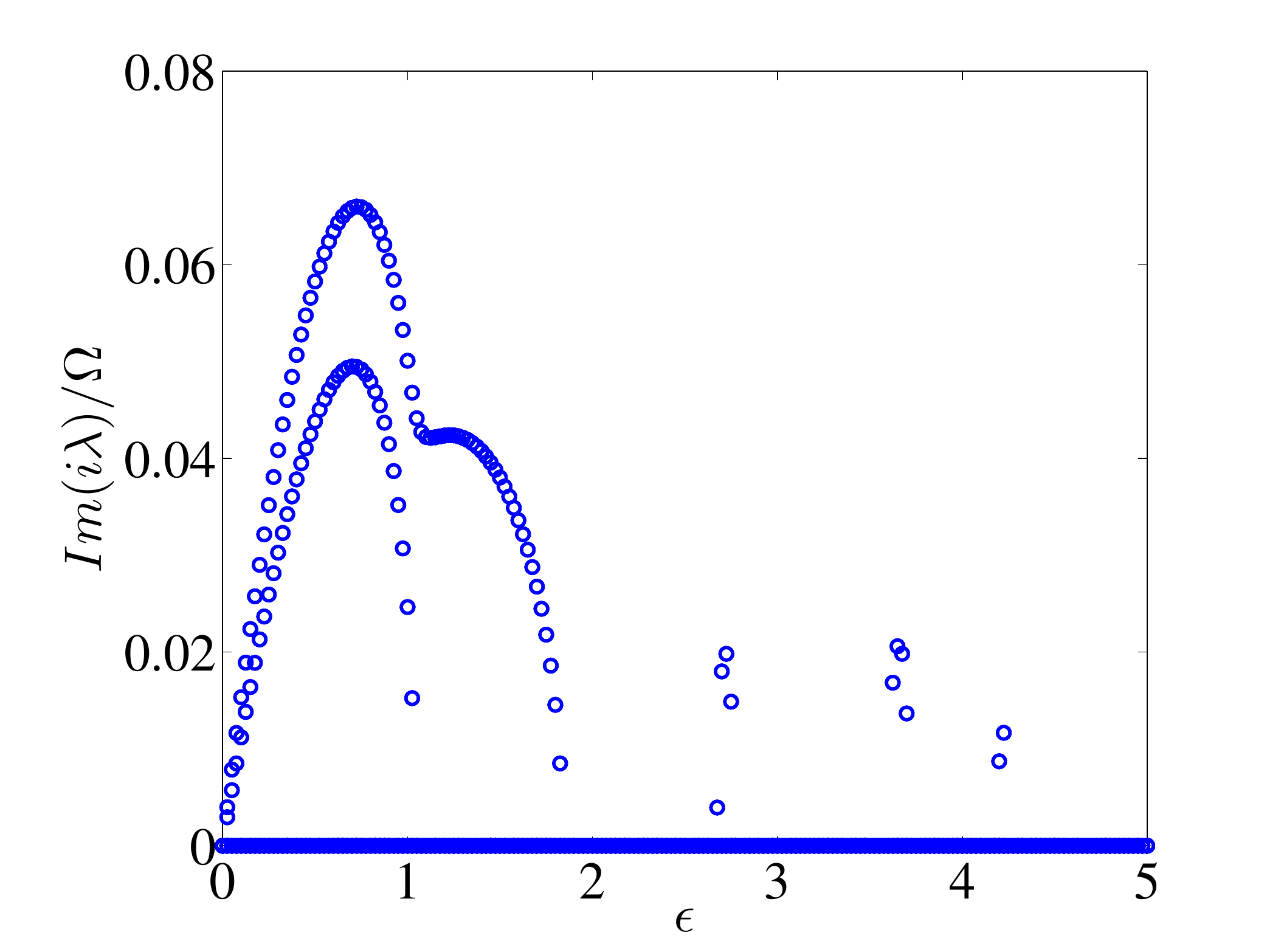}
\includegraphics[width=5cm]{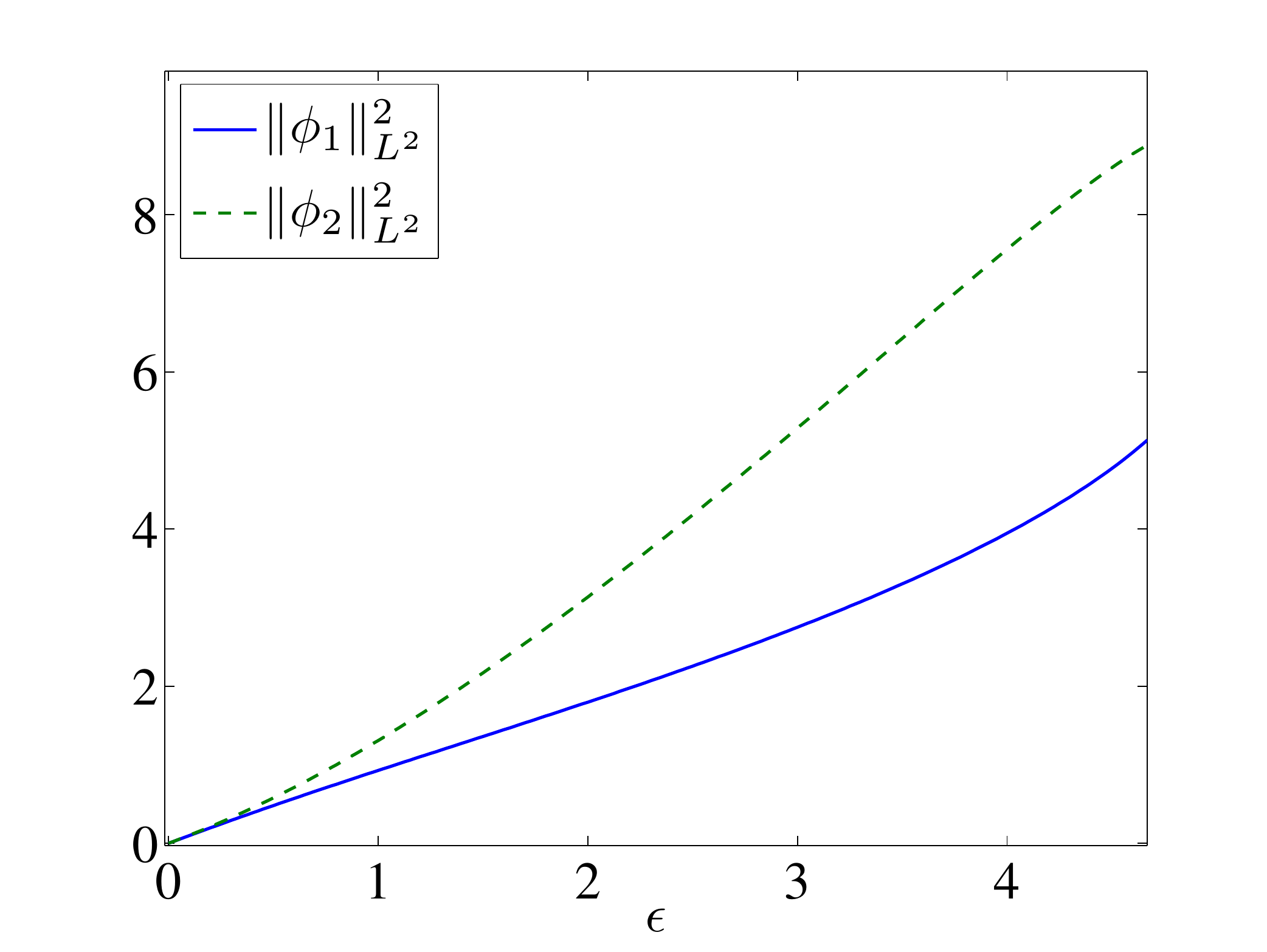}
\end{tabular}
\caption{ The left (middle) panel shows the change of imaginary (real) parts of the eigenvalues
for the $(1,2)$ branch of solutions with $a=b=1, g_{11}=1, g_{12}=1.2, g_{22}=1, \Omega=0.1$. In this
extended parametric continuation, the
signature of the instabilities is evident in the middle panel of
the figure. The right panel shows the change of the $L^2$-norm of the solution of $\phi_j$ over $\epsilon$.}
\label{fig1_5_1}
\end{figure}

For the numerical calculations we let $a=b=1, g_{11}=1, g_{12}=1.2, g_{22}=1,
\Omega=0.1$ and we compare our analytical predictions for the leading order
corrections to the eigenvalues up to $O(\epsilon)$ against the corresponding
numerical eigenvalues in Fig.~\ref{fig1_5}. We find two eigenvalue pairs (one
pair each near $\pm i\Omega$ and $\pm i2\Omega$) introducing respective
instability eigendirections.
In Fig.~\ref{fig1_5_1}, we see that the pairs near $-i\Omega$ and $-i2\Omega$
will eventually return to the imaginary axis, over considerably wider
parametric continuations in $\epsilon$, splitting along the axis.
Among these returned imaginary eigenvalues, the one that stems from
$-i2\Omega$ and goes upward will meet the eigenvalue coming from $-i4\Omega$
to generate another pair of complex eigenvalues at $\epsilon\approx 2.7$.
Shortly after (parametrically), these complex eigenvalues will come back to
the axis and split into two eigenvalue pairs, with one of them going up to
further repeat this process at $\epsilon\approx 3.6$ and $\epsilon\approx 4.2$.

\begin{figure}[!htbp]
\begin{tabular}{cc}
\includegraphics[width=7cm]{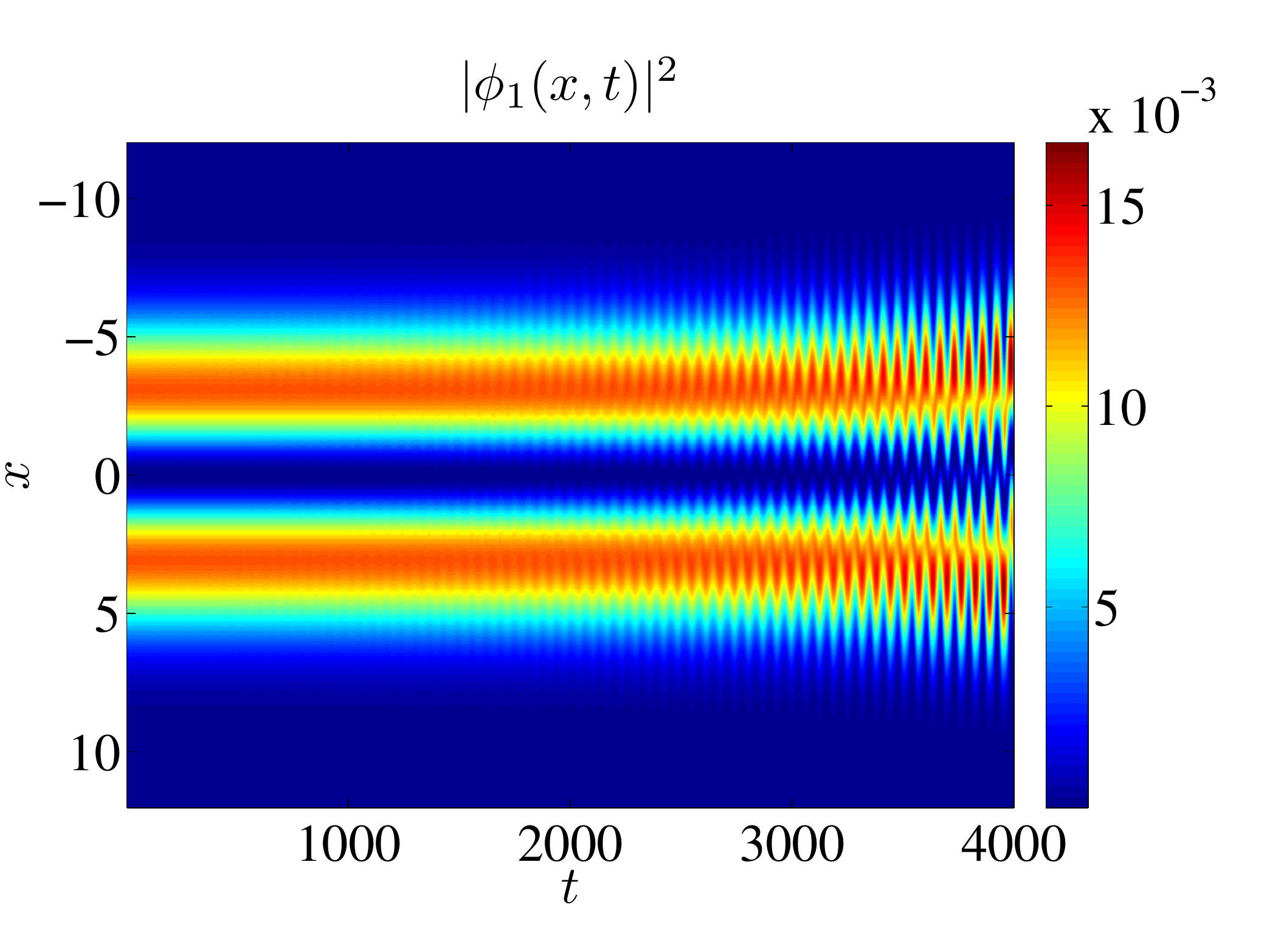}
\includegraphics[width=7cm]{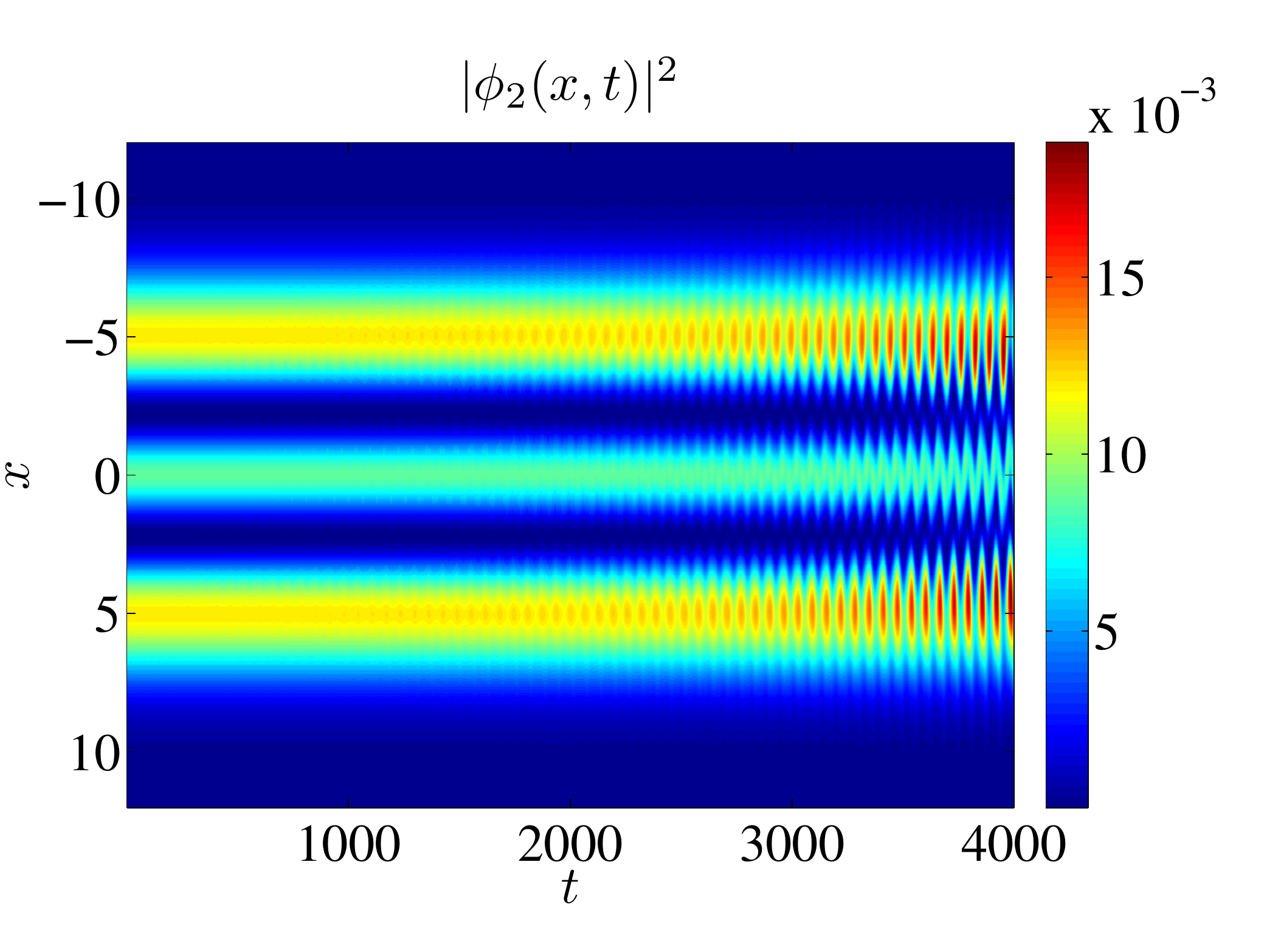}
\end{tabular}
\caption{ The left (right) panel shows an example of the dynamics of $|\phi_1|^2$ ($|\phi_2|^2$) for the $(1,2)$ state
with $a=b=1, g_{11}=1, g_{12}=1.2, g_{22}=1$, $\Omega=0.1$ and $\epsilon=0.1$. Again, the oscillatory
nature of the associated instability
eventually kicks in over longer time scales.}
\label{fig1_5_2}
\end{figure}

In Figure~\ref{fig1_5_2}, we illustrate the numerical evolution of the
unstable configuration shown in Figure \ref{fig1_5} with $\epsilon=0.1$. With
a small initial perturbation, the oscillation around the stationary solution
gradually grows and the instability becomes apparent in the dynamics.

\subsection{$(m,n)=(2,2)$}

When $\mu_1^{(0)}=\mu_2^{(0)}=5\Omega/2$ we consider $({\psi}_1^{(0)},
{\psi}_2^{(0)})=(a u_{2}, b u_{2})$, where
\[
a^2=\frac{64}{\sqrt{41}}\sqrt{\frac{2\pi}{\Omega}}\,
\frac{g_{22}\mu_1^{(1)}-g_{12}\mu_2^{(1)}}{g_{11}g_{22}-g_{12}^2},\quad
b^2=\frac{64}{\sqrt{41}}\sqrt{\frac{2\pi}{\Omega}}\,
\frac{g_{11}\mu_2^{(1)}-g_{12}\mu_1^{(1)}}{g_{11}g_{22}-g_{12}^2}.
\]
Regarding spectral stability we have $K_\mathrm{Ham}=8$, and the dangerous
eigenvalues are $\lambda^{(0)}=\pm i\Omega,\pm i2\Omega$. It is possible for
two pairs of eigenvalues with nonzero real part to emerge from each of the
dangerous eigenvalues.

First consider the perturbation calculation associated with
$\lambda^{(0)}=-i\Omega$. We have case (a), and the matrix $M_a$ is
\[
M_a=\frac{1}{128}\sqrt{\frac{\Omega}{2\pi}}\left(
    \begin{array}{cccc}
        20 g_{11} a^2 - 31 g_{12}b^2 & 10\sqrt{6}\,g_{11}a^2  & 51 g_{12}a b & 10\sqrt{6}\, g_{12}a b \\
        -10\sqrt{6}\,g_{11}a^2 & -30 g_{11} a^2 +26 g_{12}b^2 & -10\sqrt{6}\,g_{12}a b & -56 g_{12}a b \\
        51 g_{12}a b  & 10\sqrt{6}\,g_{12}a b & 20 g_{22}b^2 - 31 g_{12}a^2 & 10\sqrt{6}\, g_{22} b^2 \\
        -10\sqrt{6}\, g_{12}a b & -56 g_{12}a b & -10\sqrt{6}\, g_{22} b^2 & -30 g_{22}b^2 + 26 g_{12}a^2
    \end{array}
\right).
\]
One of the eigenvalues is zero, with associated eigenvector
$(-\sqrt{6}a/(2b),a/b,-\sqrt{6}/2,1)^\mathrm{T}$, for the same
(dipolar) symmetry reasons as before. Consequently, $M_a$ can
have at most one pair of eigenvalues with nonzero imaginary part, so at most
one pair of eigenvalues with nonzero real part can emerge from $\pm i\Omega$.
As an example, if we assume $a=b=1$ and $g_{11}=g_{22}=1$, then the other
eigenvalues of $M_a$ are
\[
-\frac5{64}\sqrt{\frac{\Omega}{2\pi}}\,(1+g_{12}),\quad
\sqrt{\frac{\Omega}{2\pi}}\,
\frac{-5\pm\sqrt{25-2900 g_{12}+6124g_{12}^2}}{128}
\]
where the instability (nonzero imaginary parts) will emerge for
\[
g_{12}\in (\frac{725-285\sqrt{6}}{3062}, \frac{725+285\sqrt{6}}{3062} )\approx (0.0088, 0.4648).
\]
Now consider the perturbation calculation associated with
$\lambda^{(0)}=-i2\Omega$. We again have case (a), and the matrix $M_a$ is
now
\[
M_a=\frac{1}{1024}\sqrt{\frac{\Omega}{2\pi}}\left(
    \begin{array}{cccc}
        2 g_{11} a^2 - 327 g_{12}b^2 & 24\sqrt{6}\,g_{11}a^2  & 329 g_{12}a b & 24\sqrt{6}\, g_{12}a b \\
        -24\sqrt{6}\,g_{11}a^2 & -112 g_{11} a^2 +272 g_{12}b^2 & -24\sqrt{6}\,g_{12}a b & -384 g_{12}a b \\
        329 g_{12}a b  & 24\sqrt{6}\,g_{12}a b & 2 g_{22}b^2 - 327 g_{12}a^2 & 24\sqrt{6}\, g_{22} b^2 \\
        -24\sqrt{6}\, g_{12}a b & -384 g_{12}a b & -24\sqrt{6}\, g_{22} b^2 & -112 g_{22}b^2 + 272 g_{12}a^2
    \end{array}
\right).
\]
If we set $a=b=1$ and $g_{11}=g_{22}=1$, then the eigenvalues of $M_a$ are
\[
\sqrt{\frac{\Omega}{2\pi}}\,
\frac{-55(1+g_{12}) \pm i3\sqrt{23}\,|1+g_{12}|}{1024},\quad
\sqrt{\frac{\Omega}{2\pi}}\,
\frac{-55\pm \sqrt{-207-67872 g_{12}+426880g_{12}^2}}{1024}.
\]
Minimally one pair of eigenvalues will gain nonzero imaginary part, and if
\[
g_{12}\in (\frac{3(1414-599\sqrt{6})}{53360}, \frac{3(1414+599\sqrt{6})}{53360}) \approx (-0.0030, 0.1620)
\]
two pairs of eigenvalues with nonzero imaginary part will emerge.

\begin{figure}[!htbp]
\begin{tabular}{cc}
\includegraphics[width=8cm]{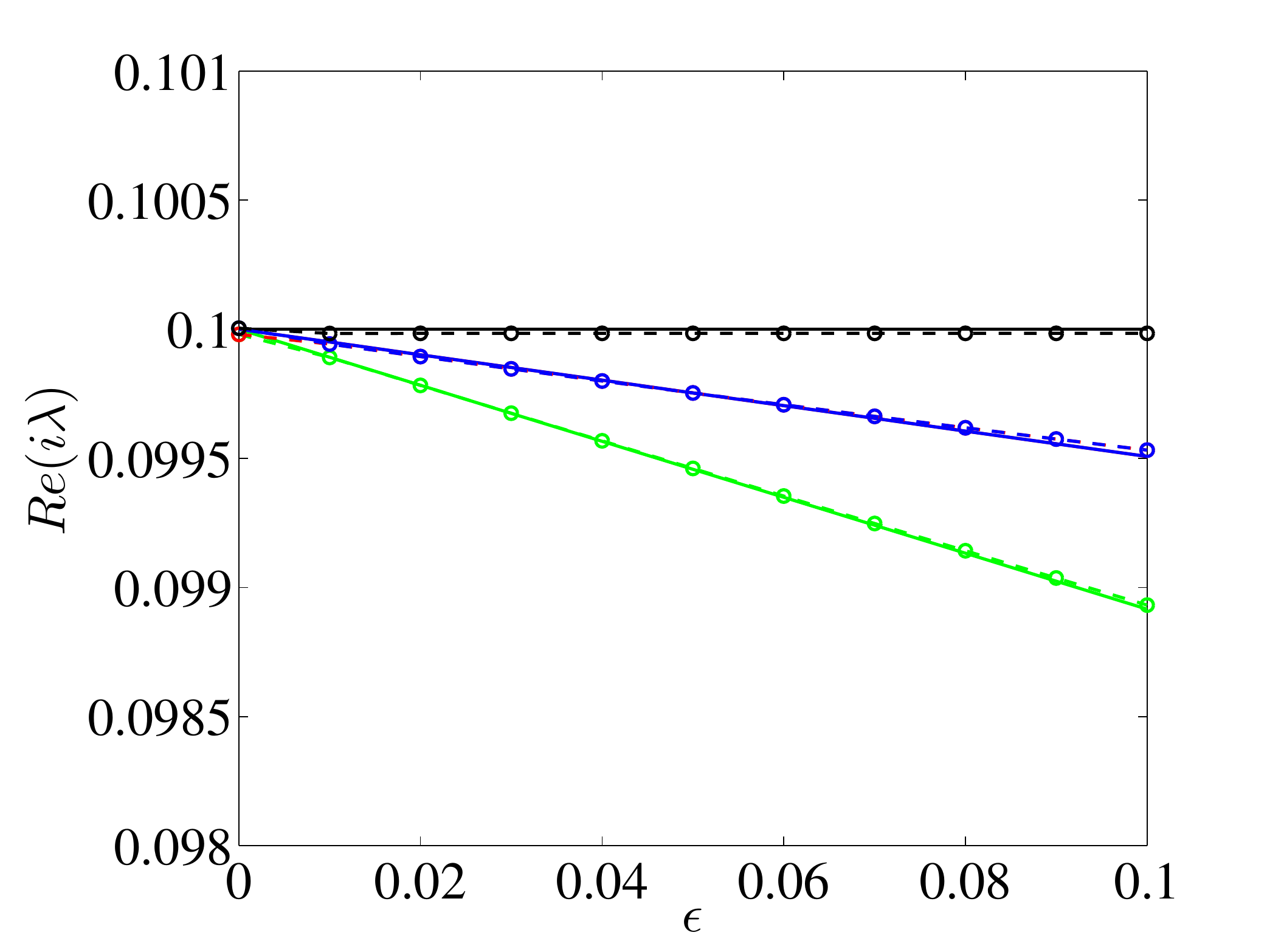}
\includegraphics[width=8cm]{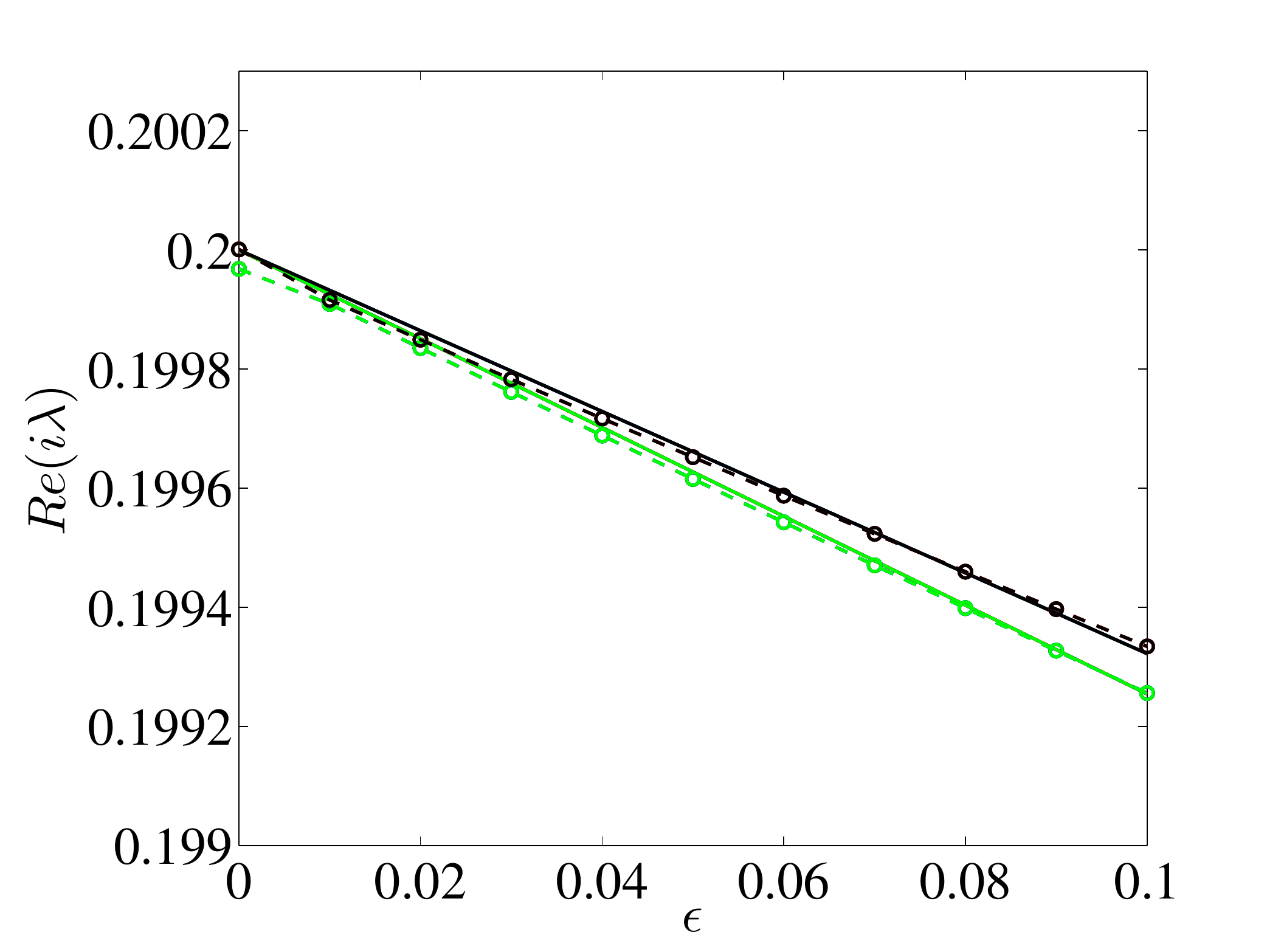}\\
\includegraphics[width=8cm]{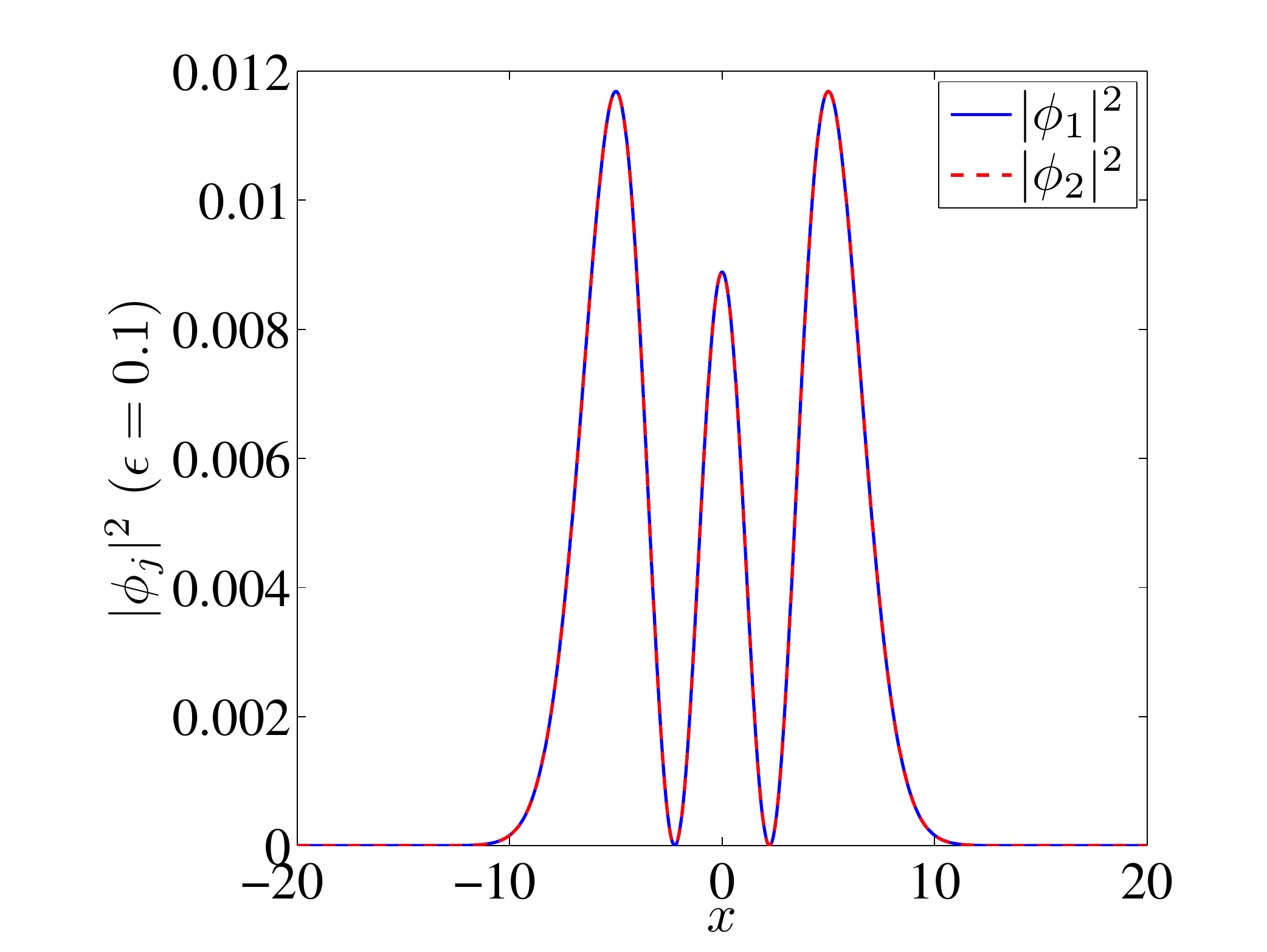}
\includegraphics[width=8cm]{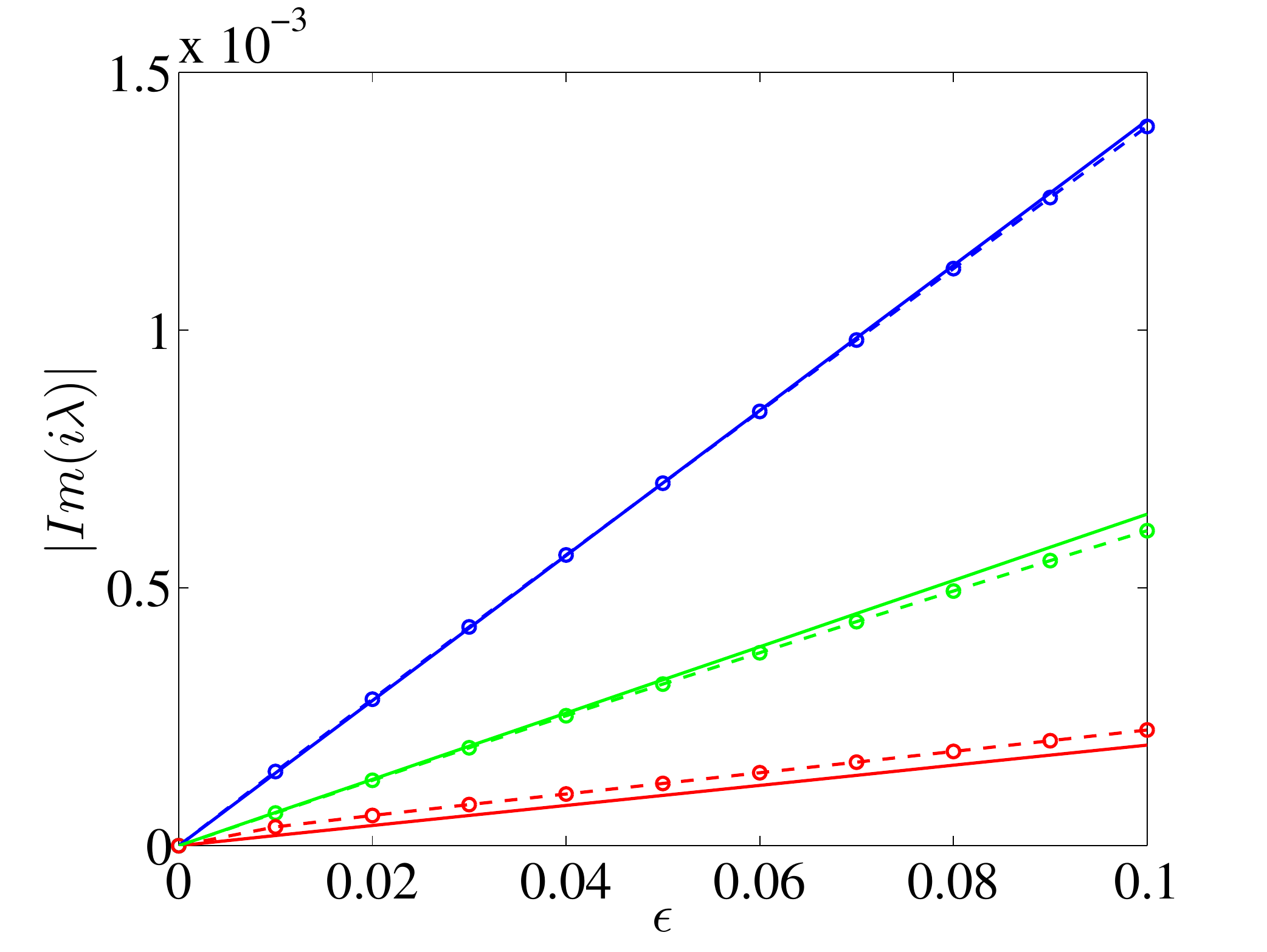}
\end{tabular}
\caption{ Case of
$a=b=1, g_{11}=1, g_{12}=0.1, g_{22}=1, \Omega=0.1$ for the
$(2,2)$ branch: The top left panel shows the imaginary parts of the eigenvalues around $-i\Omega$ as functions of
$\epsilon$ with $O(\epsilon)$ corrections (solid lines) and
corresponding numerical results for comparison
(dashed lines with circles). The red and green lines are almost identical
since one pair of the eigenvalues of $M_a$ for $\lambda^{(0)}=-i\Omega$ corresponds
to complex conjugates. The top right panel shows the imaginary parts of the eigenvalues around $-i2\Omega$.
It should be noticed again that the red lines and green lines (solid and dashed) are essentially
identical and the blue lines and black lines (solid and dashed) are almost the same since two
quartets of eigenvalues of $M_a$ for $\lambda^{(0)}=-i2\Omega$ arise in this case. Moreover,
the nonzero real parts
of $\lambda$'s also imply the instability of the solution, as shown in the bottom
right panel (solid lines for the $O(\epsilon)$ correction using this pair of complex
conjugates and the dashed lines for the numerical computation of the real parts of the
eigenvalues; green is used for $M_a$ for $\lambda^{(0)}=-i\Omega$ while red and blue
denote the imaginary parts for the $M_a$ with $\lambda^{(0)}=-i2\Omega$ eigenvalues).
The densities of $\phi_1$ and $\phi_2$ at $\epsilon=0.1$ are given in the
bottom left panel, showcasing the second excited state nature of both
fields. }
\label{fig1_6}
\end{figure}

\begin{figure}[!htbp]
\begin{tabular}{cc}
\includegraphics[width=5cm]{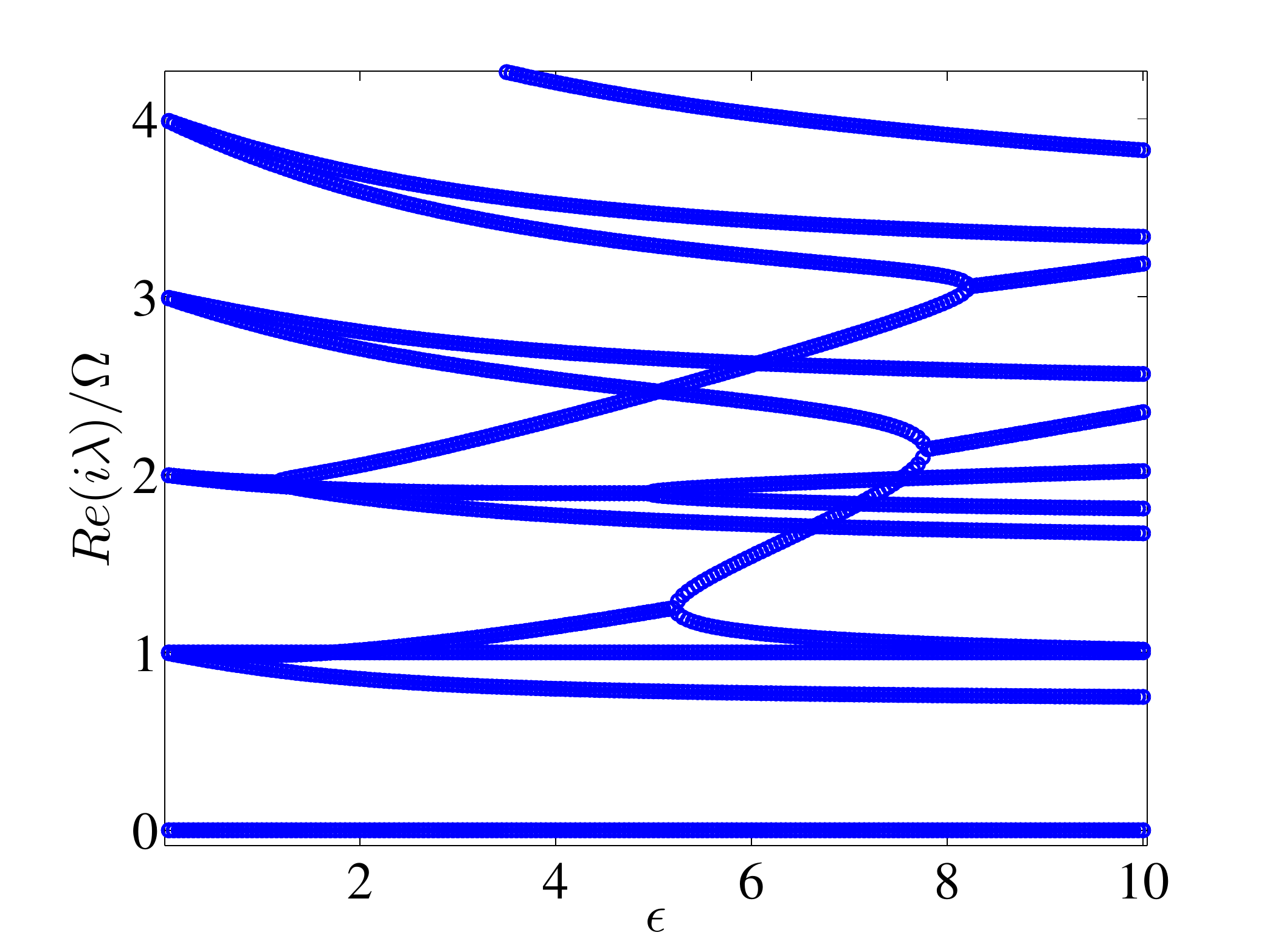}
\includegraphics[width=5cm]{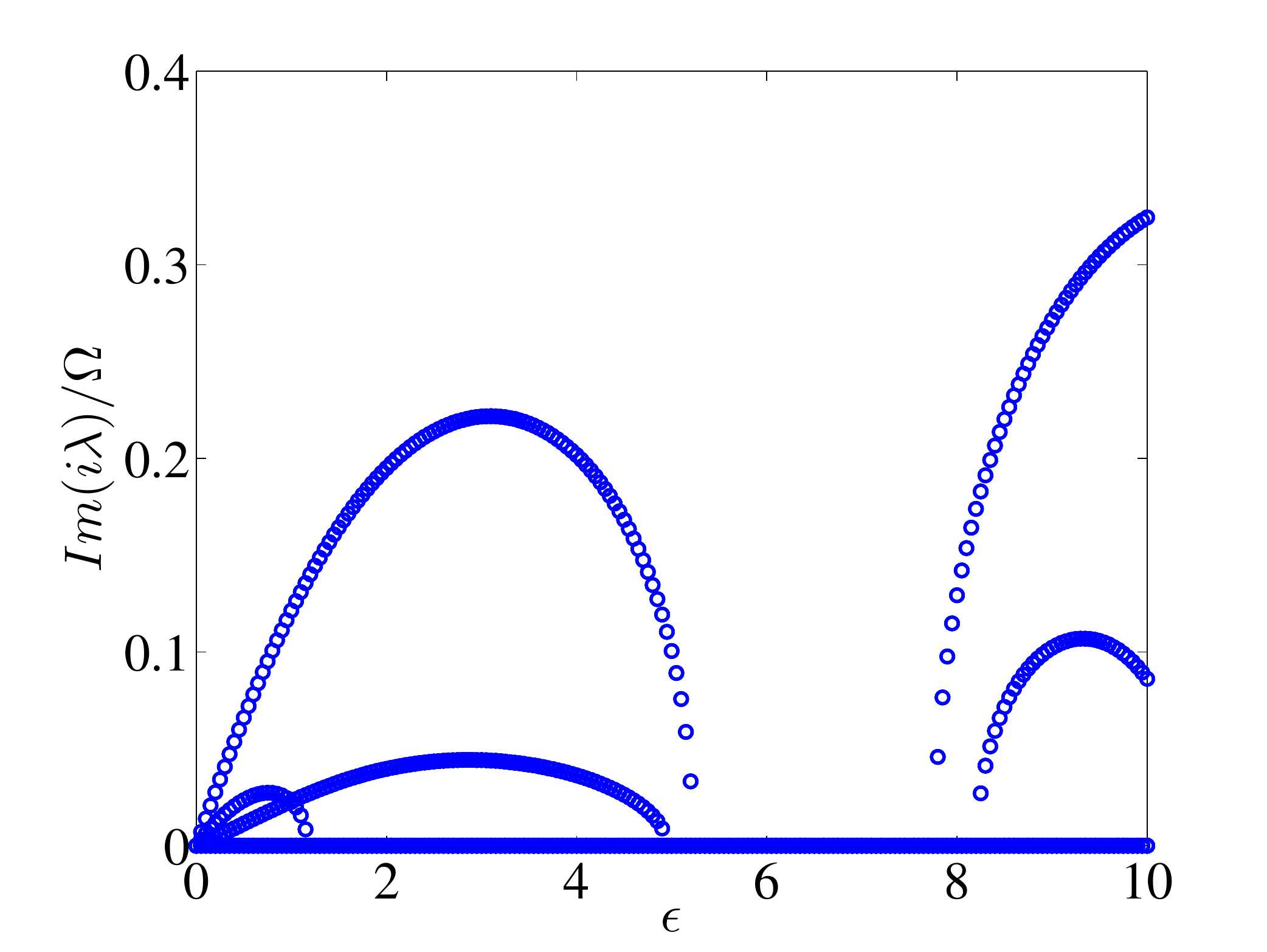}
\includegraphics[width=5cm]{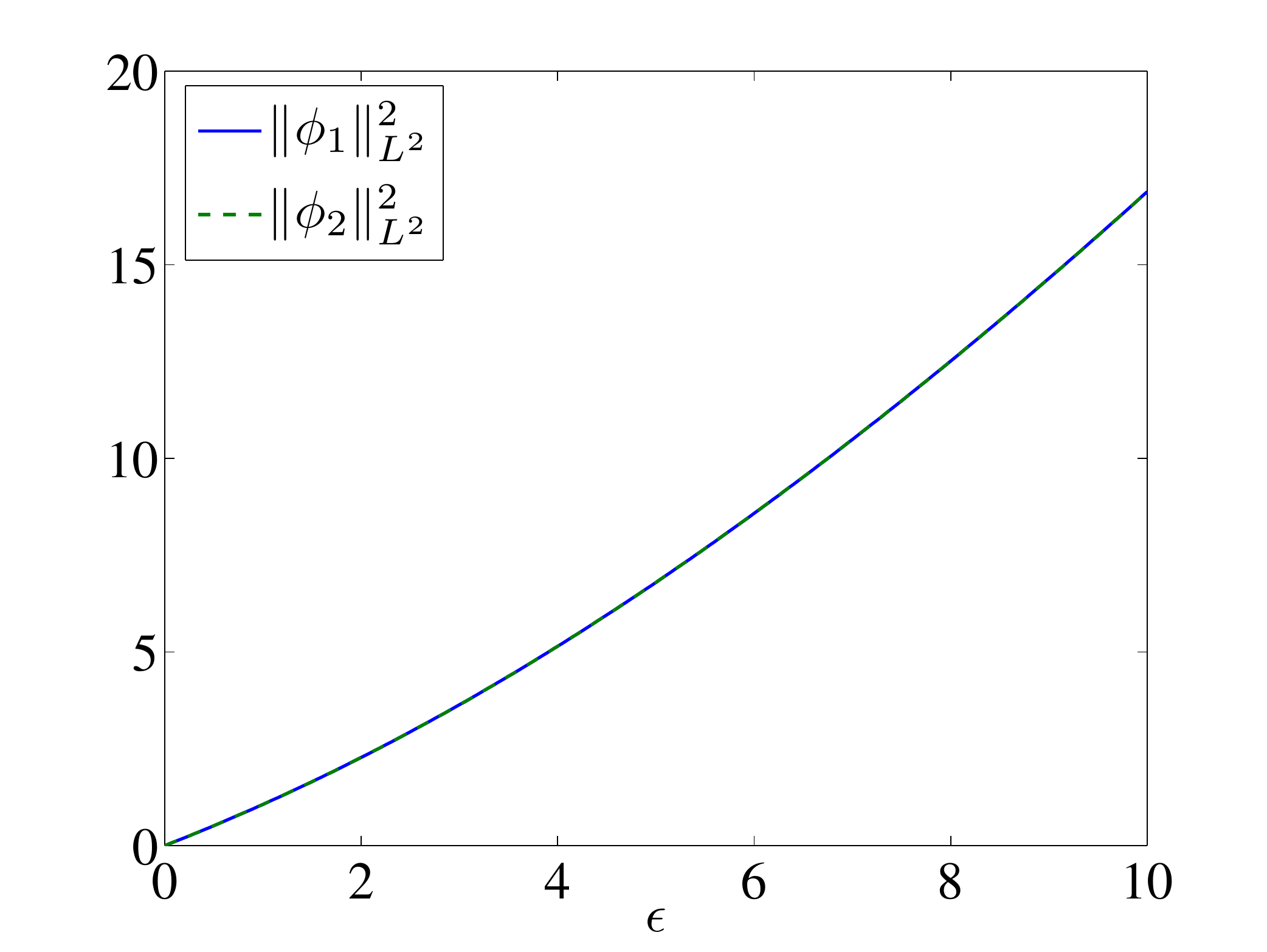}
\end{tabular}
\caption{ The left (middle) panel shows the change of imaginary (real) parts of the
eigenvalues for the $(2,2)$ branch of solutions with $a=b=1, g_{11}=1, g_{12}=0.1, g_{22}=1, \Omega=0.1$.
In this case, the middle panel illustrates the potential
of the configuration for 3 instabilities all of which are manifested
(possibly even concurrently for small $\epsilon$). Interestingly though,
for sufficiently large $\epsilon$, there exists a potential
parametric interval of spectral stability. In the right panel, we plot the $L^2$-norm of the solution of $\phi_j$ as a function of $\epsilon$.}
\label{fig1_6_1}
\end{figure}

\begin{figure}[!htbp]
\begin{tabular}{cc}
\includegraphics[width=7cm]{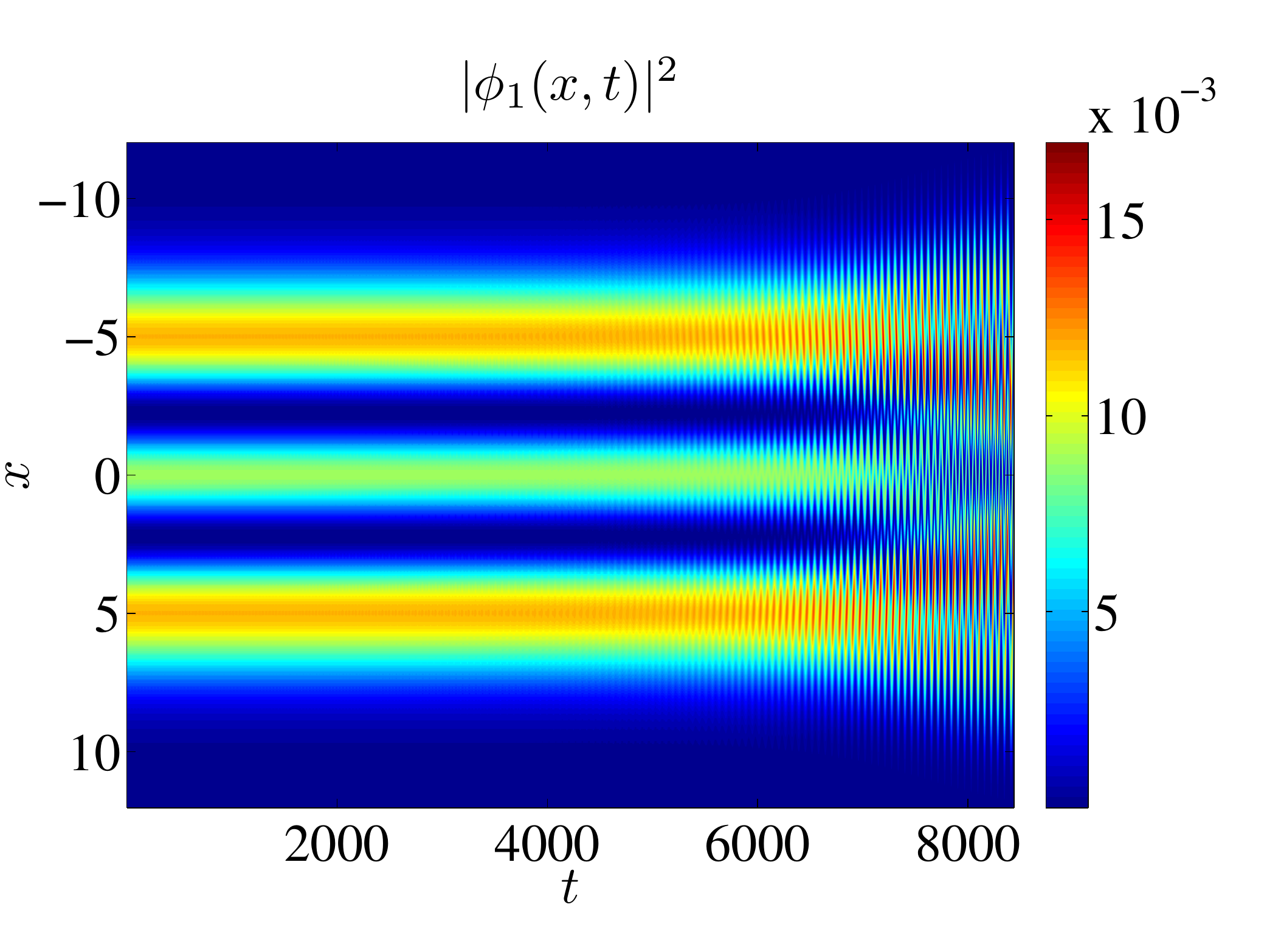}
\includegraphics[width=7cm]{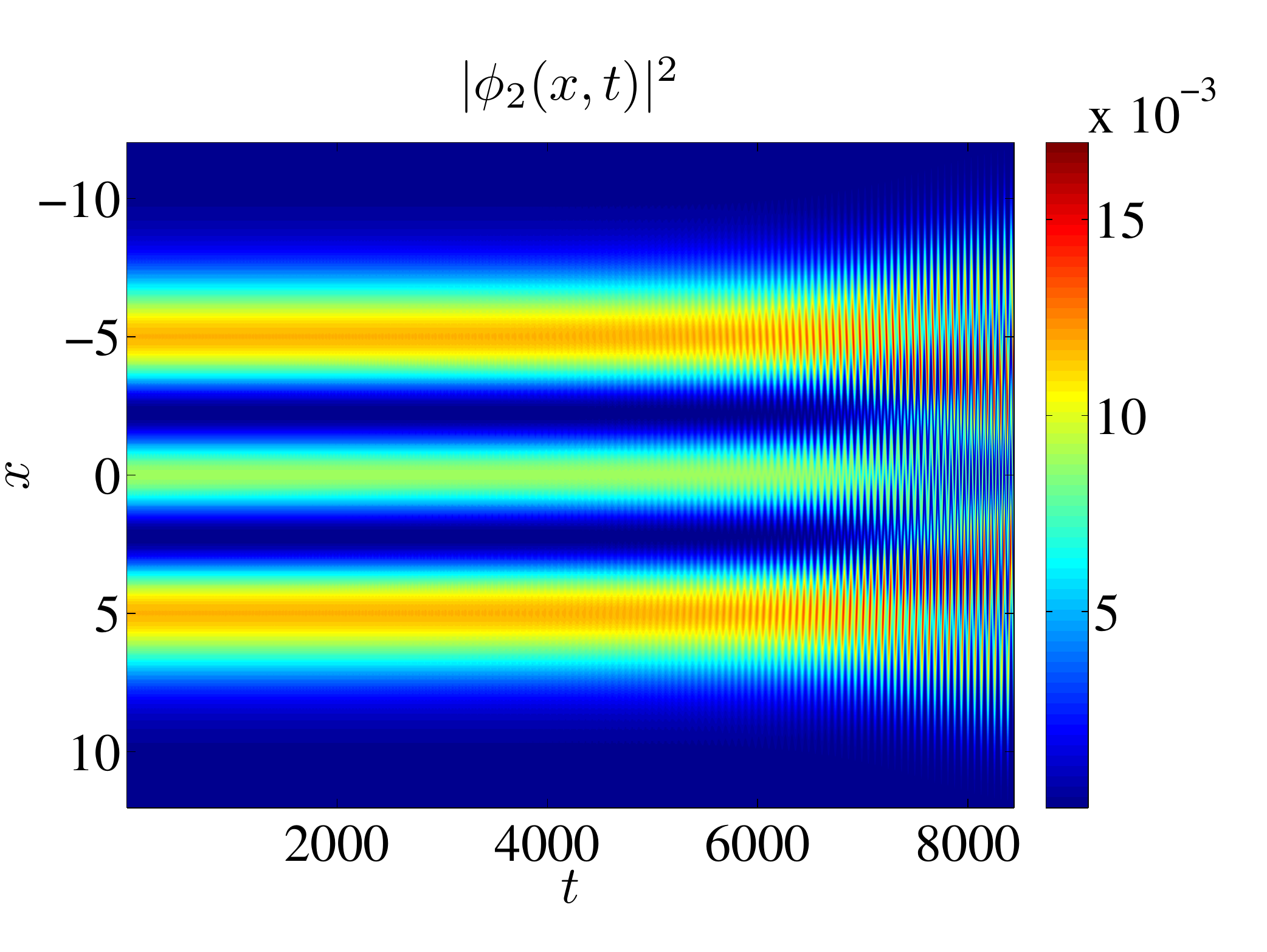}
\end{tabular}
\caption{ The top left (right) panel shows an example of the dynamics of $|\phi_1|^2$
($|\phi_2|^2$) with $a=b=1, g_{11}=1, g_{12}=0.1, g_{22}=1$, $\Omega=0.1$ and $\epsilon=0.1$. Over
longer time scales, the oscillatory instability
sets in re-arranging the atomic distribution in both components.}
\label{fig1_6_2}
\end{figure}

If $a=b=1, g_{11}=1, g_{12}=0.1, g_{22}=1, \Omega=0.1$ we
provide the relevant comparison of analytical predictions
and numerically computed eigenvalues in Fig.~\ref{fig1_6}.
In this case, we identify three quartets of unstable eigenvalues (one near
$\pm i\Omega$ and two near $\pm i2\Omega$). As $\epsilon$ increases, we see
that the complex eigenvalues near $-i\Omega$ (at $\epsilon\approx 5.2$) and
the ones near $-i2\Omega$ (at $\epsilon\approx 1.2$ and $\epsilon\approx 4.9$)
will return to the imaginary axis and split along it as shown in
Fig.~\ref{fig1_6_1}. Additionally, the split eigenvalues from $-i\Omega$ and
$-i2\Omega$ going upward will collide with the eigenvalues from $-i3\Omega$
and $-i4\Omega$, respectively, to generate new eigenvalues with nonzero real
part, a feature illustrated in the extended parametric continuation of
Fig.~\ref{fig1_6_1}. In Figure~\ref{fig1_6_2}, we illustrate the numerical
evolution of this unstable configuration $\epsilon=0.1$. {The instability
settles in an oscillatory manner after a long time evolution, redistributing
the atoms within the condensate and resulting in the recurrence of different states. We note that this is a more complicated case than the breathing case in Figure~\ref{fig1_2_4} since there are more possible unstable eigendirections in this case.}

\subsection{Summary of the spectral stability results}
\label{section:summary-stability}

In Section~\ref{section:stability-catologue}, we have examined solutions with different $(m, n)$ pairs for $0\leq m, n \leq
2$. The spectral stability of each solution has been studied both
analytically and topologically via perturbation theory for small $\epsilon$.
The topological results are robust, and valid for all pairs of $(m,n)$.
In order to upgrade the stability results for general $(m,n)$, we first introduce several identities for $m,n\ge 1$ as follows:
\begin{eqnarray}
\label{eqn_claim_1_1}
\alpha_m:=
\frac{ A_m - 2B_{m,m+1} }{ D_{m,m,m-1,m+1} }=\frac{D_{m,m,m-1,m+1}}{ A_m - 2B_{m,m-1} }=-\sqrt{\frac{m}{m+1}}
\end{eqnarray}
\begin{eqnarray}
\label{eqn_claim_1_2}
\frac{D_{m,n,m+1,n+1}+\alpha_n\cdot D_{m,n,m+1,n-1}}{B_{m,n}-B_{m+1,n}} = \frac{D_{m,n,m-1,n+1}+\alpha_n\cdot D_{m,n,m-1,n-1}}{\alpha_m (B_{m,n}-B_{m-1,n})}
=\sqrt{\frac{m+1}{n+1}}
\end{eqnarray}
\begin{eqnarray}
\label{eqn_claim_1_3}
\frac{B_{0,n-1}-B_{0,n}}{D_{0,1,n-1,n}}=-\frac{1}{\alpha_n} \frac{B_{0,n}-B_{0,n+1}}{D_{0,1,n,n+1}} =
\frac{1}{\sqrt{n}},\quad
B_{0,n}=\frac{(2n-1)!}{(n-1)!\, n!\, 2^{2n-1}} \sqrt{\frac{\Omega}{2 \pi}}.
\end{eqnarray}
Though no analytical proofs for these identities are provided here, we have verified them for general $(m,n)$ pairs through extensive numerical experiments.
Then for the stability results:
\begin{itemize}
\item When $m>0$, there are four eigenvalues of $-iJ \mathcal{L}$ near
    $-i\ell\Omega$ for $1 \leq\ell \leq m$. According to the
    Hamiltonian-Krein index, at most two eigenvalues among four can have
    positive real parts (leading to instability). However, the perturbation
    calculation suggests that not all of the four eigenvalues can enter the
    complex plane for $m\ge 1$.
    \begin{remark}
    \label{remark_4_1} For $1\leq m\leq n$, it can be directly checked
    using (\ref{eqn_claim_1_1})--(\ref{eqn_claim_1_2}) that $M_a$ for
    $\ell=1$ has an eigenvalue $0$ with eigenvector $\displaystyle{\left(1,
    \sqrt{\frac{m}{m+1}}, \frac{b}{a}\sqrt{\frac{n+1}{m+1}},
    \frac{b}{a}\sqrt{\frac{n}{m+1}}\right)^\mathrm{T}}$.
    \end{remark}
    Thus, near $-i\ell \Omega$, two eigenvalues will always stay on the imaginary axis (one of them is $0$) and there are at most one pair of complex eigenvalues. We discussed previously the physical origin of the corresponding
(dipolar) symmetry removing the potential for one among the
pertinent instability eigendirections.
%

\item When $m<n$, there are three eigenvalues near $-i\ell\Omega$ for
    $m<\ell\leq n$. At most one pair of these eigenvalues will have nonzero
    real part. Similar to Remark~\ref{remark_4_1}, we particularly notice that one of eigenvalues near $-i\Omega$ will always be zero.
    \begin{remark}
    \label{remark_4_1a} For $m=0$ and $1\leq n$, it can be directly checked
    using (\ref{eqn_claim_1_1})--(\ref{eqn_claim_1_3}) that $M_b$ for
    $\ell=1$ has an eigenvalue $0$ with eigenvector
    $\displaystyle{\left(\frac{a}{b}\frac{1}{\sqrt{n}},
    \sqrt{\frac{n+1}{n}}, 1\right)^\mathrm{T}}$.
    \end{remark}


\item The Hamiltonian-Krein index, $K_\mathrm{Ham}=2(m+n)$, gives an upper
    bound for the number of pairs of eigenvalues that can leave the
    imaginary axis and bring about an instability. In the examined
    examples, this upper bound can be reached only when $m=0$. For $m>0$,
    the exact upper bound will be $2(m+n-1)$, given the presence of the
    symmetry/invariance associated with dipolar motion of the condensate
    removing one of the potentially unstable associated eigendirections

\item When $(m,n)=(0,1)$, an instability will arise if and only if
    $g_{12}>g_{22}>0$, i.e., the inter-component nonlinear interactions are
    stronger than the nonlinear interactions within the ``dark" species.
\end{itemize}

As $\epsilon$ grows away from $0$, we notice that the eigenvalue starting
from $-i\Omega$ can collide with the eigenvalues from $-i3\Omega$,
$-i5\Omega,\dots$ on the imaginary axis to generate eigenvalues with nonzero
real part. Similarly, the eigenvalue from $-i2\Omega$ can meet with the
eigenvalues from $-i4\Omega,\,-i6\Omega,\dots$ on the imaginary axis to
produce new pairs of eigenvalues with nonzero real part.
Our numerical results suggest that (given their respective
parities) eigenmodes at odd multiples of $\Omega$ interact with
other ones such and similarly even ones interact with even.
While our analysis does not lend itself to the
consideration of this wide parametric regime in $\epsilon$, numerical
computations reveal the corresponding potential (oscillatory) instabilities
and their customary restabilization for some interval of wider
parametric variations of
$\epsilon$.

\section{Conclusions \& Future Challenges}

In the present work, we illustrated the usefulness of Lyapunov-Schmidt
reductions, as well as of Hamiltonian-Krein index theory, in acquiring a
systematic understanding of bifurcations from the linear limit of the
multi-component system of atomic gases. Here, we have focused on the
two-component case, yet it should be evident from the analysis how general
multi-component cases will modify the specifics yet not the overall
formulation of the present setting. Once again, this mean-field limit may be
of somewhat limited applicability to the atomic case for very small atom
numbers (mathematically, squared $L^2$ norms), as there additional (quantum)
effects may skew the picture. Nevertheless, optical settings (with suitably
tailored refractive index profiles) can lend themselves to the analysis
presented herein. Moreover, and arguably more importantly, the
topological nature of the tools developed provides insights on the number of
potentially unstable eigendirections even far from the linear limit, where
the mean field model has been successfully used to monitor different
multi-component excited states, such as most notably e.g. dark-bright
solitons and their close relatives (such as dark-dark ones).  We have found a
number of surprising results in the process, such as the fact that $(0,1)$
states (involving one fundamental and one excited state) may be unstable
provided that inter- to intra-component interaction ratios are suitably
chosen. Another intriguing feature is that the presence of additional
symmetry (embedded in the dipolar motion inside the trap) may prevent
particular instability eigendirections from manifesting themselves.

It would be interesting to extend the present considerations
to spinor systems that are intensely studied over the past few
years in atomic experiments~\cite{kawueda,stampueda}. Additionally,
higher dimensional settings, both two-dimensional ones where
vortex-bright and related states have been devised~\cite{VB},
but also three-dimensional ones involving vortex-rings~\cite{rings} and
skyrmions~\cite{ruost} or related patterns would be especially interesting
to attempt to explore through this methodology, as traditionally
the complexity of such states limits the potential for
analytical results. Lastly, it does not escape us that an
equally interesting and analytically tractable (at least to some
degree) limit is that of large chemical potentials where the solitary
waves can be treated as particles. Developing a general theory
of that limit and connecting that with the low amplitude
limit presented herein, would be of particular interest.
This would also allow to showcase the connection between
the two tractable limits via numerical computations and to confirm the
robustness of the topological tools in revealing the potential
for instability while traversing the continuum from one to
the other limit. Such studies are currently in progress and
will be reported in future publications.

{\it Acknowledgements.}\\
P.G.K. gratefully acknowledges support from NSF-DMS-1312856. T.K. gratefully acknowledges support from Calvin College through a Calvin Research Fellowship.

\end{document}